\documentclass[useAMS,usenatbib]{mn2e}  
\usepackage {graphicx,subfig,aas_macros,float}
\newcommand{\equ}[1]{eq.~(\ref{eq:#1})}

\newcommand{\se}[1]{\S\ref{sec:#1}}
\newcommand{\fig}[1]{Fig.~\ref{fig:#1}}

\newcommand{\Fig}[1]{Figure~\ref{fig:#1}}

\newcommand{\tab}[1]{Table~\ref{tab:#1}}
\newcommand{\be}{\begin{equation}}
\newcommand{\ee}{\end{equation}}
\newcommand{\bea}{\begin{eqnarray}}
\newcommand{\eea}{\end{eqnarray}}

\newcommand{\no}{\noindent}

\newcommand{\msun}{{\rm M}_\odot}
\newcommand{\Msun}{M_\odot}

\newcommand{\ifm}[1]{\relax\ifmmode#1\else$\mathsurround=0pt #1$\fi}
\newcommand{\kms}{\ifmmode\,{\rm km}\,{\rm s}^{-1}\else km$\,$s$^{-1}$\fi}

\newcommand{\kpc}{\,{\rm kpc}}
\newcommand{\pc}{\,{\rm pc}}
\newcommand{\Gyr}{\,{\rm Gyr}}

\newcommand{\Myr}{\,{\rm Myr}}

\newcommand{\ltsima}{$\; \buildrel < \over \sim \;$}
\newcommand{\lsim}{\lower.5ex\hbox{\ltsima}}
\newcommand{\gtsima}{$\; \buildrel > \over \sim \;$}
\newcommand{\gsim}{\lower.5ex\hbox{\gtsima}}

\def\la{\langle}

\def\omm{\Omega_{\rm m}}
\def\oml{\Omega_{\Lambda}}

\def\sy{\,M_\odot\, {\rm yr}^{-1}}

\def\cmc{\,{\rm cm}^{-3}}

\def\Mv{M_{\rm v}}
\def\Rv{R_{\rm v}}

\def\Mg{M_{\rm g}}

\def\Md{M_{\rm d}}
\def\Mc{M_{\rm c}}
\def\Rc{R_{\rm c}}
\def\Sc{S_{\rm c}}
\def\nc{n_{\rm c}}
\def\tc{t_{\rm c}}
\def\tcm{t_{\rm c,\,max}}
\def\Rd{R_{\rm d}}
\def\Hd{H_{\rm d}}

\def\Ms{M_*}

\def\td{t_{\rm dyn}}

\def\Fn{F_{\rm N}}
\def\Fw{F_{\rm W}}
\def\delrho{{\delta}_{\rm {\rho}}}
\def\delmin{{\delta}_{\rm {\rho}}^{\rm min}}

\def\rd0{r_{{\rm d}0}}
\def\Vdi0{V_{{\rm d}0}}

\def\vela{{\texttt{VELA}\ }}

\def\fg{ {\rm f}_{\rm g} }

\newcommand{\Halpha}{H${\alpha}$}

\usepackage{color}

\begin{document} 

\large 

\title[Clumps with and without Radiative Feedback]
{Giant Clumps in Simulated High-z Galaxies: Properties, Evolution and Dependence on Feedback}
%{The Lifetimes and Properties of Giant Clumps in Simulations with and without Radiative Feedback}

\author[Mandelker et al.] 
{\parbox[t]{\textwidth} 
{ 
Nir Mandelker$^1$\thanks{E-mail: nir.mandelker@mail.huji.ac.il },
Avishai Dekel$^1$, %\thanks{E-mail: dekel@phys.huji.ac.il},
Daniel Ceverino$^2$, 
Colin DeGraf$^{1,3}$, 
Yicheng Guo$^4$, 
Joel Primack$^5$
} 
\\ \\  
$^1$Center for Astrophysics and Planetary Science, Racah Institute of Physics, The Hebrew University, Jerusalem 91904, Israel\\ 
$^2$Universit${\ddot{a}}$t Heidelberg, Zentrum f${\ddot{u}}$r Astronomie, Institut f${\ddot{u}}$r Theoretische Astrophysik, Albert-Ueberle-Str. 2, 69120 Heidelberg, Germany\\
$^3$Institute of Astronomy and Kavli Institute for Cosmology, University of Cambridge, Madingley Road, Cambridge CB3 0HA, UK\\
$^4$UCO/Lick Observatory, Department of Astronomy and Astrophysics, University of California, Santa Cruz, CA, USA\\
$^5$Department of Physics, University of California, Santa Cruz, CA 95064, USA} 
\date{} 
 
\pagerange{\pageref{firstpage}--\pageref{lastpage}} \pubyear{0000} 
 
\maketitle 
 
\label{firstpage} 
 
\begin{abstract} 
We study the evolution and properties of giant clumps in high-$z$ disc galaxies using AMR cosmological simulations at redshifts $z\sim 6-1$. Our sample consists of 34 galaxies, of halo masses $10^{11}-10^{12}\msun$ at $z=2$, run with and without radiation pressure (RP) feedback from young stars. While RP has little effect on the sizes and global stability of discs, it reduces the amount of star-forming gas by a factor of $\sim 2$, leading to a similar decrease in stellar mass by $z\sim 2$. Both samples undergo extended periods of violent disc instability (VDI) continuously forming giant clumps of masses $10^7-10^9\msun$ at a similar rate, though RP significantly reduces the number of long-lived clumps (LLCs). When RP is (not) included, clumps with circular velocity $\lsim 40\:(20) \kms$, baryonic surface density $\lsim 200\:(100) \msun\pc^{-2}$ and baryonic mass $\lsim 10^{8.2}\:(10^{7.3}) \msun$ are short-lived, disrupted in a few free-fall times. More massive and dense clumps survive and migrate toward the disc centre over a few disc orbital times. In the RP simulations, the distribution of clump masses and star-formation rates (SFRs) normalized to their host disc is similar at all redshifts, exhibiting a truncated power-law with a slope slightly shallower than $-2$. The specific SFR (sSFR) of the LLCs declines with age as they migrate towards the disc centre, producing gradients in mass, stellar age, gas fraction, sSFR and metallicity that distinguish them from the short-lived clumps which tend to populate the outer disc. Ex situ mergers comprise $\sim 37\%$ of the mass in clumps and $\sim 29\%$ of the SFR. They are more massive and with older stellar ages than the in situ clumps, especially near the disc edge. Roughly half the galaxies at redshifts $z=4-1$ are clumpy, with $\sim 3-30\%$ of their SFR and $\sim 0.1-3\%$ of their stellar mass in clumps.
\end{abstract} 
 

%%%%%%%%%%%%%%%%%%%%%%%%%%%%% 
\section{Introduction} 
\label{sec:intro}
\smallskip
The peak phase of galaxy formation is at redshifts $z=1-4$, 
when star-formation is at its peak and most of the mass 
is assembled into galaxies \citep{Madau98,hopkins06,Madau14}. 
Observations of massive star-forming galaxies (SFGs) of 
$\sim 10^{11}\msun$ in baryons at this epoch reveal high 
star-formation rates (SFRs) of order $100\sy$ \citep{Genzel06,
Forster06,Elmegreen07,Genzel08,Stark08}. A large fraction of 
these galaxies have been spectroscopically confirmed to be 
rotating discs \citep{Genzel06,Shapiro08,Forster09,wisnioski15}. 
They are perturbed, thick and turbulent, with high velocity 
dispersions of $\sigma\sim 20-80 \kms$, and low rotation to 
dispersion ratios of $V/{\sigma} \sim 2-7$, as opposed to 
$10-20$ in today's spiral galaxies \citep{Elmegreen05b,Genzel06,
Forster06,Forster09,Cresci09}. Estimates of their gas fractions 
from CO measurements are in the range $0.2-0.8$ \citep{Daddi10,
Tacconi10,Tacconi13,Genzel15}, much higher than the 
fractions of $0.05-0.1$ in today's discs \citep{Saintonge11}. Many of 
these galaxies exhibit irregular morphologies, in both rest-frame 
UV and rest-frame optical emission \citep{Genzel08,Forster09,Forster11b},
with much of the UV light concentrated in a few large ``clumps", each 
a few percent of the disc mass and on the order of a $\kpc$ in size, 
much larger than the star-forming complexes observed in local galaxies. 
Recent observations of over 3000 galaxies in the \textit{Cosmic Assembly 
Near-IR Deep Extragalactic Legacy Survey} (CANDELS, \citealp{Grogin11,
Koekemoer11}) reveal that roughly $60\%$ of the SFGs at $z\gsim 2$ are 
clumpy, with clumps accounting for $\sim 20-40\%$ of their UV light and 
$\gsim 10\%$ of their SFR \citep{Guo15}.

\smallskip
According to our developing theoretical framework of high 
redshift galaxy formation, these SFGs are in a perpetual 
state of violent-disc-instability (VDI, \citealp{DSC}). 
Intense inflow of cold gas in narrow streams along the 
filaments of the cosmic web \citep{bd03,Keres05,db06,
Ocvirk08,Dekel09} maintains high gas fractions throughout 
the disc, repleneshing losses to star-formation, outflows, 
and compaction events which drive large amounts of gas into 
the disc centre \citep{Bournaud11,Forbes13,DekelBurkert14,
Zolotov15,Tacchella16a,Tacchella16b}. The high gas fraction 
together with the high density in the early Universe, leads 
to a violent gravitational disc instability \citep{toomre64}, 
which involves giant clumps and operates on short, orbital 
timescales \citep{DSC,CDB}, as opposed to the slow, secular 
instability in today's discs. 

\smallskip
The basic idea, summarized in \citet{DSC}, is that during VDI 
the high surface density of gas and ``cold" young stars, $\Sigma$, 
drives the Toomre $Q$ parameter below unity, $Q\sim \sigma\Omega/
(\pi G\Sigma)\lsim 1$, where $\sigma$ is the one-dimensional velocity 
dispersion and $\Omega$ is the angular frequency, a proxy to the 
epicyclic frequency $\kappa$, which is related to  the potential 
well \citep{toomre64}. It has been established that under such 
conditions the disc will fragment and produce large star-forming 
clumps. This has been shown using idealized simulations of isolated 
galaxies \citep{Noguchi99,Gammie01,Immeli04a,Immeli04b,Bournaud07,
Bournaud09,Elmegreen08,Hopkins12a}, as well as cosmological simulations 
\citep{Agertz09,CDB,Ceverino12,Genel12a,M14,Oklopcic16}. The ratio of clump mass 
to the mass of the cold disc scales as $\Mc/\Md\propto \delta^2$, where 
$\delta= \Md/M_{\rm tot}$ is the ratio of the cold disc mass to the total 
mass within the disc radius, which includes the bulge and dark matter halo 
\citep[e.g.][]{DSC}. This leads to much larger clumps at $z\sim 2$ than the 
low-redshift giant molecular clouds (GMCs). Gravitational interactions 
in the perturbed disc drive turbulence causing the disc to self-regulate in 
a marginally stable state with $Q\lsim 1$ \citep{DSC,CDB,krumholz_burkert10,
Cacciato12,Forbes12,Forbes13} that can last for more than a Gyr so long as 
the accretion is not interrupted. Some recent works have called into question 
the validity of linear Toomre analysis in the context of these highly non-linear 
galaxies \citep{Behrendt15,Tamburello15,Inoue16} and others have suggested 
alternate fragmentation mechanisms related to turbulence \citep[e.g.][]{Hopkins13}. 
However, since clump formation is largely determined by the balance between self-gravity, 
turbulent pressure and the centrifugal force, the largest clumps are always roughly 
at the Toomre scale. Larger clumps would be disrupted due to the shear and/or tidal 
forces within the disc, or would not collapse in the first place due to the centrifugal 
force. Therefore, regardless of the full validity of linear Toomre analysis, it is 
plausible that the Toomre $Q$ parameter can serve as a crude criterion for instability, 
possibly with a critical value that is larger than unity.

\smallskip
If clumps survive stellar feedback, their large masses cause them 
to migrate to the galactic centre on short timescales of $\sim 2-3$ 
orbital times at the disc edge \citep{DSC,Ceverino12,Bournaud14,M14}. 
This was proposed as a mechanism for the formation of galactic spheroids, 
in parallel with the traditional merger scenario \citep{Bournaud07,
Elmegreen08,Genzel08,DSC,Ceverino15a} and the fueling of a central AGN 
\citep{Bournaud11,Bournaud12}. However, simulations show that most 
of the gas inflowing to the galactic centre is in fact inter-clump 
gas \citep[e.g.][]{Hopkins12a,Zolotov15}. Indeed, wet inflows to the 
galactic centre induced by gravitational torques are a generic feature 
of VDI, not necessarily associated with clump migration 
\citep{krumholz_burkert10,Forbes12,Forbes13,DekelBurkert14}. While clump 
migration can contribute to the dry growth of the bulge, cosmological 
simulations indicate that this contribution is limited to $\lsim 25\%$ of 
the bulge mass at $z\sim 2$ \citep{Zolotov15}. Nevertheless, the giant clumps 
themselves, their properties, and their evolution, are all major observational 
indicators of VDI. Clump migration is related to the overall inflow, and if 
signatures of this migration are observed, they would indicate the validity of 
VDI and VDI-driven gas inflow. 

\smallskip
There is much debate regarding the ability of these giant clumps to 
survive feedback. Based on rather crude arguments, massive clumps of 
$\Mc\gsim 10^8\msun$ are not expected to be disrupted by supernova 
feedback alone \citep{DSC,CDB,Ceverino12,Bournaud14,M14}. However, 
winds are observed from giant high-$z$ clumps \citep{Genzel11,Newman12}, 
with mass loading factors of order unity. If these outflows are very 
intense on timescales of $\sim 1-2$ free fall times, they could lead 
to significant mass loss and maybe even disruption of the clump 
\citep{Genel12a,Hopkins12a}. \citet{murray10} argued that the high-z 
clumps are likely to be disrupted by momentum driven feedback, similar 
to GMCs at low redshift, but \citet{KrumholzDekel} showed that this was 
unlikely, unless the star-formation efficiency per free fall time in the 
clumps is much higher than observed locally, which does not appear to be 
the case \citep{Freundlich13}. Furthermore, \citet{DK13} argued that a 
steady wind generated by radiation pressure was not expected to unbind 
the clump before it had turned most of its mass into stars. Their argument 
was based on simulations by \citet{KT12,KT13} which showed that radiation 
trapping within the giant clumps is negligible because it destabilizes the 
wind. Thus, the total momentum injection from radiation pressure, stellar 
winds and supernovae is limited to $\lsim 5L/c$, significantly lower than 
the values of $\sim 20-30 L/c$ advocated by \citet{Hopkins12a}.

\smallskip
In previous work, we studied the formation and evolution of giant clumps at 
redshifts $z=1-4$ in the weak feedback limit \citep[][hereafter M14]{M14}, 
using a large suite of 29 cosmological zoom-in simulations of $\sim 10^{12}\msun$ 
haloes at $z\sim2$, run with adaptive mesh refinement (AMR) with a maximal resolution 
of $35-70\pc$. Using a three-dimensional clump finder on the gas, we identified 
clumps in the simulations, distinguishing in situ clumps formed during VDI from 
ex situ clumps that joined the disc as minor mergers. The simulations included 
only thermal feedback from supernovae and stellar winds, under which the massive 
clumps survived and migrated towards the disc centre. Clump evolution during 
migration generated gradients in clump properties across the disc, so that 
clumps closer to the centre tended to be more massive, with older stellar 
populations, lower specific SFR (sSFR), lower gas fractions and higher metalicities. 
Observational indications for similar gradients \citep{Forster11b,Guo12} provided 
support for the survival of massive clumps. 

\smallskip
In a preliminary attempt to study the effect of radiation pressure (RP) feedback 
on clump survival, we examined a small subset of the simulations used in this work 
\citep{Moody14}, consisting of 8 pairs of cosmological zoom-in simulations of 
$\sim (2-5)\times10^{11}\msun$ haloes at $z\sim2$, with a maximal AMR resolution 
of $17-35\pc$. Each pair consisted of two simulations run from the same 
initial conditions, one implementing only thermal feedback from supernovae and 
stellar winds and the second adding a model for RP feedback from UV photons, with 
no IR trapping \citep{Ceverino14}. Using a two-dimensional clump finder on stellar 
mass maps, we found that the number of clumps with stellar mass less than $5\%$ 
of the disc stellar mass was drastically reduced with the inclusion of RP, but that 
more massive clumps were largely unaffected. We also found that clump counts 
in mock HST observations of the RP simulations roughly matched observational 
estimates. However, this study had several limitations. Firstly, we did not 
account for the significant gas content of clumps, studying only the stellar 
clumps. Second, while identifying clumps in two-dimensional stellar mass maps 
is closer to what is done observationally, it limits the analysis of the 
three-dimensional physical properties of the clumps and the role played by 
RP in their formation, evolution and disruption. 

\smallskip
In this work, we expand on M14 and \citet{Moody14}. We use a large suite of 
34 cosmological zoom-in simulations of galaxies, most evolved to $z\sim 1$, whose 
halo masses are in the range $\sim 10^{11}-10^{12}\msun$ at $z\sim 2$. These 
34 simulations include the same model for RP feedback, in addition to supernova 
feedback and stellar winds, as in \citet{Ceverino14} and \citet{Moody14}, and 29 of 
them have counterparts run from the same initial conditions without the inclusion 
of RP. We identify clumps in three-dimensions, in both gas and stars, and study their 
physical properties. By tracking clumps through time using their stellar particles, 
we study the effect of RP feedback on the formation and evolution of giant clumps and 
find the conditions leading to clump disruption or survival. Our two main goals are to 
asses theoretically the conditions for clump survival when RP is included, and to find 
observational tests that can distinguish short-lived clumps from long-lived clumps and 
place constraints on models of stellar feedback and clump evolution.

% Outline:
\smallskip
This paper is organized as follows: In \se{sim} we 
provide an overview of the simulations, including their 
limitations. In \se{analysis} we describe our analysis 
method and how we define galaxy and clump properties. 
In \se{gal_comp} we compare global properties of the 
two galaxy samples, with and without radiative feedback. 
In \se{clump_comp} we address the effect of radiative 
feedback on clumps. In \se{obs} we address properties 
of clumps, observational signatures of clump migration 
and disc clumpiness. We discuss our results in \se{disc} 
and summarize our conclusions in \se{conc}.

%%%%%%%%%%%%%%%%%%%%%%%%%%%%%%%%%%%%%%%%%%%%%%%%  
\section{Simulations} 
\label{sec:sim} 
\smallskip
We use zoom-in hydro-cosmological simulations of 34 
moderately massive galaxies that comprise the \vela 
simulation suite \citep{Ceverino14,Zolotov15}. The 
fiducial sample includes an implementation of RP 
feedback from young stars (RP simulations, hereafter 
RPs), and 29 of the 34 galaxies were resimulated without 
RP (NoRP simulations, hereafter NoRPs), but with all 
other sub-grid physics unchanged. Several recent works 
have used the \vela suite to study a variety of issues 
relating to high redshift galaxy formation. These include 
the stellar to halo mass relation \citep{Moody14}; galaxy 
clumpiness and morphological evolution \citep{Moody14,Snyder15}; 
the evolution of galaxy shapes \citep{Ceverino15b,Tomassetti16}; 
the link between metal inhomogeneities and gas inflows 
\citep{Ceverino16a}; galaxy outflows \citep{Ceverino16b}; 
compaction, quenching and the formation of blue and red 
nuggets \citep{Zolotov15} and their relation to the evolution 
of density profiles \citep{Tacchella16a} and the confinement 
of the main sequence of star formation \citep{Tacchella16b}; 
and the relation between clump formation and Toomre instability 
\citep{Inoue16}. In this section, we give an overview of the key 
aspects of the simulations. Further details regarding the numerical 
methods can be found in \citet{Ceverino14}.

%--------------------------
\subsection{Simulation Method and Sub-Grid Physics}
\label{sec:art}
\smallskip
The \texttt{VELA} simulations utilize the Adaptive Refinement 
Tree (ART) code \citep{kravtsov97, kravtsov03, ceverino09}, 
which accurately follows the evolution of a gravitating 
$N$-body system and the Eulerian gas dynamics, with an 
AMR maximum resolution of $17.5-35~\mathrm{pc}$ in 
physical units at all times\footnote{The minimum 
cell size is set to $17.5\pc$ in physical units at expansion 
factor $a=0.16$ ($z=5.25$). Due to the expansion of the 
whole mesh while the refinement level remains fixed, the 
minimum cell size grows in physical units, becoming $35\pc$ 
at $a=0.32$ ($z=2.125$). At this time we add a new level 
to the comoving mesh, so the minimum cell size becomes 
$17.5\pc$ again, and so on.}. The dark matter particle 
mass is $8.3\times 10^4 \msun$ and the stellar particles have 
a minimum mass of $10^3 \msun$, similar to the stellar mass of 
an Orion-like star cluster. Each AMR cell is split into 8 cells 
once it contains a mass in stars and dark-matter higher than 
$2.6\times 10^5 \msun$, equivalent to 3 dark-matter particles, 
or a gas mass higher than $1.5\times 10^6 \msun$. 
This quasi-Lagrangian strategy ends at the highest level of 
refinement that marks the minimum cell size at each redshift. 
We often refine based on stars and dark-matter particles rather 
than gas, so within the central halo and the star-forming disc 
the highest refinement level is reached for gas densities between 
$\sim 10^{-2}-100\cmc$, and occasionally for densities as low as 
$\sim 10^{-3}\cmc$. In the outer circumgalactic medium, near the 
halo virial radius, the median resolution is $\sim 500 \pc$.

\smallskip
Beyond gravity and hydrodynamics, the code incorporates the 
physics of gas and metal cooling, UV-background photoionization, 
stochastic star formation, gas recycling, stellar winds and 
metal enrichment, and thermal feedback from supernovae 
\citep{CDB,Ceverino12}, plus an implementation of feedback 
from radiation pressure \citep{Ceverino14}. A subsample of the 
full \texttt{VELA} suite, 29 of the 34 galaxies, was run using 
the same initial conditions and sub-grid physics, but without 
the implementation of radiative feedback.

\smallskip
We use the CLOUDY code \citep{Ferland98} to calculate the cooling 
and heating rates for a given gas density, temperature, metallicity, 
and UV background, assuming a slab of thickness 1 kpc. We assume a 
uniform UV background, following the redshift-dependent \citet{HaardtMadau96} 
model, except for gas densities higher than $0.1~\mathrm{cm}^{-3}$ where we 
use a substantially suppressed UV background ($5.9\times10^ 6~\mathrm{erg}~\mathrm{s}^{-1}~\mathrm{cm}^{-2}~\mathrm{Hz}^{-1}$) in order 
to mimic the partial self-shielding of dense gas. This allows dense gas to 
cool down to temperatures of $\sim300~\mathrm{K}$. The equation of state is 
assumed to be that of an ideal monoatomic gas. Artificial fragmentation on 
the cell size is prevented by introducing a pressure floor, ensuring 
that the Jeans scale is resolved by at least $N=7$ cells \citep{CDB}. 
The pressure floor is given by
\be 
\label{eq:Pfloor}
P_{\rm floor}=\frac{G \rho^2 N^2 \Delta^2}{\pi \gamma}
\ee
{\no}where $\rho$ is the gas density, $\Delta$ is the cell size, 
and $\gamma=5/3$ is the adiabatic index of the gas.

\smallskip
Star formation is allowed to occur at densities above a threshold 
of $1~\mathrm{cm}^{-3}$ and at temperatures below $10^4~\mathrm{K}$. 
Most stars ($>90~\%$) form at temperatures well below $10^3~\mathrm{K}$, 
and more than half of them form at $300~\mathrm{K}$ in cells where 
the gas density is higher than $10~\mathrm{cm}^{-3}$. New stellar 
particles are generated with a timestep of $dt_{\rm SF}\sim 5\Myr$. We 
implement a stochastic model, where the probability to form a stellar 
particle in a given timestep is 
\be 
\label{eq:PSF}
\textbf{P}={\rm min}\left(0.2,\:\sqrt{\frac{\rho_{\rm gas}}{1000\:\cmc}}\right)
\ee
{\no}In the formation of a single stellar particle, its mass is equal to 
\be 
\label{equ:mstar}
m_{*} = m_{\rm gas}\frac{dt_{\rm SF}}{\tau}\sim 0.42m_{\rm gas}
\ee
{\no}where $m_{\rm gas}$ is the mass of gas in the cell where the particle 
is being formed, and $\tau=12\Myr$ is a parameter of the simulations which 
was calibrated to match the empirical Kennicutt-Schmidt law 
\citep{Kennicutt98}. We assume a \citealt{Chabrier03} stellar 
initial mass function. Further details can be found in \citet{Ceverino14}.

\smallskip
The thermal stellar feedback model releases energy from stellar winds 
and supernova explosions as a constant heating rate over $40~\mathrm{Myr}$ 
following star formation. The heating rate due to feedback may or may 
not overcome the cooling rate, depending on the gas conditions in the 
star-forming regions \citep{Dekel86,ceverino09}, as we do not 
explicitly switch off cooling in these regions. The effect of runaway 
stars is included by applying a velocity kick of $\sim10~\kms$ to $30~\%$ 
of the newly formed stellar particles. The code also includes the later 
effects of Type Ia supernova and stellar mass loss, and it follows the 
metal enrichment of the ISM. 

\smallskip
In the fiducial (RP) models, radiation pressure is incorporated through 
the addition of a non-thermal pressure term to the total gas pressure in 
regions where ionizing photons from massive stars are produced and may 
be trapped. This ionizing radiation injects momentum in the cells 
neighbouring massive star particles younger than $5~\mathrm{Myr}$ 
whose column density exceeds $10^{21}~\mathrm{cm}^{-2}$, isotropically 
pressurizing the star-forming regions. The expression for the radiation 
pressure is 
\be 
\label{eq:RP}
P_{\rm rad} = \frac{\Gamma m_*}{R^2 c}
\ee
{\no}where $R$ is set to half the cell size for the cell hosting a 
stellar mass $m_*$, and to the cell size for its closest neighbours. 
The value of $\Gamma$ is taken from the stellar population synthesis 
code \texttt{STARBURST99} \citep{Starburst99}. We use a value of 
$\Gamma=10^{36}~{\rm erg~s^{-1}~\msun^{-1}}$ which corresponds to the 
time averaged luminosity per unit mass of the ionizing radiation 
during the first $5\Myr$ of the evolution of a single stellar population. 
After $5\Myr$, the number high mass stars and ionizing photons declines
signifcantly. Since the significance of radiation pressure also depends 
on the optical depth of the gas within a cell, we use a hydrogen column
density threshold of $N = 10^{21}~\mathrm{cm}^{-2}$, above which ionizing
radiation is effectively trapped and radiation pressure is added to the 
total gas pressure. See the ``RadPre" model of \citet{Ceverino14} for 
further details.

\smallskip
The initial conditions for the simulations are based on dark-matter 
haloes that were drawn from dissipationless N-body simulations at 
lower resolution in three comoving cosmological boxes (box-sizes of 
10, 20, and 40 Mpc/h). We assume the standard $\Lambda$CDM cosmology 
with the WMAP5 cosmological parameters, namely $\Omega_m=0.27$, 
$\Omega_{\Lambda}=0.73$, $\Omega_b=0.045$, $h=0.7$ and $\sigma_8=0.82$ 
\citep{WMAP5}. Each halo was selected to have a given virial mass 
at $z = 1$ and no ongoing major merger at that time. This latter 
criterion eliminates less than $10~\%$ of the haloes, which tend 
to be in dense proto-cluster environments at $z\sim1$. The target 
virial masses at $z=1$ were selected in the range 
$\Mv = 2\times10^{11}-2\times10^{12}~\Msun$, with a median 
of $5.6\times10^{11}~\Msun$. If left in isolation, the median 
mass at $z=0$ would be $\sim10^{12}~M_{\odot}$. In practice, 
the actual mass range is broader, with some of the haloes merging into 
more massive haloes that host groups at $z=0$.

%--------------------------
\subsection{Galaxy Samples}
\label{sec:sample}
\smallskip
More than half the sample was evolved with RP to $z\le 1$ and 
all but 5 were evolved to $z\le 2$. The NoRP runs were often 
stopped at higher redshifts, due to an overproduction of stars 
which slowed the simulations down considerably. The simulation 
outputs were stored and analyzed at fixed intervals in the 
cosmic expansion factor $a=(1+z)^{-1}$, $\Delta a=0.01$, which 
at $z=2$ corresponds to about $100\Myr$. \tab{sample} lists 
the final available snapshot for each of the 34 RP runs and the 
29 corresponding NoRP runs in terms of expansion factor, 
$a_{\rm fin}$, and redshift, $z_{\rm fin}$. We detect the main 
progenitor in the final available output by using the 
\texttt{AdaptaHOP} group finder \citep{Aubert04,Tweed09} on the 
stellar particles. It is then traced back in time until it contains 
fewer than 100 stellar particles, typically between $a=0.10-0.13$ 
($z=6.5-9$).

\smallskip
In \se{gal_comp} and \se{clump_comp} we directly compare the RP 
and NoRP samples using only snapshots available in both sets of 
simulations. In these cases, visual inspection of each snapshot 
is performed to ensure that the same galaxy is being traced in 
both samples. Occasionally, when there is a roughly equal mass 
merger, the automated algorithm may trace different progenitors 
in the RP and NoRP samples. Whenever this happens, we change the 
NoRP version to track the same galaxy as in the RP sample. In 
\se{obs} we analyze the RPs independently of their 
NoRP counterparts, making use of the full sample.

%--------------------------
\subsection{Limitations of the Current Simulations}
\label{sec:limitations}
\smallskip
The cosmological simulations used in this paper are state-of-the-art 
in terms of high-resolution AMR hydrodynamics and the treatment of key 
physical processes at the subgrid level, highlighted above. 
%These simulations trace the cosmological streams that feed galaxies at high redshift, including mergers and smooth flows \citep[e.g.][]{Danovich12,Danovich15,Goerdt15}, they resolve the VDI that governs high-$z$ disc evolution and bulge formation (M14; \citealp{Moody14, Ceverino15a, Inoue16}), they capture the formation of blue and red nuggets via compaction of the gaseous disc followed by quenching \citep{Zolotov15} and they reproduce observed gradients accross the star-forming main-sequence and changes in the surface density profiles of galaxies that are associated with compaction and quenching \citep{Tacchella16a,Tacchella16b}. \smallskip
However, like other simulations, they are not yet perfect in their treatment of 
the star formation and feedback processes. While the SFR recipe was calibrated 
to reproduce the KS relation and a realistic SFR efficiency per free fall time 
\citep{Ceverino14}, the code does not yet follow in detail the formation of 
molecules and the effect of metallicity on SFR \citep{KD12}. Additionally, the 
resolution does not allow the capture of Sedov-Taylor adiabatic phase of supernova 
feedback. The radiative stellar feedback assumed no infrared trapping, in the 
spirit of low trapping advocated by \citet{DK13} based on \citet{KT12,KT13}. 
Other works assume more significant trapping \citep{murray10,Hopkins12a,Hopkins12b}, 
which makes the strength of the radiative stellar feedback here lower than in other 
simulations\footnote{On the other hand, unresolved inhomogeneities and substructure 
within star-forming clumps could create low-density ``chimneys" from which photons 
would leak out, thus lowering the effective boost-factor. For reasonable parametrizations 
of the unresolved density distributions within typical star-forming regions and 
expected values of the cloud-averaged optical depth, photon-leakage could easily 
lower the effective trapping factor by a factor of $\gsim 2$ \citep[][appendix B]{Hopkins11}. 
Thus, while our model for RP feedback is not formally a ``lower limt", it is still 
weaker than what is assumed in other simulations, with both more significant trapping 
and no accounting of photon-leakage.}. 
Finally, AGN feedback and feedback associated with cosmic rays and 
magnetic fields are not yet incorporated. Nevertheless, the star formation rates, 
gas fractions, and stellar to halo mass ratios are all in the ballpark of the 
estimates deduced from abundance matching and provide a better match to observations 
than earlier simulations (\citealp{Ceverino14,Moody14} and \fig{Ms_Mh}).

\smallskip
The uncertainties, and any possible remaining mismatches by a factor of order 2, are 
comparable to the observational uncertainties. For example, the stellar-to-halo mass 
fraction is not well constrained observationally at $z\sim 2$. Recent estimates by 
\citet{Burkert15} (see their Fig. 5) based on the observed kinematics of $z \sim 0.6 - 2.8$ 
SFGs reveal significantly larger ratios than the estimates based on abundance matching 
\citep{Conroy09,Moster10,Moster13,Behroozi10,Behroozi13} at $\Mv < 10^{12} \msun$. 
In \se{gal_comp}, we present a detailed comparison of the stellar-to-halo mass relation 
in our simulations and the observational data (\fig{Ms_Mh}). We conclude that our 
simulations produce values that are in the ballpark of the observational estimates, 
and within the observational uncertainties.

\smallskip
If the SFR at very high redshifts is still overestimated, then this may 
lead to more clumps forming at earlier times when the gas fractions are 
still high, and less efficient clump formation at $z\lsim 2$ when 
gas fractions are underestimated. However, we find that the fraction 
of clumpy galaxies as a function of redshift roughly matches observations 
(\se{obs}). Additional sources of feedback, from infrared trapping and 
AGN, may affect the lifetimes and properties of clumps. However, our 
simulations with intermediate feedback have the advantage of forming both 
long-lived bound clumps and short-lived disrupting clumps in the same 
galactic environment. By studying these two populations, we deduce observable 
differences between them (\se{obs}), which will place tighter constraints on models for 
clump evolution and for feedback. Furthermore, \citet{Ceverino14} showed that 
photon trapping with boost factors of $\sim 3$ above the UV radiation, as 
advocated by \citet{DK13}, changes the final stellar mass by $\lsim 20\%$ 
and has only a minor effect on the density distribution of ISM gas. We therefore 
suspect that our current simulations capture the majority of the effect.

%%%%%%%%%%%%%%%%%%%%%%%%%%%%%%%%%%%%%%%%%%%%%%%%  
\section{Analysis} 
\label{sec:analysis} 

\begin{table*}
\centering
\begin{tabular}{@{}lcccccccccc}
\multicolumn{10}{c}{{\bf The \vela suite of 34 simulated galaxies with radiative feedback (RP)}} \\
\hline
Galaxy & $\Rv$ & $\Mv$ & $M_{\rm s,\,0.15\Rv}$ & $M_{\rm g,\,0.15\Rv}$ & ${\rm SFR}_{\rm 0.15\Rv}$ & $a_{\rm fin}$ & $z_{\rm fin}$ & $a_{\rm fin,\,NoRP}$ & $z_{\rm fin,\,NoRP}$ \\
  & $\kpc$ & $10^{12}~\Msun$ & $10^{10}~\Msun$ & $10^{10}~\Msun$ & $\sy$ &  &  &  & \\
\hline
\hline
01      & 58.25  & 0.16 & 0.22 & 0.20  & 2.65  & 0.50 & 1.00 & 0.50 & 1.00     \\
02      & 54.50  & 0.13 & 0.19 & 0.24  & 1.81  & 0.50 & 1.00 & 0.50 & 1.00     \\
03      & 55.50  & 0.14 & 0.42 & 0.15  & 3.72  & 0.50 & 1.00 & 0.50 & 1.00     \\
04      & 53.50  & 0.12 & 0.09 & 0.13  & 0.48  & 0.50 & 1.00 & ------ & ------ \\
05      & 44.50  & 0.07 & 0.09 & 0.11  & 0.58  & 0.50 & 1.00 & 0.50 & 1.00     \\
06      & 88.25  & 0.55 & 2.19 & 0.49  & 20.60 & 0.37 & 1.70 & 0.30 & 2.33     \\
07      & 104.25 & 0.90 & 6.23 & 1.62  & 25.86 & 0.54 & 0.85 & 0.35 & 1.86     \\
08      & 70.50  & 0.28 & 0.35 & 0.21  & 5.69  & 0.57 & 0.75 & 0.42 & 1.38     \\
09      & 70.50  & 0.27 & 1.07 & 0.43  & 3.93  & 0.40 & 1.50 & 0.39 & 1.56     \\
10      & 55.25  & 0.13 & 0.63 & 0.18  & 3.22  & 0.56 & 0.79 & 0.44 & 1.27     \\
11      & 69.50  & 0.27 & 0.92 & 0.70  & 14.59 & 0.46 & 1.17 & 0.46 & 1.17     \\
12      & 69.50  & 0.27 & 2.03 & 0.28  & 2.88  & 0.44 & 1.27 & 0.44 & 1.27     \\
13      & 72.50  & 0.31 & 0.78 & 0.90  & 13.88 & 0.51 & 0.96 & 0.51 & 0.96     \\
14      & 76.50  & 0.36 & 1.33 & 0.70  & 25.40 & 0.42 & 1.38 & 0.37 & 1.70     \\
15      & 53.25  & 0.12 & 0.55 & 0.15  & 1.57  & 0.56 & 0.79 & 0.56 & 0.79     \\
16 $^*$ & 62.75  & 0.50 & 4.27 & 0.67  & 20.26 & 0.24 & 3.17 & 0.24 & 3.17     \\
17 $^*$ & 105.75 & 1.13 & 8.97 & 1.55  & 64.82 & 0.31 & 2.23 & ------ & ------ \\
19 $^*$ & 91.25  & 0.88 & 4.52 & 0.88  & 40.78 & 0.29 & 2.45 & 0.26 & 2.85     \\
20      & 87.50  & 0.53 & 3.84 & 0.62  & 7.15  & 0.44 & 1.27 & ------ & ------ \\
21      & 92.25  & 0.62 & 4.21 & 0.68  & 9.50  & 0.50 & 1.00 & 0.41 & 1.44     \\
22      & 85.50  & 0.49 & 4.50 & 0.32  & 12.07 & 0.50 & 1.00 & ------ & ------ \\
23      & 57.00  & 0.15 & 0.82 & 0.24  & 3.28  & 0.50 & 1.00 & 0.50 & 1.00     \\
24      & 70.25  & 0.28 & 0.92 & 0.42  & 4.31  & 0.48 & 1.08 & ------ & ------ \\
25      & 65.00  & 0.22 & 0.73 & 0.13  & 2.30  & 0.50 & 1.00 & 0.50 & 1.00     \\
26      & 76.75  & 0.36 & 1.61 & 0.40  & 9.63  & 0.50 & 1.00 & 0.50 & 1.00     \\
27      & 75.50  & 0.33 & 0.83 & 0.61  & 7.92  & 0.50 & 1.00 & 0.46 & 1.17     \\
28      & 63.50  & 0.20 & 0.24 & 0.32  & 5.70  & 0.50 & 1.00 & 0.50 & 1.00     \\
29      & 89.25  & 0.52 & 2.56 & 0.49  & 18.49 & 0.50 & 1.00 & 0.36 & 1.78     \\
30      & 73.25  & 0.31 & 1.67 & 0.52  & 3.80  & 0.34 & 1.94 & 0.30 & 2.33     \\
31 $^*$ & 38.50  & 0.23 & 0.81 & 0.24  & 16.18 & 0.19 & 4.26 & 0.19 & 4.26     \\
32      & 90.50  & 0.59 & 2.71 & 0.56  & 14.89 & 0.33 & 2.03 & 0.26 & 2.85     \\
33      & 101.25 & 0.83 & 5.04 & 0.63  & 32.68 & 0.39 & 1.56 & 0.30 & 2.33     \\
34      & 86.50  & 0.52 & 1.66 & 0.62  & 14.66 & 0.35 & 1.86 & 0.29 & 2.45     \\
35 $^*$ & 44.50  & 0.23 & 0.59 & 0.35  & 23.60 & 0.22 & 3.55 & 0.22 & 3.55     \\ \hline
\end{tabular}
\caption{Quoted are the virial radius, $\Rv$, the total virial 
mass, $\Mv$, the stellar mass, $\Ms$, the gas mass, $\Mg$ and 
the star formation rate, SFR, for the 34 \vela simulations 
with RP. $\Ms$, $\Mg$ and SFR are quoted within $0.15\Rv$. 
Also listed are the final simulation scale factor, $a_{\mathrm{fin}}$, 
and redshift, $z_{\mathrm{fin}}$, as well as the final scale 
factor and redshift of the NoRP counterpart if one exists, 
$a_{\rm fin,\,NoRP}$ and $z_{\rm fin,\,NoRP}$. All physical 
properties are quoted at $z=2$, except for the five cases 
marked $^*$, where they are quoted at the final simulation 
output, $z_{\rm fin}>2$.}
\label{tab:sample}
\end{table*}

\smallskip
In this section we describe how physical properties of the virial 
haloes, the galactic discs and the clumps are defined. We also 
describe the clump finding algorithm, the method for tracking 
clump histories and the distinction between in situ, ex situ and 
bulge clumps.

%--------------------------
\subsection{Physical Quantities}
\label{sec:measure}
\smallskip
The virial mass, $\Mv$, is the total mass within a sphere of radius 
$\Rv$ that encompasses an overdensity of 
$\Delta(z)=(18\pi^2-82\oml(z)-39\oml(z)^2)/\omm(z)$, where $\oml(z)$ 
and $\omm(z)$ are the cosmological parameters at $z$ \citep{Bryan98}. 
The virial properties for the 34 RP galaxies are listed in \tab{sample}, 
together with the stellar mass, $\Ms$, gas mass, $\Mg$, and star-formation 
rate, SFR, within $0.15\Rv$. These are quoted at $z=2$ except for the 5 
galaxies that were stopped at higher redshift, marked by $*$, for 
which we quote the properties at $z=z_{\rm fin}$. 

\smallskip
The stellar mass, $\Ms$, is the instantaneous mass in stars, after 
accounting for stellar mass loss. The simulation calculates stellar 
mass loss using an analytic fitting formula, where $10\%$, $20\%$ 
and $30\%$ of the initial mass of a stellar particle is lost after 
$30\Myr$, $260\Myr$ and $2\Gyr$ respectively. We refer to the birth 
mass of stars, \textit{without} accounting for stellar mass loss, as 
$M_{\rm *,\,i}$.

\smallskip
The SFR is obtained by 
${\rm SFR}=\langle M_{\rm *,\,i}(t_{\rm age}<t_{\rm max})/t_{\rm max} \rangle_{t_{\rm max}}$, 
where $M_{\rm *,\,i}(t_{\rm age}<t_{\rm max})$ is the mass at birth in stars younger than 
$t_{\rm max}$. The average $\langle\cdot\rangle_{t_{max}}$ is obtained for $t_{\rm max}$ in 
the interval $[40,80]~\Myr$ in steps of 0.2 Myr in order to reduce fluctuations due to the 
$\sim5 \Myr$ discreteness in stellar birth times in the simulation. The $t_{\rm max}$ in 
this range are long enough to ensure good statistics. This represents the SFR on $\sim 60\Myr$ 
timescales, which is a crude proxy for \Halpha~ based SFR measurements, while UV based 
measurements are sensitive to stars younger than $\sim 100\Myr$. 

\smallskip
We define the specific SFR as ${\rm sSFR}=SFR/\Ms$\footnote{Note that this 
is not the inverse formation timescale of the stellar system, which is 
$SFR/M_{\rm *,\,i}$. However, since the mass loss rate is small compared 
to the SFR in our galaxies, this makes only a small difference.} 
and the gas fraction as $\fg=\Mg/(\Mg+\Ms)$. The stellar age of a system is 
defined as the instantaneous mass-weighted mean age of its stellar particles.

%--------------------------
\subsection{The Galactic Disc}
\label{subsec:disc}
\smallskip
We define a disc in each snapshot using gas with temperatures 
$T<1.5\times10^4~{\rm K}$ and stars younger than $100\Myr$. 
Varying these thresholds in the range $(1-5)\times10^4~{\rm K}$ and
$50-150\Myr$ has no significant impact on any of our results. 
We hereafter refer to this as the ``cold" component of the system. 
The disc plane is determined by the angular momentum vector of the 
cold component. This vector, along with the disc radius, $\Rd$, and 
height (half thickness), $\Hd$, are computed iteratively until all 
converge to within $5\%$, following the prescription in Appendix B 
of M14. Briefly, the disc radius, $\Rd$, contains $85\%$ of 
the cold mass within a cylinder of radius $r_0=0.15\Rv$ and height 
$h_0=1\kpc$. The disc thickness, $2\Hd$, contains $85\%$ of 
the cold mass within a cylinder of radius and height $r_1=h_1=\Rd$. 
The disc angular momentum is that of the cold material within a 
cylinder of radius $\Rd$ and height $\Hd$. We varied $r_0$ in the 
range $0.1-0.2\Rv$, $h_0$ in the range $0.5-1.5\kpc$ as well as 
the value of $\Hd$ from the previous iteration, and the cold mass 
fraction in the range $80-90\%$, and found no significant effect 
on our main results. See M14 for further details.

\smallskip
All gas within the final cylinder is assigned to the disc. However, 
a stellar particle is assigned to the disc only if it meets an 
additional kinematic criterion, whereby the $z$-component of its 
specific angular momentum (parallel to the disc angular momentum), 
$j_{\rm z}$, is higher than a fraction $f_J$ of the maximum specific 
angular momentum for the same orbital energy, $j_{\rm max}=|v|r$. Here, 
$|v|$ is the magnitude of the velocity of the stellar particle and $r$ 
is its cylindrical distance from the galactic centre. We adopt $f_J=0.7$ 
(\citealp{CDB}, M14). Whenever there is a well defined disc, the vast 
majority of the cold mass obeys this criterion (see \citealp{Ceverino15b} 
and \citealp{Tomassetti16} for the typical stellar mass and redshift when 
galaxies develop well defined discs).

%--------------------------
\subsection{Clump Analysis}
\label{sec:clump_find}
\smallskip
We identify clumps in 3D using a method similar to that of 
M14. The main difference is that in that work we only identified 
clumps in gas, while here we identify clumps in both the cold 
component and the stellar component. As was highlighted in M14 
and in \citet{Moody14}, there is typically not a one-to-one 
correspondence between gas and stellar clumps, making it important 
to study both populations. We briefly review the clump finding method 
here and refer the reader to M14 for further details. 

\begin{figure*}
\centering
\subfloat{\includegraphics[width =0.95 \textwidth]{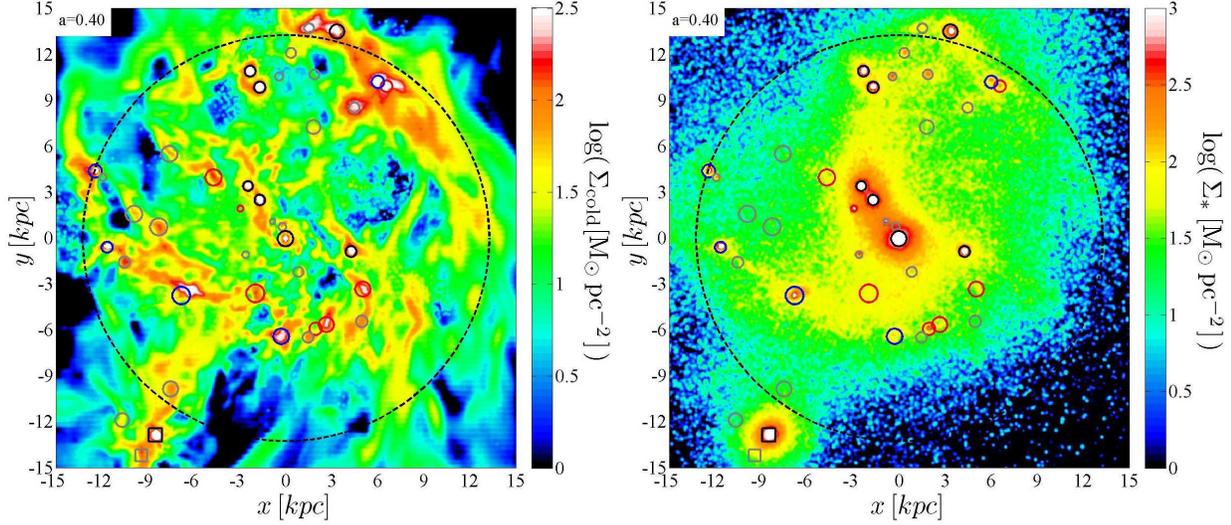}}
\caption{Clump identification. Shown is a face-on view of the RP version 
of V13 at $a=0.40$ ($z=1.5$). The disc radius, $\Rd \simeq 13.3\kpc$, is 
marked by a dashed circle. The integration depth is $\pm 2\Rd$. 
The left panel shows the surface density of the cold component 
and the right panel shows the stellar surface density. Note the 
different scale of the color bar in both panels. All clumps with 
baryonic densities $n_{\rm c}> 1\cmc$ and masses $\Mc > 10^7 \msun$ 
are marked in both panels, in situ clumps with circles and ex situ 
clumps with squares. The size of the marker shows the clump radius, 
$\Rc$. Clumps with masses $10^7<\Mc<10^{7.5}\msun$ are marked in 
grey, clumps with $10^{7.5}<\Mc<10^8\msun$ are marked in red, 
clumps with $10^8<\Mc<10^{8.5}\msun$ are marked in blue, and 
clumps with $10^{8.5}<\Mc$ are marked in black. The marked clumps 
agree with a by-eye inspection of the galaxy, and highlight the 
fact that clumps identified in the cold component (or in gas) 
do not necessarily match clumps identified in stars.
}
\label{fig:clump_finder} 
\end{figure*} 

% Clump finder
\subsubsection{Clump finder}
\smallskip
We detect clumps within a cube of side $L={\rm max}(4\Rd, 10\kpc)$ 
centred on the main galaxy. Using a cloud-in-cell (CiC) interpolation, 
we deposit mass in a uniform grid with a cell size of $\Delta=70\pc$, 
so between 2 and 4 times the maximal AMR resolution, whose $\textit{\textbf{z}}$ 
axis is alligned with the normal to the disc plane. We then smooth this 
grid with a spherical Gaussian filter whose full width at half maximum 
(FWHM) is $\Fw={\rm min}(2.5\kpc, 0.5\Rd)$, and calculate 
the density residual at each point 
$\delrho = (\rho - \rho_{\rm W})/\rho_{\rm W}$, where 
$\rho$ and $\rho_{\rm W}$ are the unsmoothed and smoothed density 
fields respectively. We do this once for the cold mass and once 
for the stellar mass, and at each point take the maximum of the 
two residual values. We define clumps as connected regions 
containing at least 8 grid cells, corresponding to between 64 and 
512 cells at the maximal AMR refinement, above a residual threshold 
of $\delmin=10$. This technique selects regions 
that are at least 10 times denser than their local surroundings 
in cold and/or stellar mass. 
%The average density residual among clump cells, ${\bar {\delta}}_{\rm c}\equiv \left<\delrho\right>_{\rm c}$, can be used to distinguish compact clumps from diffuse false positives (see \se{clump_comp}, and also figure 5 in M14). 
A discussion on the sensitivity of our results to variations in 
the parameters of the algorithm, $\Delta$, $\Fw$ and $\delmin$, 
is provided in \se{params}.

% Mass, centre and radius
\smallskip
Gas, stars and dark matter particles are deposited in the clump cells using 
CiC interpolation, and no attempt was made to remove unbound material from 
the clumps. We remove from the sample all clumps with baryonic mass $\Mc<10^6\msun$. 
Hereafter, when we discuss clump mass we refer to the baryonic mass unless 
explicitly stated otherwise. 

\smallskip
We define the clump centre as the baryonic density peak and the clump radius, 
$\Rc$, as the radius of a sphere with the same volume as the clump: 
$(4\pi/3)\Rc^3=N\times(70\pc)^3$, where $N\ge 8$ is the total number of grid 
cells within the clump. This translates to a minimum clump diameter of 
$\sim 180 \pc$, which is on the order of 7 cells at the maximal AMR resolution 
of $\sim 25\pc$. 
Since the pressure floor ensures the Jeans length is always resolved by at least 
7 cells, this is the minimal clump size where we could in principle have a complete 
sample.

% Shape
\smallskip
We assign a shape parameter to each clump, $S_{\rm c}$, using its inertia tensor, 
$I_{\rm i,j} = \sum_{n=1}^{N} m_{\rm n}(r_{\rm n}^2\delta_{\rm i,j} - r_{\rm n,i}r_{\rm n,j})$. 
The sum is over all cells in the clump, $m_{\rm n}$ is the baryonic mass in 
the $n$-th cell, ${\vec {r}}_{\rm n}$ is the displacement vector of the $n$-th 
cell relative to the clump centre, 
$r_{\rm n}^2 = {\vec {r}}_{\rm n}\cdot{\vec {r}}_{\rm n}$, and $\delta_{\rm i,j}$ 
is the Kronecker Delta. We diagonalize the tensor and find the three eigenvalues, 
$I_{\rm 1}\ge I_{\rm 2}\ge I_{\rm 3}$. Finally, we define 
$S_{\rm c}\equiv I_{\rm 3}/I_{\rm 1}$. For a perfect sphere, $S_{\rm c}=1$. 
For a flattened, oblate clump, $S_{\rm c}\simeq 0.5$. For a very prolate 
or filamentary clump, $S_{\rm c}<<1$.

% Figures
\smallskip
The clump finder is illustrated in \fig{clump_finder}, for one of the 
galaxies in the RP sample, V13 at $a=0.40$ ($z=1.5$). On the left we 
plot the surface density of the cold component in a face on projection 
and on the right we plot the stellar surface density in the same projection. 
Besides being a large, clumpy, star-forming disc, this example highlights the 
importance of detecting clumps in multiple tracers, since there is not 
a one-to-one correspondence between stellar and cold clumps.

\smallskip
The disc radius, $\Rd \simeq 13.3\kpc$, is marked in both panels, as are all 
identified clumps with baryonic volume densities $n_{\rm c}>1\cmc$ and masses 
$\Mc>10^7 \msun$. These thresholds were introduced here for clarity, because 
many low-mass, low-density clumps are not visible in 2D projection plots (recall 
that clumps are identified in 3D), and in any case for most of our analysis in 
\se{clump_comp} and \se{obs} we will focus on clumps above these thresholds. 
There are two ex situ clumps (defined below) in this galaxy, which are marked 
with squares. The rest of the off-centre clumps are all in situ and are marked 
with circles, as is the central bulge clump (defined below). The size of the marker 
corresponds to the clump radius, $\Rc$, while the color represents its mass. Clumps 
with $10^7<\Mc<10^{7.5}\msun$ are marked in grey, clumps with $10^{7.5}<\Mc<10^8\msun$ 
are marked in red, clumps with $10^8<\Mc<10^{8.5}\msun$ are marked in blue, and clumps 
with $10^{8.5}\msun<\Mc$ are marked in black. The disc mass is 
$\Md\simeq 2.5\times 10^{10}\msun$ and the gas fraction is $\sim 40\%$.

% Tracker
\subsubsection{Tracking individual clumps through time}
\smallskip
The time between consecutive outputs is typically about half 
an orbital time at the disc edge. Since the expected migration 
time for clumps is roughly 2 disc orbital times, we are 
occasionally able to track individual clumps across several 
snapshots, which allows us to directly study the evolution and 
lifetimes of clumps. As the simulations were run with an Eulerian, 
grid based, code we cannot trace gas elements through time, and 
therefore track clumps using their stellar particles. We only attempt 
to track clumps that contain at least 10 stellar particles. For each 
such clump in a given snapshot, we search the preceding snapshot 
for all ``progenitor clumps", defined as clumps that contributed at 
least $25\%$ of their stellar particles to the clump in question\footnote{As 
stellar particles are deposited on the grid using CiC interpolation, 
we only consider particles that have contributed at least half their 
mass to the clump for the purpose of clump tracking. This is to ensure 
that no stellar particle is associated with more than one clump in a 
given snapshot.}. If a given clump has more than one progenitor clump 
in the same snapshot, we rank them by baryonic mass, consider the most 
massive one the main progenitor and the others as having merged, thus 
creating a clump merger tree. If a clump has no progenitors in the 
preceding snapshot, we search the previous snapshots until we either 
find a progenitor or reach the initial timestep of the simulation.

\smallskip
We define the clump time, $\tc$, as the time since clump formation. In the 
first snapshot when the clump is identified, we set $\tc$ to the stellar 
age of the clump, provided this is less than the time since the previous 
snapshot, when the clump was not identified. Otherwise, we set $\tc=0$. In 
later snapshots when the clump is identified, we add to $\tc$ the physical 
time that has elapsed since its initial identification. Note that clump 
\textit{time} differs from clump \textit{age}, which is given by the mean 
stellar age and is thus affected by ongoing star-formation and tidal stripping 
of older stars (see \fig{age_time} and \citealp{Bournaud14}). The clump 
\textit{lifetime}, $\tcm$, is the clump time at the final snapshot where 
the clump is identified. We refer to clumps that were identified in only 
one snapshot and have $\tcm=0$ as zero-lifetime-clumps, or ZLCs. In 
\se{clump_comp} we will show that most of these are false positives, 
characterised by low densities and masses. 

% in situ, ex situ, bulge
\subsubsection{In situ, ex situ and bulge clumps}
\label{sec:Is}
\smallskip
Following M14 we divide clumps into central ``bulge" clumps, 
and off-centre clumps. A bulge clump forms the nucleus of the galactic 
bulge and is defined as a clump whose centre is within 2 grid cells 
of the galaxy centre. If there is more than one such clump, the most 
massive is considered the bulge clump and the rest are considered 
off-centre clumps. We discuss the prevalence of bulge clumps in the 
simulations in \se{clump_comp}.

\begin{figure}
\subfloat{\includegraphics[width =0.48 \textwidth]{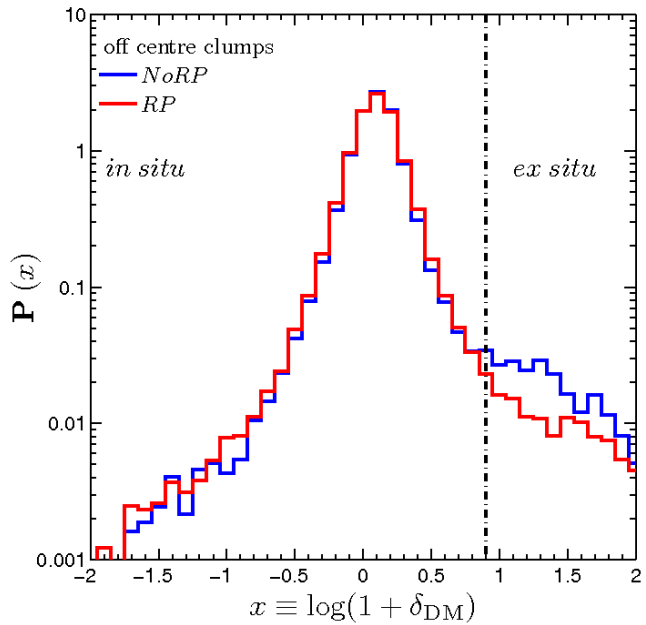}}
\caption{In situ vs ex situ clumps. Probability density of 
$x={\rm log}(1+\delta_{\rm DM})$ (see text for details) for 
the off-centre clumps in the NoRPs (blue) and RPs (red). The 
vertical dash-dotted line marks the threshold between 
in situ and ex situ clumps at $x=0.9$. 
}
\label{fig:ES} 
\end{figure} 

\smallskip
The off-centre clumps are divided into ``in situ" clumps which 
formed within the disc through VDI, and ``ex situ" clumps which 
are external mergers. As in M14 we distinguish in situ 
from ex situ clumps by their dark matter content. We compare the 
mean dark matter density within the clump, $\rho_{\rm DM,\,c}$, 
to the mean dark matter density in a spherical shell (concentric 
with the host halo) $1\kpc$ thick around on the clump, $\rho_{\rm DM,\,s}$, 
and define the dark matter overdensity 
$1+\delta_{\rm DM} \equiv \rho_{\rm DM,\,c}/\rho_{\rm DM,\,s}$. 
\Fig{ES} shows the probability density\footnote{The probability for 
a clump to have a value in the interval $\Delta x$ about $x$ is 
$\textbf{P}(x)\Delta x$. }, $\textbf{P}(x)$ where 
$x={\rm log}(1+\delta_{\rm DM})$, for the off-centre clumps in the 
NoRPs (blue line) and the RPs (red line). In both samples, the in 
situ clumps exhibit a roughly log-normal distribution centred on 
$x\sim 0.1$ with a dispersion of $\sigma_x\sim 0.13$, while the 
ex situ clumps form an extended tail to high values. We define ex 
situ clumps as clumps with $x_{\rm max}>0.9$, where $x_{\rm max}$ 
is the maximal value of $x$ during the clump lifetime, and mark 
this threshold in \fig{ES}. We discuss the contribution of ex situ 
clumps to the total clump population in \se{clump_comp}.

\smallskip
An alternate definition of ex situ clumps could have been based on 
the birth place of the stellar particles within the clump. Among clumps 
with $\Mc>10^7\msun$ and ${\bar {\delta}}_{\rm c}>25$, where our sample 
is complete and ZLCs are rare (see \se{clump_comp}), these two definitions 
agree well. In $\sim 78\%\:(88\%)$ of ex situ clumps selected by dark matter 
in the RPs (NoRPs), more than half the baryonic mass consists of stars formed 
outside the disc, as defined at the snapshot closest to the star particle's 
birth. On the other hand, in both the RPs and the NoRPs, only $\sim 9\%$ of 
in situ clumps selected by dark matter have more than half their baryonic mass 
in ex situ stars. 

%\subfloat{\includegraphics[width =0.6 \textwidth]{Ms_Mh_panel.eps}}
\begin{figure*}
\centering
\subfloat{\includegraphics[trim={0 1.76cm 0 0}, clip, width =0.40 \textwidth]{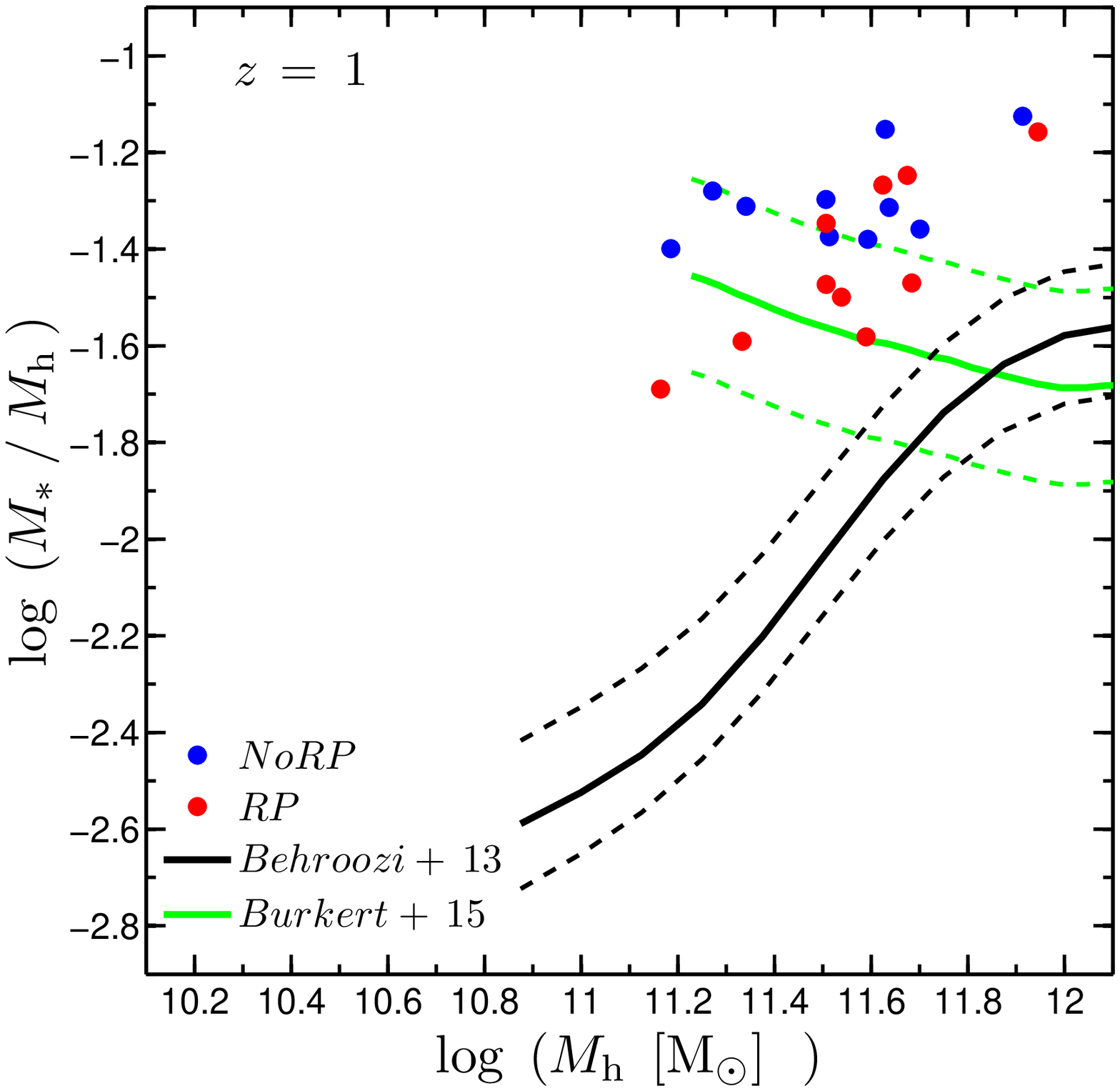}}
\subfloat{\includegraphics[trim={2.30cm 1.76cm 0 0}, clip, width =0.347 \textwidth]{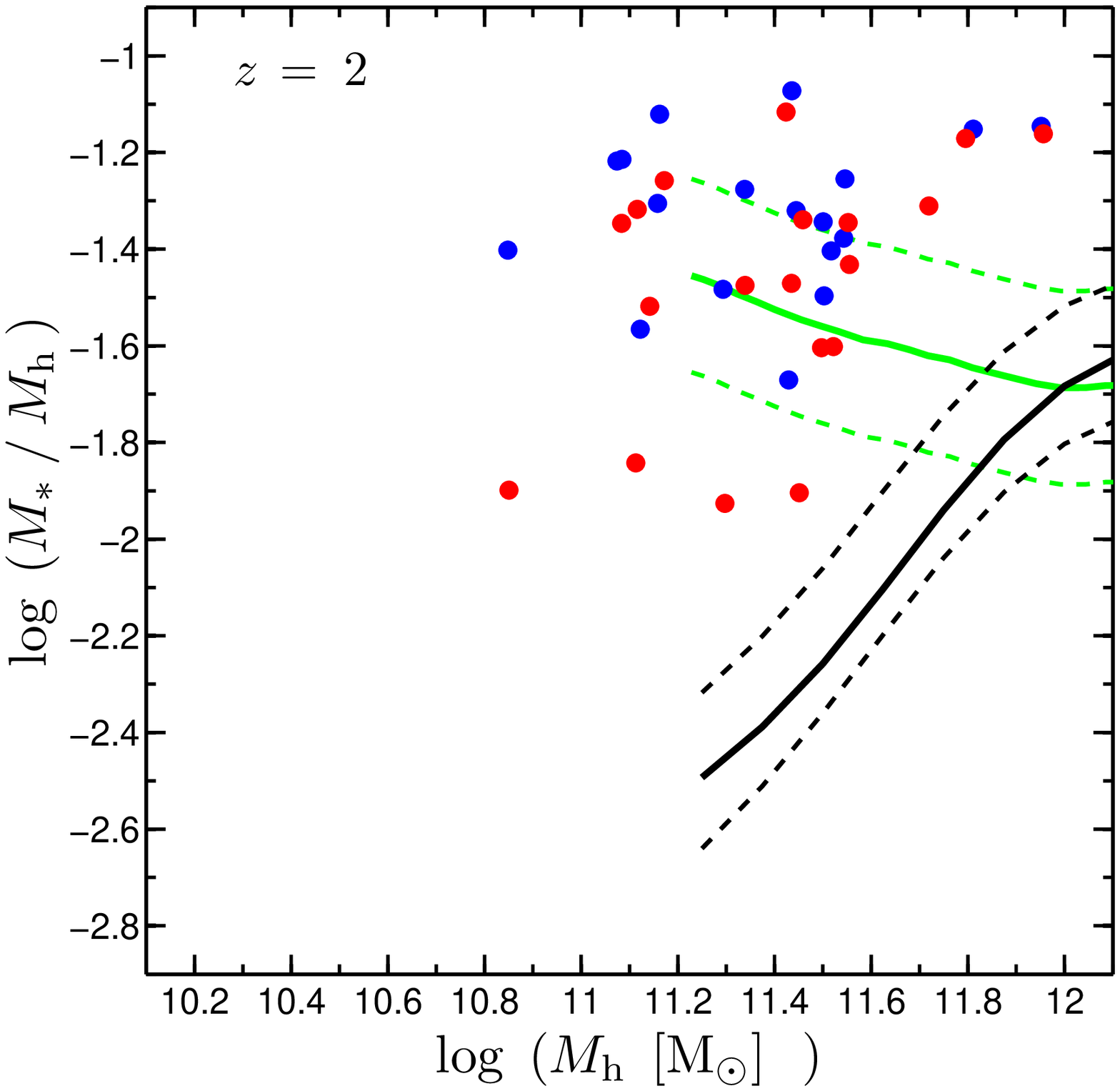}}\\
\vspace{-0.57mm}
\hspace{-1.00mm}
\subfloat{\includegraphics[trim={0 0 0 0}, clip, width =0.398 \textwidth]{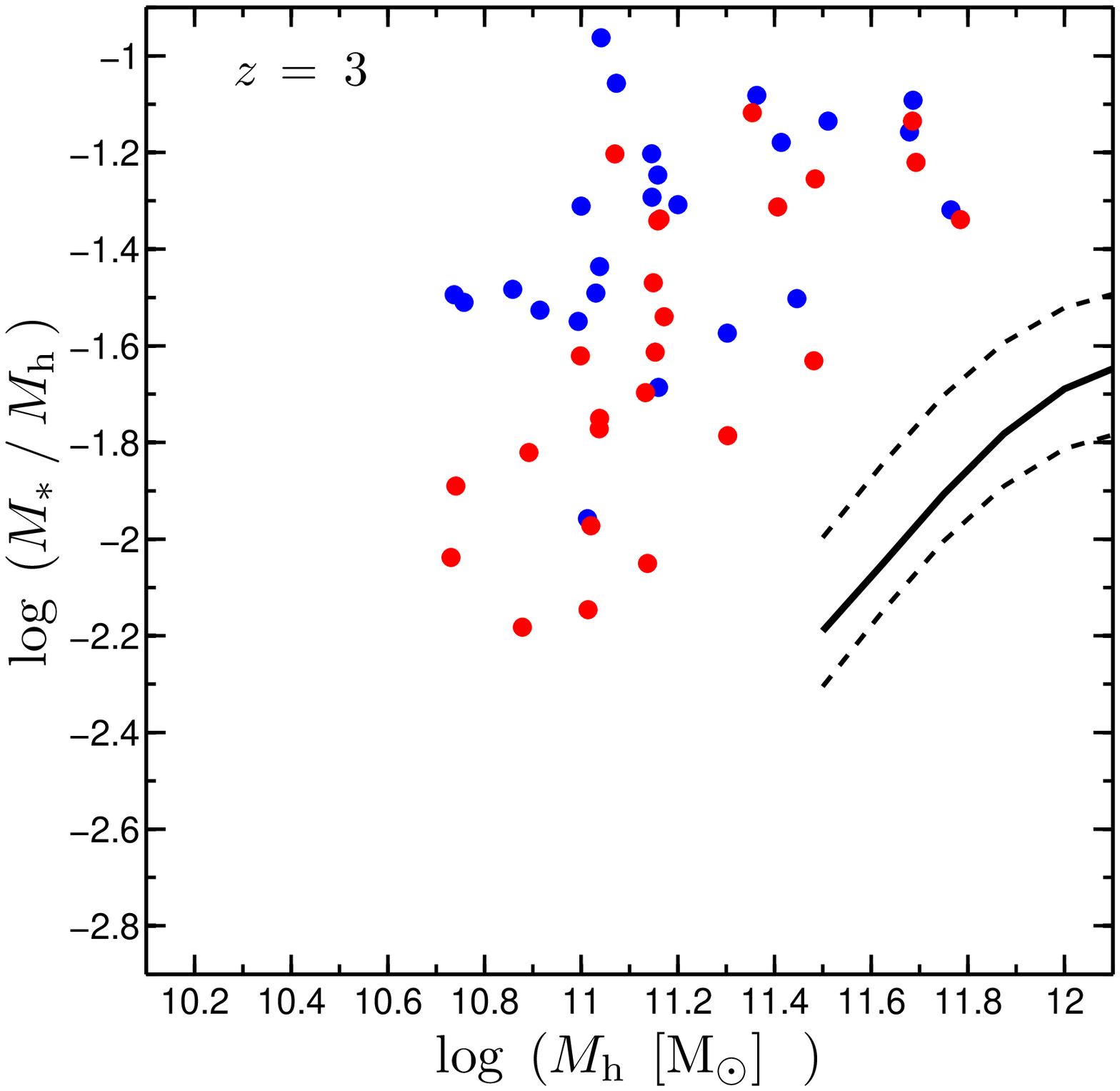}}
\hspace{-0.60mm}
\subfloat{\includegraphics[trim={2.30cm 0 0 0}, clip, width =0.347 \textwidth]{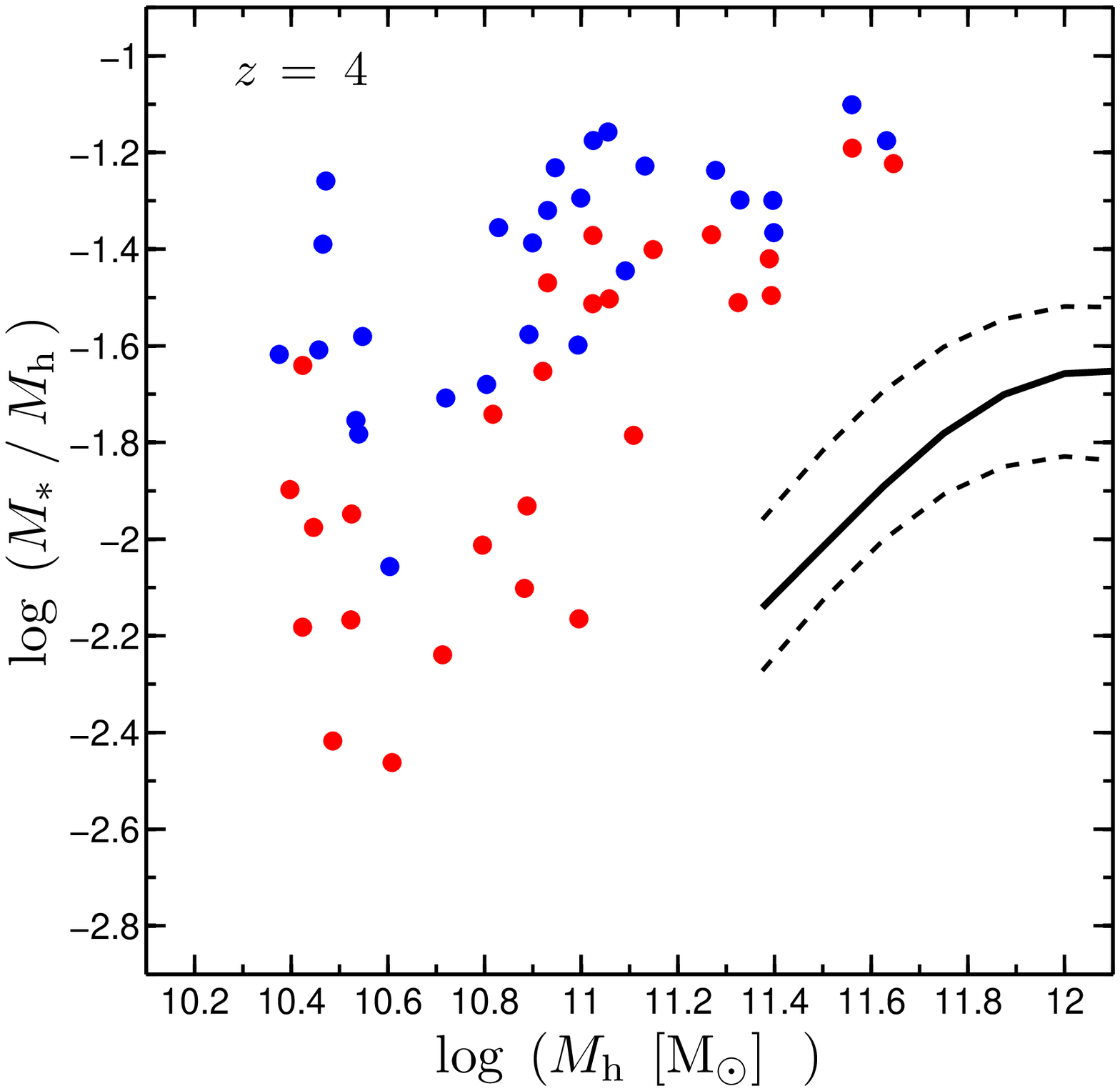}}\\
\caption{Ratio of stellar mass to halo mass in the NoRPs 
(blue points) and the RPs (red points), at redshifts $z=1,2,3,4$, 
in clockwise order from the top left. The solid black line is the mean 
relation from \citet{Behroozi13} and the dashed black lines are the 
1-$\sigma$ scatter. The green solid and dashed lines are the median 
relation and 1-$\sigma$ scatter from the ``full" sample of 359 SFGs 
at $0.8\lsim z \lsim 2.6$ from \citet{Burkert15}. The stellar mass is 
reduced in the RPs, typically by a factor of $\sim 2$, and by up to a 
factor of $\sim 10$ in low mass haloes at early times. The RPs are still 
a factor of $\lsim 2$ above the \citet{Behroozi13} relation at $z\sim 1-2$, 
but are in much better agreement with \citet{Burkert15}, 
reflecting a factor $\sim 2$ uncertainty in observational estimates.
}
\label{fig:Ms_Mh} 
\end{figure*} 

%%%%%%%%%%%%%%%%%%%%%%%%%%%%%%%%%%%%%%%%%%%%%%%%  
\section{Global Properties of the Galaxy Samples} 
\label{sec:gal_comp} 
\smallskip
In this section we examine the global effect of RP 
feedback on the galaxies in our sample, before studying 
its effect on giant clumps in \se{clump_comp}. We limit 
our analysis here to snapshots that have both a RP and 
a NoRP version, which leaves 798 snapshots: 255 at 
$4 \le z$, 131 at $3 \le z < 4$, 168 at $2 \le z < 3$, 
244 at $0.8 \le z < 2$. 

\begin{figure*}
\centering
\subfloat{\includegraphics[width =0.4 \textwidth]{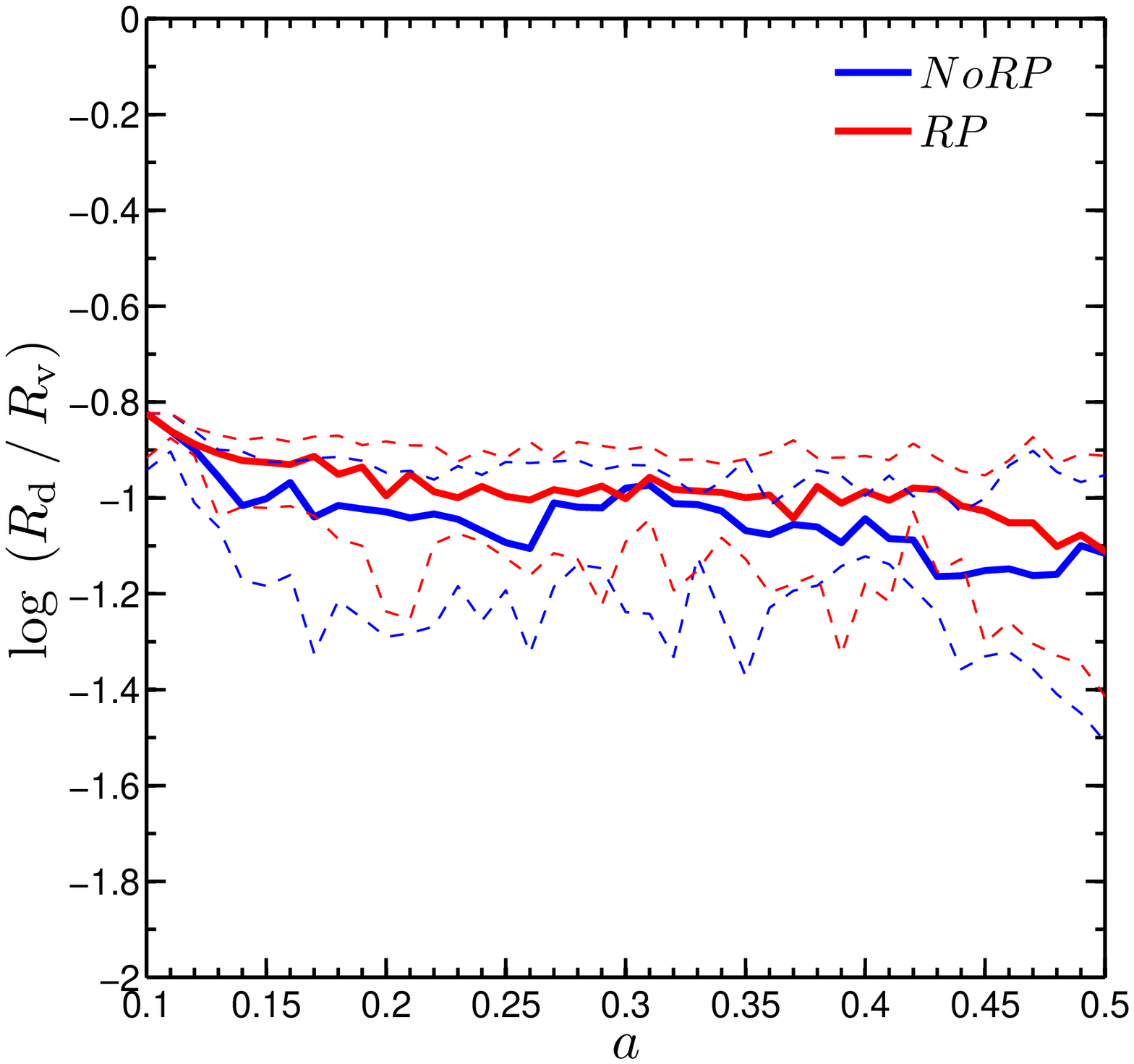}}
\subfloat{\includegraphics[width =0.4 \textwidth]{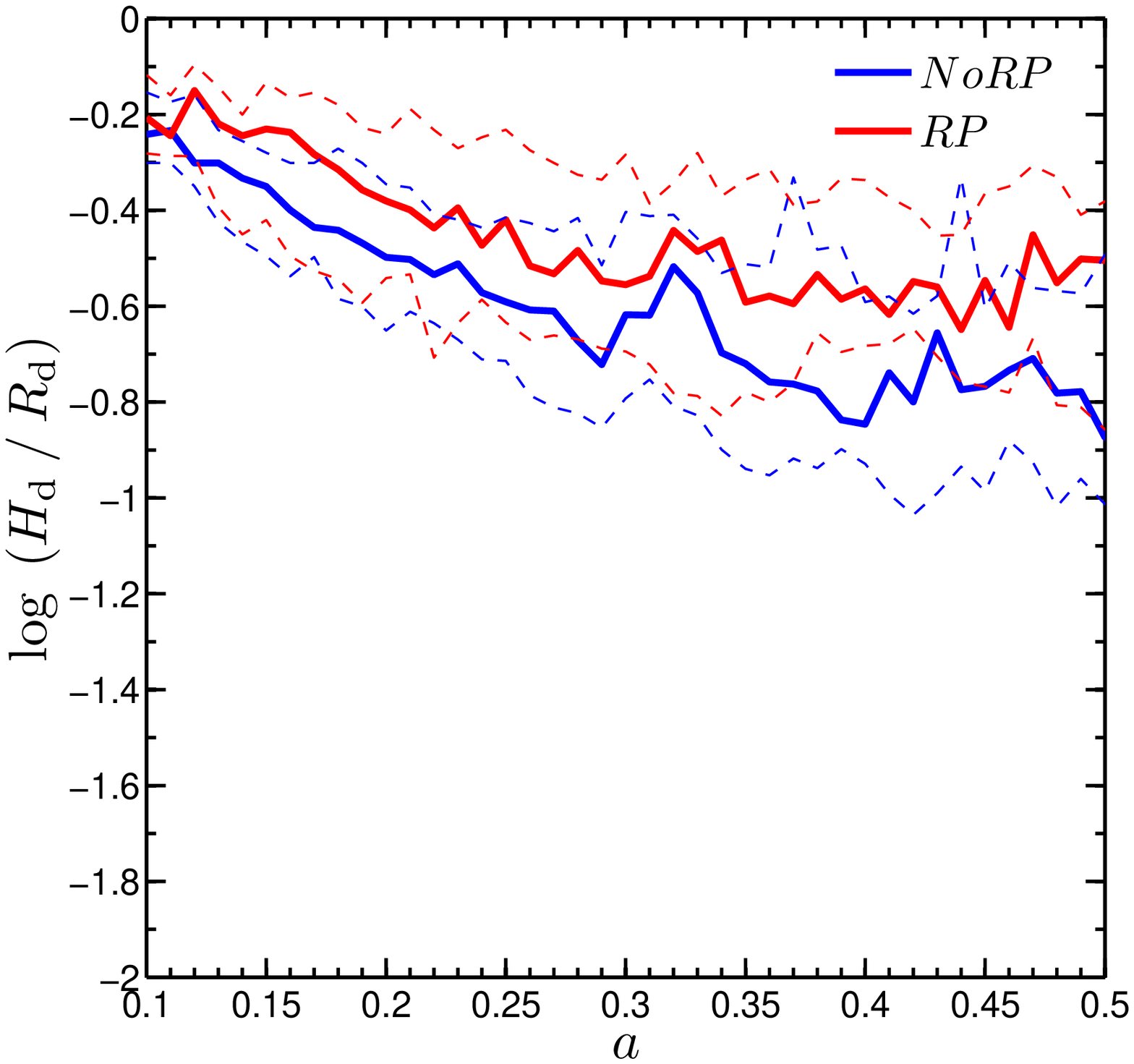}}\\
\vspace*{-0.4cm}
\subfloat{\includegraphics[width =0.4 \textwidth]{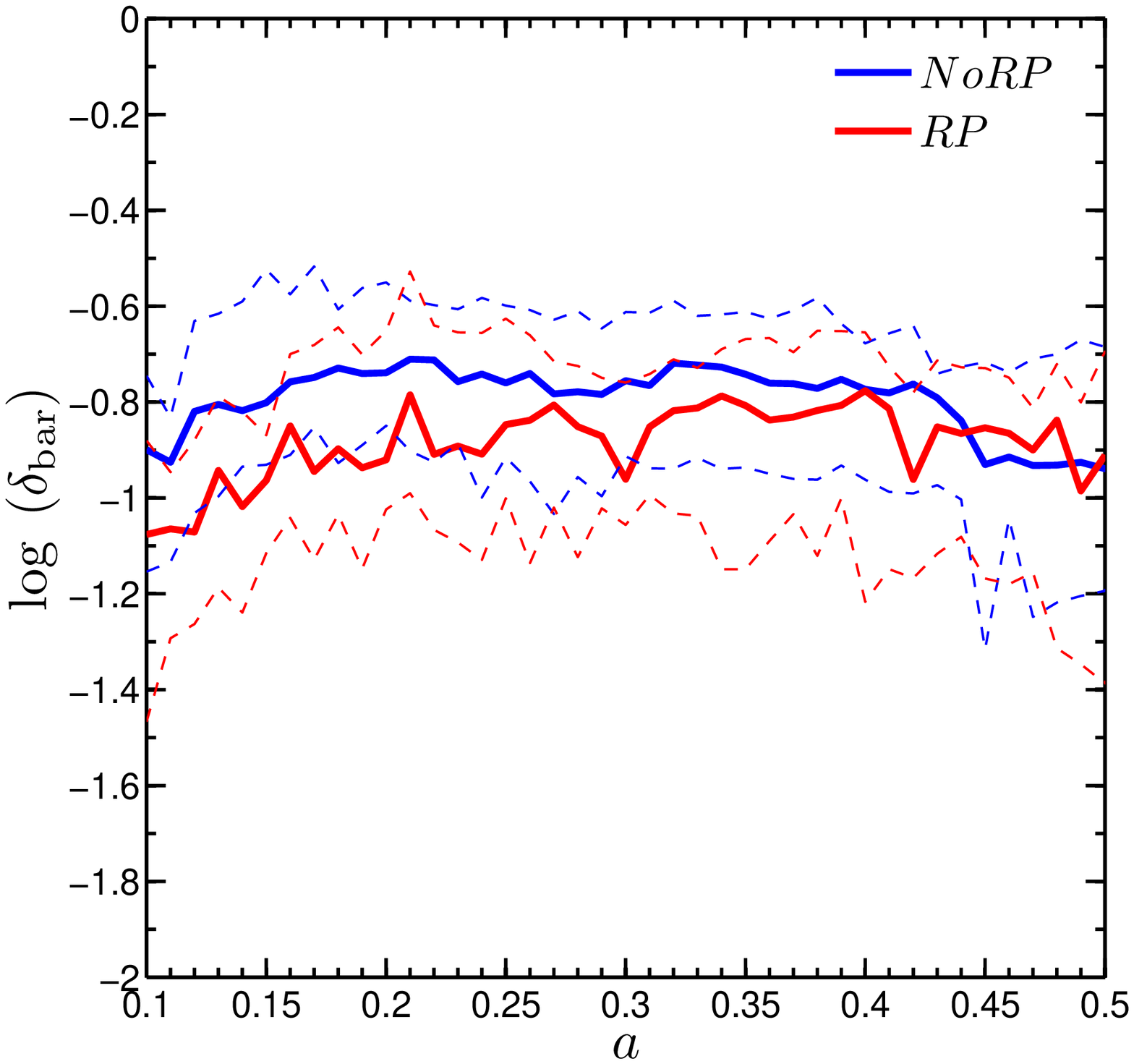}}
\subfloat{\includegraphics[width =0.4 \textwidth]{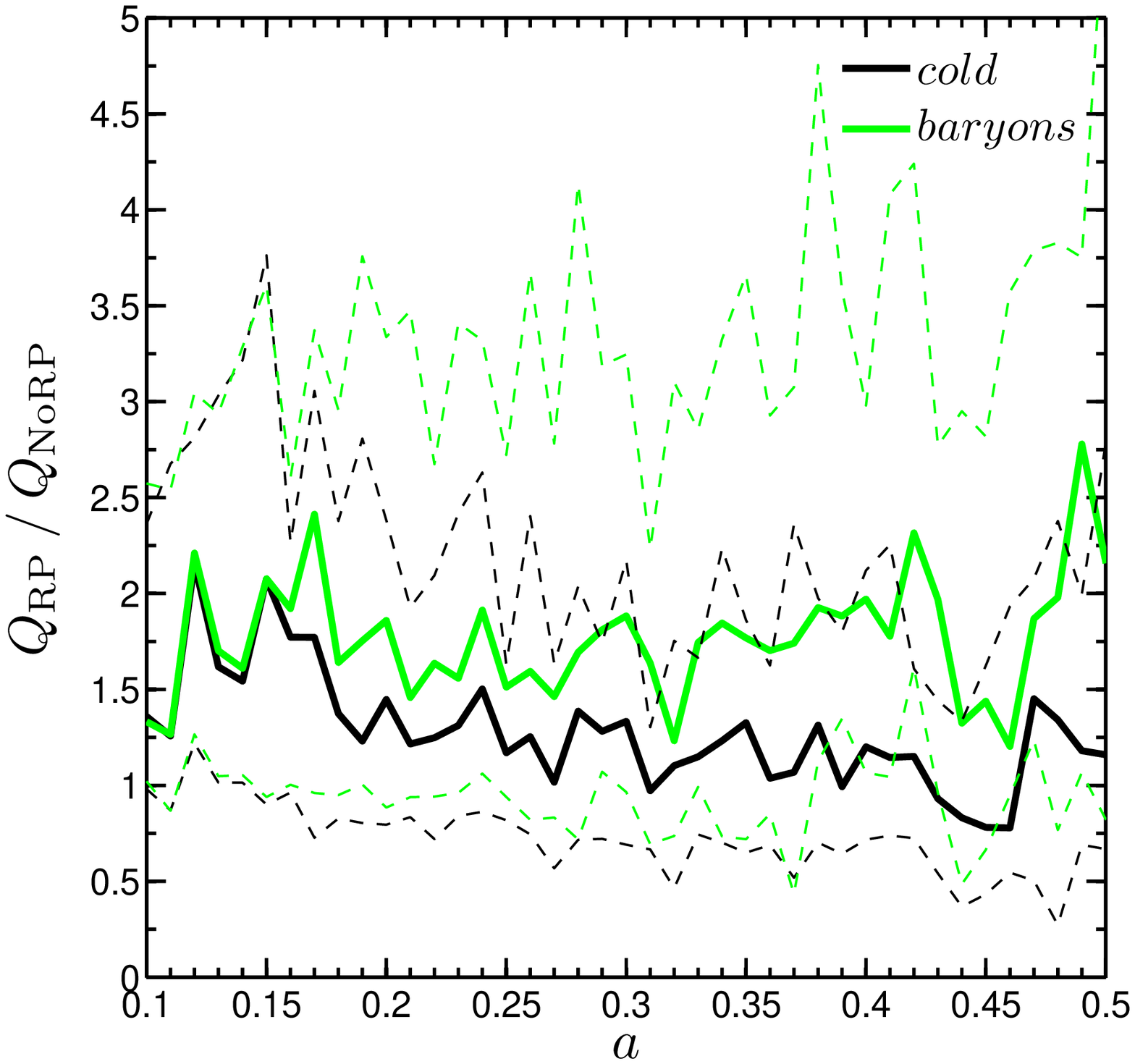}}
\caption{Disc properties in RPs and NoRPs. Each panel 
examines a different property related to the disc structure or 
stability: the disc radius normalized to the virial radius, $\Rd/\Rv$ 
(top left), the disc height normalized to its radius, $\Hd/\Rd$ 
(top right), and the baryonic disc mass normalized to the total mass 
within the disc radius $\delta_{\rm bar}=\Md/M_{\rm tot}(\Rd)$ (bottom 
left). We show as a function of time represented by the expansion factor, 
$a$, the evolution of each property for the NoRP sample (blue) and the RP 
sample (red). The thick solid lines show the median and the thin dashed 
lines show the $67\%$ scatter. In the bottom right we show the ratio of 
the global Toomre $Q$ parameter between the two samples, using 
$Q \propto \delta^{-1}\Hd/\Rd$. In this panel, the green line refers to 
$\delta$ calculated using all baryons in the disc as shown in the bottom 
left panel, while the black line refers to $\delta$ calculated using only 
the cold mass in the disc. Both simulations produce thinner, more relaxed 
discs at later times. The NoRPs have a slight tendency for smaller, thinner 
discs, but are a higher fraction of the total mass within the disc radius.
The global Toomre parameter in the RPs is typically $\lsim 70\%$ larger than 
in the NoRPs, indicating that while RP may affect VDI and clump formation, 
the effect is not dramatic.
}
\label{fig:disc_comp} 
\end{figure*} 

\subsection{Stellar Mass}
\smallskip
\Fig{Ms_Mh} shows the stellar-to-halo mass ratio as a function of 
halo mass for the two galaxy samples at four redshifts, $z=1,2,3,4$.
\footnote{In practice, for each simulation we use the closest 
snapshot to each redshift, in the range $|\Delta z|<0.1$.} 
Stellar mass is defined here as the total within $0.15\Rv$. For 
comparison, we show the abundance matching predictions 
of \citet{Behroozi13}, and recent estimates from \citep{Burkert15} which 
are based on observed kinematics of several hundred SFGs at redshifts 
$0.8\lsim z\lsim 2.6$ from \texttt{KMOS$^{\rm 3D}$}, with similar masses 
to our simulated galaxies. RP feedback typically lowers the stellar mass 
by a factor of $\gsim 2$ (see also \citealp{Moody14}). At low halo masses 
at high redshifts, the stellar mass can be decreased by up to a factor of 
$\sim 10$, suggesting that radiative feedback is more efficient in low mass 
galaxies. 

\smallskip
At redshifts $z\le 2$, the RPs are still a factor of $\lsim 2$ above 
the \citet{Behroozi13} curve, though they appear consistent with 
the estimates of \citet{Burkert15}, at least for halos with 
$\Mv<4\times 10^{11}\msun$, where most of our data lies. This reflects 
a factor of $\sim 2$ uncertainty in the observational estimates of the 
stellar-to-halo mass ratio for these halo masses and redshifts, and our 
simulations produce values within the observational uncertainties. At 
$3\le z$, most of our sample lies at low halo masses where no direct 
comparison with existing predictions can be made. By extrapolating the 
\citet{Behroozi13} relation towards lower halo masses, the off-set between 
our sample and the prediction is comparable to that at $z\le 2$. We conclude 
that the RPs produce stellar masses that are in the ballpark of realistic 
values, within the observational uncertainties of a factor $\sim 2$.

\subsection{Disc Properties and Toomre Q}
\smallskip
In \fig{disc_comp} we compare the galaxy samples in terms of 
global disc structure and stability. We examine as a function 
of time, represented by the expansion factor $a$, the disc radius, 
thickness, mass fraction with $\Rd$ and Toomre $Q$ parameter, as 
described below. For each of these properties we show the median 
and $67\%$ scatter among all snapshots at a given expansion factor. 

\smallskip
In the top left panel we examine the disc radius normalized to the 
virial radius, $\Rd/\Rv$. On average, the disc sizes are very similar 
in the two galaxy samples. The NoRPs have a very slight tendency to 
produce smaller discs than the RPs, by $\sim 0.05~{\rm dex}$ at all 
times. However, given the scatter in the data this does not appear 
significant. Most of the scatter is caused by small differences 
in the times of major mergers and compaction events between the 
two samples, which mostly occur after $z\sim 4$ \citep{Zolotov15}. 

\smallskip
The top right panel examines the disc thickness, normalized to 
its radius $\Hd/\Rd$. To first order, this represents the velocity 
dispersion within the disc, $\Hd/\Rd \sim \sigma_{\rm z}/V$. As $\Rd$ 
and $\Hd$ were both calculated using the cold component, this refers 
to the gas velocity dispersion, rather than the stellar value. The 
stronger feedback in the RPs results in thicker and more turbulent 
discs, especially at later times. The median ratio between individual 
snapshots paired in both sets of simulations increases monotonically 
from $\sim 0.08~{\rm dex}$ at $4<z$ to $\sim 0.19~{\rm dex}$ at $z<2$. 
However, the scatter is large, $\sigma\sim 0.14~{\rm dex}$ at $4<z$ and 
$\sim 0.19~{\rm dex}$ at $z<2$. It is unclear how much of this trend is 
the result of instantaneous feedback driving additional turbulence, and 
how much is the result of decreased gas fractions in NoRPs due to excessive 
star-formation at early times. Surely the latter contributes to the increased 
discrepancy at later times. However, this distinction is beyond the scope of 
the current paper.

\smallskip
The bottom left panel examines the ratio of the disc mass to the total 
mass within the disc radius, including the spheroid and dark matter halo 
components, $\delta_{\rm bar} \equiv \Md/M_{\rm tot}(\Rd)$. This is an 
important indicator of disc instability \citep[e.g.][]{DSC}. At early 
times, the NoRPs have a more prominent disc, which results from earlier 
settling of the galaxies into oblate structures \citep{Tomassetti16}. 
Detailed analysis of the shapes of galaxies in the RPs indicate 
that galaxies tend to maintain a prolate shape at early times, developing 
a disc only after the first compaction event, when the baryons dominate 
the mass and potential at the inner radii \citep{Ceverino15b}. This happens 
earlier in the NoRP galaxies \citep{Tomassetti16}. At late times, the two 
sets of simulations have comparable $\delta_{\rm bar}$. It is interesting 
to characterise the cold component in the disc by defining instead 
$\delta_{\rm cold} = M_{\rm cold,\,d}/M_{\rm tot}(\Rd)$. At $4<z$, the NoRPs 
and the RPs have ${\rm log}(\delta_{\rm cold})\sim -0.9$ and $-1$ respectively. 
At $z\sim 3.5$, both have comparable values of ${\rm log}(\delta_{\rm cold})\sim -1.1$. 
By $z<1.5$, the NoRPs have saturated at ${\rm log}(\delta_{\rm cold})\sim -1.5$, 
while the RPs have saturated at ${\rm log}(\delta_{\rm cold})\sim -1.3$. This 
indicates that the gas fractions in the RP discs are $\sim 60\%$ higher than 
in the NoRP discs at these redshifts. 

\begin{figure*}
\centering
\subfloat{\includegraphics[width =0.411 \textwidth]{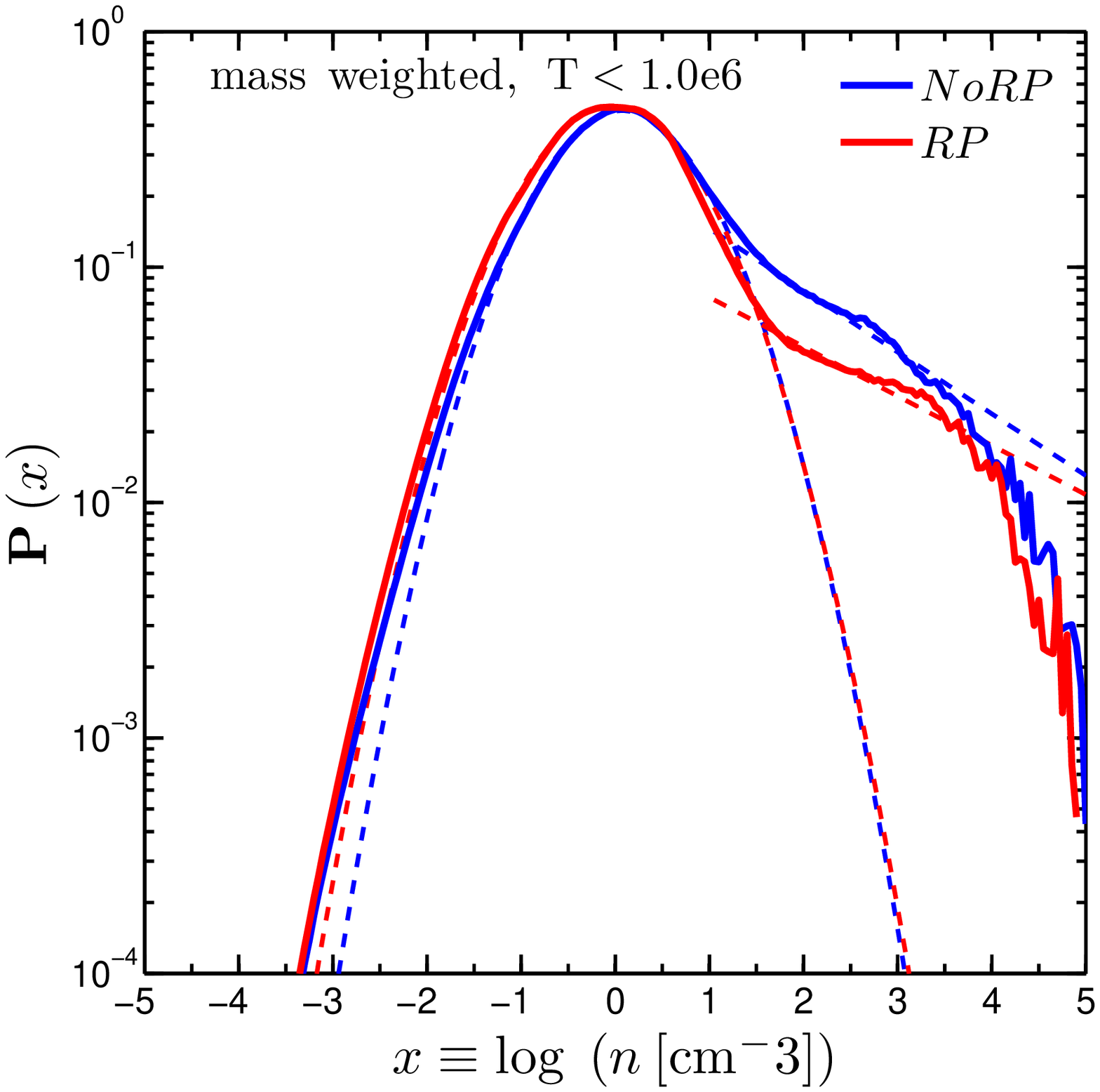}}
\hspace{-1.5mm}
\subfloat{\includegraphics[width =0.366 \textwidth]{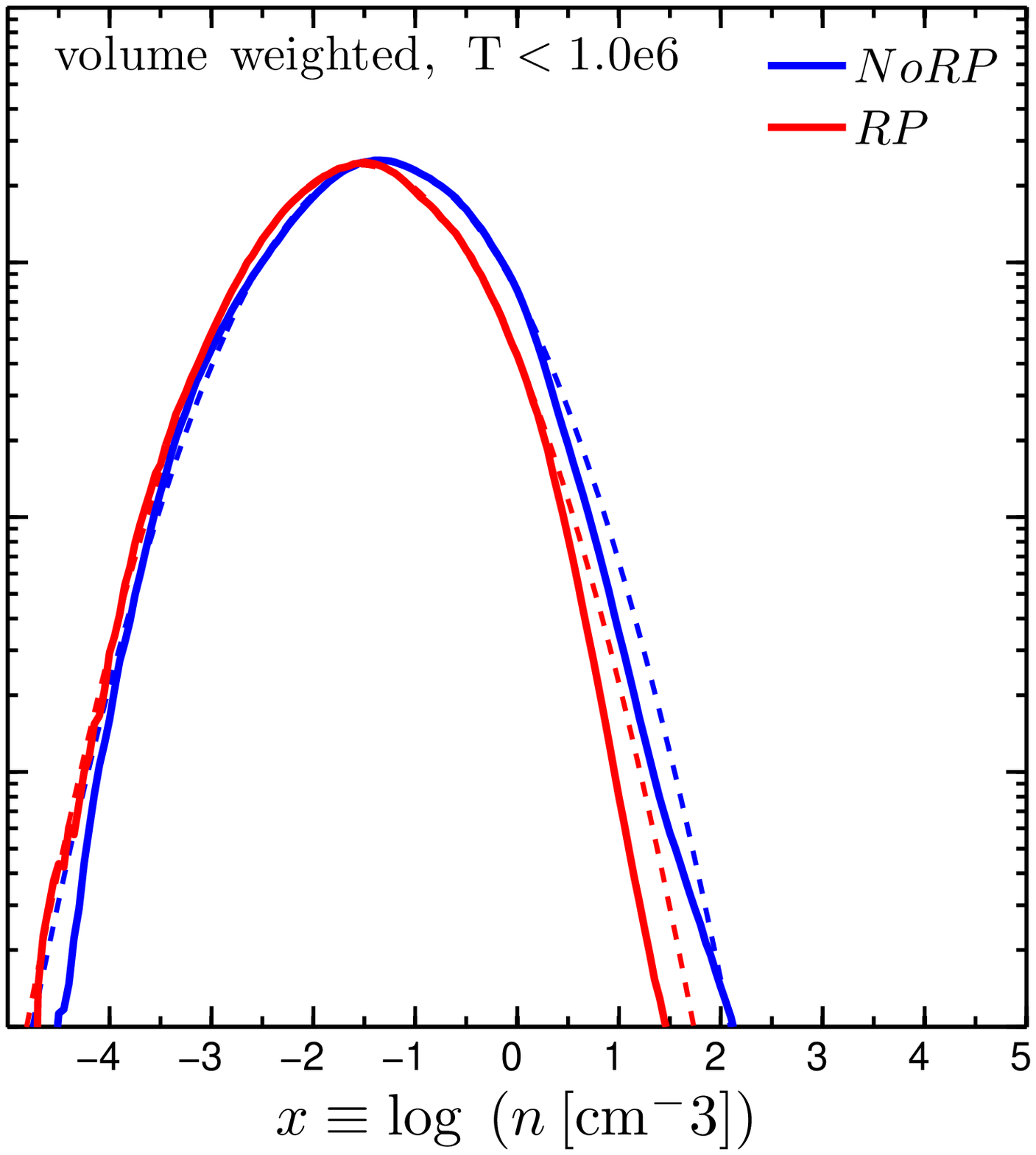}}
\caption{Density distribution of ISM gas with temperatures $T<10^6~{\rm K}$, 
stacked over the all common snapshots of NoRP (blue) and RP (red) galaxies. Solid 
lines are the data and dashed lines show best fit models. Left: mass-weighted density 
PDF, $dm_{\rm gas}/d{\rm log}(n)$. In both sets of simulations, the distribution 
of gas with $n<10\cmc$ is well fit by a log-normal distribution with a mean density of 
${\rm log}(n\cdot {\rm cm}^3)\simeq 0$ and a dispersion of $\sigma_{\rm log(n)}\simeq 0.74$ 
(dashed lines). At high densities, $n\gsim 10\cmc$, the distributions deviate from 
log-normality and are well fit by power laws, with slopes of $\sim -0.5$ and $-0.6$ 
for RP and NoRP respectively (dashed lines). While both models have a similar range 
of densities, RP decreases the amount of dense gas by roughly a factor of $\sim 2$, 
heating it to temperatures $T>10^6~{\rm K}$ and driving it to lower densities. Right: 
volume-weighted density PDF, $dV_{\rm gas}/d{\rm log}(n)$. Less than $1\%$ of the volume 
has densities $n>10\cmc$. Both models are well fit by log-normal distributions, which peak 
at ${\rm log}(n\cdot{\rm cm}^3)\simeq -1.34$ and $-1.54$ in the NoRPs and RPs respectively 
(dashed lines), though densities as high as ${\rm log}(n\cdot{\rm cm}^3)\gsim -0.5$ are common.
}
\label{fig:dens_PDF} 
\end{figure*} 

\smallskip
The bottom right panel shows the ratio of the global Toomre $Q$ parameter in 
the RPs to $Q$ in the NoRPs, using the crude approximation \citep[e.g.][]{DSC}:
\be 
\label{eq:Q}
Q\propto \frac{\sigma\Omega}{\Sigma} \propto \frac{\sigma}{V}\frac{M_{\rm tot}}{\Md} \propto \delta^{-1}\frac{\Hd}{\Rd}
\ee
{\no}We do this once for the cold component only, using $\delta_{\rm cold}$, and once 
for the baryonic disc mass, using $\delta_{\rm bar}$ (which still only includes 
stars that are co-rotating with the gas disc). This crudely approximates $Q$ for 
a one-component model, assuming the gas and stars have comparable velocity dispersions. 
Since the discs contain dynamically hot stars as well, a two component Toomre analysis 
is required to fully address the issue of disc instability. However, \equ{Q} is 
not applicable for a two component system, and a more detailed analysis is beyond 
the scope of this paper. Therefore, rather than focus on the absolute values of $Q$, 
we only address the relative values in the two simulation suites, in order to compare 
the predicted strength of VDI. We note that in the RPs, the two component $Q$ is not 
very different from $Q$ computed using the cold component only \citep{Inoue16}. 

\smallskip
The NoRPs have slightly lower values of $Q$, indicating that they may be slightly more 
prone to instability and clump formation than the RPs. However, this difference is not 
large, $\sim 70\%$ when considering all baryons in the disc and less than that when 
considering only the cold component. We conclude that RP does not fundamentally stabilize 
the discs against VDI, so clump formation is not expected to be greatly suppressed.

\subsection{ISM Structure}
\smallskip
\Fig{dens_PDF} shows the density probability distribution 
function (PDF) of the ISM gas in the NoRPs and RPs, stacked 
over all common snapshots. We consider all gas with 
temperatures $T<10^6~{\rm K}$ located within the disc cylinder 
at each snapshot\footnote{Note that this is a slightly 
larger volume in the RPs (\fig{disc_comp}).}. Hot gas, 
with $T>10^6~{\rm K}$, contains only $\sim 1\%$ of the gas mass 
in the ISM and is not important for disc instability or for clump 
detection. We show both the mass-weighted PDF, $dm_{\rm gas}/d{\rm log}(n)$ 
(left), and the volume weighted PDF, $dV_{\rm gas}/d{\rm log}(n)$ 
(right). These are the mass fraction and volume fraction per 
logarithmic density interval, respectively.
% While there are significant galaxy-to-galaxy variations in the sample, the qualitative structure of the distribution functions is similar in all cases, and the stacked profiles are qualitatively similar when taken over the full sample or cut into redshift bins. 

\smallskip
At low to intermediate densities, the mass-weighted PDF of the NoRPs 
and the RPs have a very similar shape. Both are well fit by a 
log-normal distribution which has a mean density of ${\rm log}(n\cdot{\rm cm^3})\simeq 0$ 
and a standard deviation of $\sigma_{\rm log(n)}\simeq 0.74$. The 
mean density decreases from $\sim 3\cmc$ at $4<z$ to $\sim 0.6\cmc$ 
at $z<2$, in agreement with the cosmological scaling, while the standard 
deviation decreases from $\sigma_{\rm log(n)}\simeq 0.85$ to $0.64$ in 
the same redshift range. We note that the adopted density threshold for 
star formation, $n_{\rm SF}=1\cmc$, is at the peak of the mass weighted 
PDF. A log-normal shape for the density PDF is characteristic of isothermal, 
supersonic turbulence \citep{VS94,Padoan97,Scalo98,Federrath08,Price11,Hopkins12a} 
with the width of the distribution weakly depending on the Mach number as 
$\sigma_{\rm log(\rho)} \propto \sqrt{{\rm ln}(1+0.25 M^2)}$ 
\citep{Padoan97,Federrath08}. 

\smallskip
At high densities, $n>10-30 \cmc$, the distributions deviate from log-normality 
and develop power law tails, with slopes of $\sim -0.5$ and $\sim -0.6$ in 
the RPs and NoRPs respectively. The transition density increases with redshift, 
maintaining a factor $\gsim 10$ above the mean. The development of a power law 
tail at high densities is typical when self-gravity becomes important and the 
gas can no longer be held up by turbulence \citep{VS08,Elmegreen11,Hopkins12a}. 
Radiation pressure regulates the distribution of very dense, star-forming gas. 
As a result, the mass fraction of dense, self-gravitating gas in the power law 
tail is reduced by a factor of $\gsim 2$ in the RPs, consistent with the typical 
decrease in the amount of stars formed in the simulations (\fig{Ms_Mh}). 
The mass in the hot component, with $T>10^6~{\rm K}$ (not shown), is 
roughly twice as high as in the RPs compared to the NoRPs. Similar results 
have been obtained in studies of isolated galaxy simulations \citep{Hopkins12a,Roshdal15} 
and in one of the cosmological simulations used here \citep{Ceverino14}. 

\smallskip
Both distributions are attenuated at the highest densities of ${\rm log}(n\cdot \cmc)\gsim 3.5-4$, 
with a slight tendency for a higher threshold in the RPs vs the NoRPs, and at lower 
redshifts. This is likely due to the pressure floor (\equ{Pfloor}), to which we can 
associate an effective velocity 
\be 
\sigma_{\rm floor} = \sqrt{\gamma \frac{P_{\rm floor}}{\rho}} \simeq 120\:n_{\rm 4}^{1/2}\:\Delta_{\rm 25}\:\kms
\ee
{\no}where $n_{\rm 4} = n / (10^4 \cmc)$ and $\Delta_{\rm 25}=\Delta/(25\pc)$.\footnote{The 
highest level of refinement is always reached for gas densities $n>100\cmc$ and often even at 
lower densities (\se{art}).} In the RPs, the radial velocity dispersion, $\sigma_{\rm r}$, is 
typically $\sim 50\kms$ before compaction and $\sim 100 \kms$ after compaction (\citealp{Zolotov15}, 
figure 11). So at densities of $n \sim (3\times 10^3)-10^4\cmc$, the pressure floor becomes comparable 
to the turbulent pressure, preventing cells from reaching higher densities. 

\smallskip
The volume-weighted PDF is well fit by a log-normal in both NoRPs and 
RPs, with peak densities of ${\rm log}(n\cdot{\rm cm}^3)\simeq -1.34$ and 
$-1.54$ respectively, and standard deviations of $\sigma_{\rm log(n)}\simeq 0.87$ 
and $0.72$. Dense gas, with $n\gsim 10\cmc$ comprises less than $1\%$ of the total 
volume. The volume filling factor of star-forming gas, with $n\gsim 1\cmc$, is 
reduced by a factor of $\sim 2$ in the RPs, which in turn reduces the star formation 
in the RPs, as seen in \fig{Ms_Mh}. The hot component, with $T>10^6~{\rm K}$ (not shown) 
occupies a comparable volume to the cooler component, despite containing only $\sim 1\%$ 
of the mass. The distribution of densities is also log-normal, with mean densities of 
${\rm log}(n\cdot{\rm cm}^3)\simeq -3$ and $-3.4$ in the NoRPs and RPs respectively, 
and corresponding standard deviations of $\sigma_{\rm log(n)}\simeq 0.57$ and $0.72$. 
The volume fraction of this hot gas is roughly 10 times higher in the RPs than in the 
NoRPs, showing that RP heats the gas and drives it to lower densities.

\smallskip
To summarise, we expect bound star-forming clumps to 
form at densities $n>10\cmc$, while most of the volume 
in the ISM is at densities of $10^{-3}\lsim n \lsim 1 \cmc$. 
Our threshold density residual for clump detection, $\delmin=10$ 
(\se{clump_find}), is thus well suited to detect the vast majority 
of clump candidates. Naturally, given the wide range of low densities, 
we also expect to detect many unbound, low density, transient 
structures, which form the bulk of the ZLCs discussed earlier.

\begin{figure*}
\centering
\subfloat{\includegraphics[width =0.93 \textwidth]{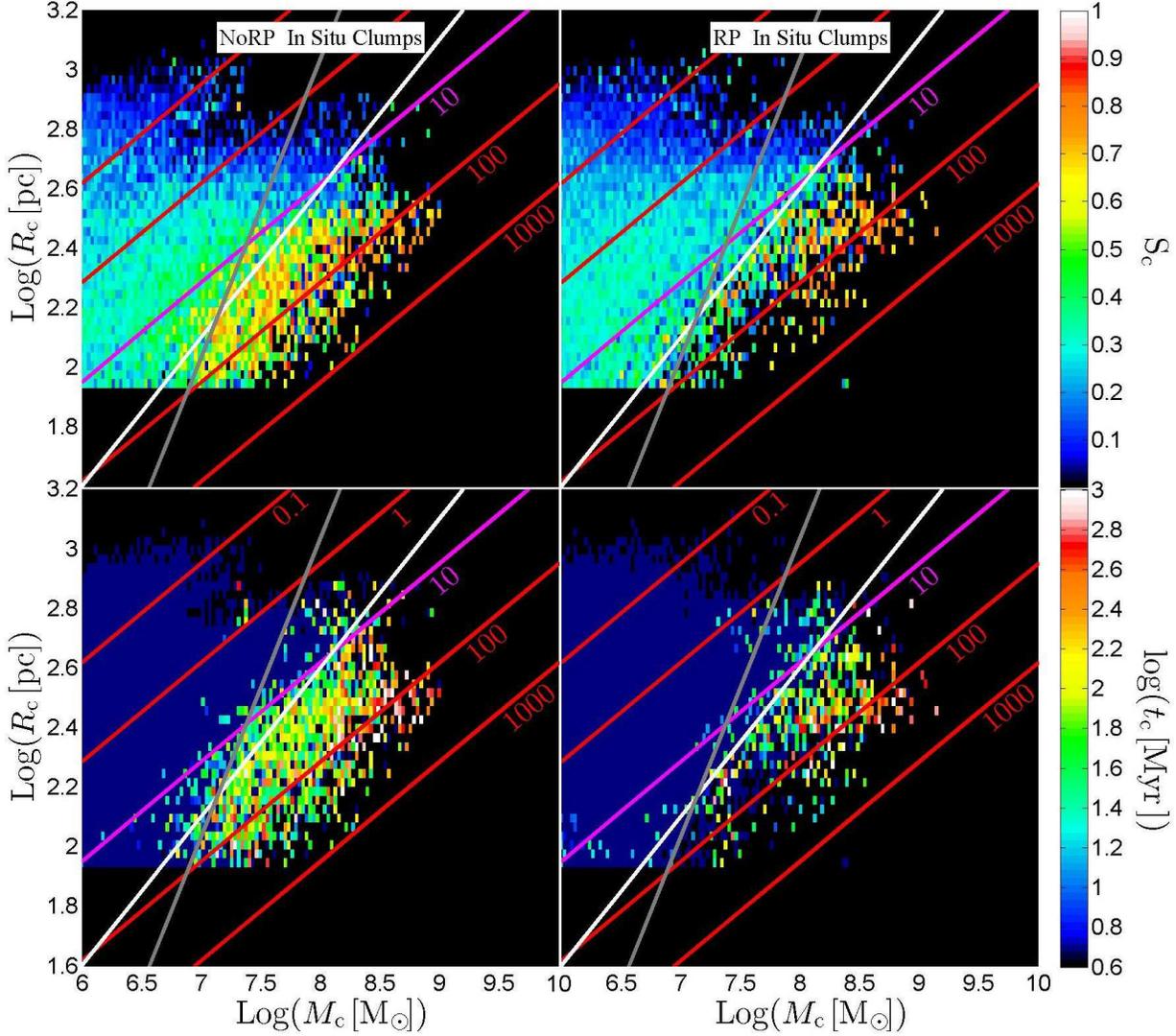}}
\caption{In situ clumps in NoRPs (left) and RPs (right). Each panel shows the 
distribution of clumps in the mass-radius plane. The red and magenta lines 
represent constant baryonic density, $n_{\rm bar}\cdot{\rm cm^3}=0.1-1000$ 
from top to bottom, as marked in the figure. In each panel, the grey line 
represents a constant circular velocity $V_{\rm circ}=20\kms$ while the white 
line represents a constant baryonic column density of $\Sigma_{\rm bar}=200\msun\,{\rm pc}^{-2}$. 
The color bar in each panel represents the median value per pixel of the 
respective property. The top row refers to the clump shape parameter, $S_{\rm c}$, 
and the bottom row to the clump time, $\tc$ in $\Myr$. Clumps with $\tc=0$ 
were artificially set to dark blue so that they show up on the plot. Clumps 
do not have densities greater than $\nc\gsim 300\cmc$, so the sample is complete 
at masses $\Mc\gsim 10^{7.5}\msun$. In both feedback models, clumps with $\tc>0$ 
are round/oblate with average densities $n_{\rm c}\gsim 10\cmc$. RP affects 
clumps with $\Mc\lsim 10^{8.3}\msun$, reducing their densities and making them more 
prolate, thus leading to their rapid disruption. More massive clumps are largely 
unaffected by the additional feedback. In NoRPs, the number of clumps with $\tc>0$ 
declines for circular velocities $V_{\rm circ}\lsim 20\kms$, while in RPs it declines 
for surface densities $\Sigma\lsim 200\msun\,{\rm pc}^{-2}$.
}
\label{fig:M_R} 
\end{figure*} 

%%%%%%%%%%%%%%%%%%%%%%%%%%%%%%%%%%%%%%%%%%%%%%%%  
\section{The Effect of Radiative Feedback on clumps} 
\label{sec:clump_comp} 

\smallskip
In this section we examine the effect of RP feedback 
on the formation, lifetime, and properties of giant 
clumps in high-$z$ disc galaxies. As in the previous 
section, we limit our analysis to the snapshots with 
both RP and NoRP versions, and discuss separately the 
effects on in situ, ex situ and bulge clumps.

\subsection{In Situ Clumps}
\label{sec:in_situ_comp} 

\subsubsection{Mass-size plane and sample completeness}
\smallskip
\Fig{M_R} shows the distribution of in situ clumps in the mass-radius 
plane, for the NoRPs (left) and the RPs (right). In all panels, the 
color represents the median value within each pixel of a respective 
property. The top row shows the clump shape, $\Sc$, and the bottom 
row shows the clump time, $\tc$. Cases where $\tc=0$ were artificially 
assigned a value of ${\rm log}(\tc)=0.7$ so that they appear 
on the plot in dark blue. The red and magenta lines represent 
constant baryonic volume densities of $n_{\rm bar}=0.1-1000 \cmc$ 
as marked. RP feedback reduces the typical density of in situ clumps by 
roughly $0.5~{\rm dex}$, similar to the factor by which the typical ISM 
density (by volume) is reduced (\fig{dens_PDF}, right). This is expected 
since we define clumps based on local overdensity. 

\smallskip
It is clear from the distribution of pixels in each diagram 
that our sample of clumps is incomplete at masses $\Mc\lsim 10^{7.5}\msun$, 
due to our minimal clump volume, $V_{\rm min}=8\times (70\pc)^3$. 
At a given mass, $\Mc$, we can only resolve clumps with densities 
$\nc<\Mc/V_{\rm min}$.  Due to the maximal ISM density we resolve, 
there is a maximal clump density, $n_{\rm max}$, so that we are 
complete for all masses $\Mc>V_{\rm min}n_{\rm max}$. For 
$\Mc=10^{7.5}$, $10^{7.2}$ and $10^7 \msun$, we can resolve all 
clumps with $\nc\lsim 340$, $200$ and $110\cmc$ respectively. We 
estimate the incompleteness at these three masses by calculating 
the fraction of clumps with $\Mc>10^{7.5}\msun$ that have densities 
above these thresholds. The corresponding fractions are $\sim 1\%$, 
$4\%$ and $9\%$, and are very similar in both sets of simulations. 
We conclude that our sample is reasonably complete for clumps with 
$\Mc\gsim 10^7\msun$. Hereafter, when we discuss the sample of 
clumps, we refer only to this mass complete sample.

\subsubsection{Clump lifetimes}
\smallskip
From \fig{M_R}, we learn that clumps with $\tc>0$ are rather round and 
oblate, $\Sc\gsim 0.5$, and dense, $\nc\gsim 10 \cmc$. This corresponds 
to the scale where the ISM becomes self-gravitating, allowing clumps 
to collapse to bound, oblate, long-lived structures. At lower densities, 
turbulence prevents the clumps from collapsing, and they remain extended, 
typically prolate structures. Such structures are rapidly mixed back into 
the ISM through a combination of turbulence and shear, or are torn apart 
by tidal forces. The oldest clumps are typically $\sim 300-500\Myr$ old, 
comparable to the expected migration time for clumps in high redshift discs 
\citep{DSC}. This will be discussed further in \se{grad}. 

\begin{figure}
\subfloat{\includegraphics[width =0.48 \textwidth]{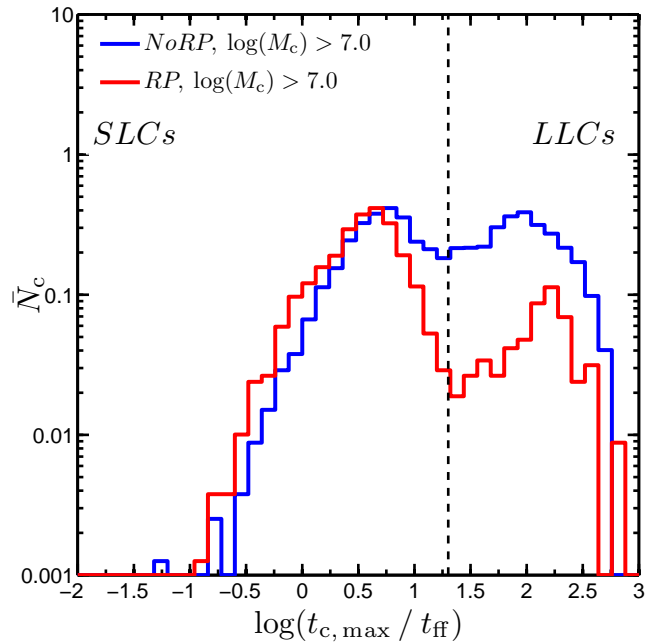}}
\caption{Distribution of clump lifetimes, $\tcm$, normalized to the 
(mass-weighted average) clump free fall time, $t_{\rm ff}\simeq 0.54/\sqrt{G\rho}$. 
Blue (red) curves show the distribution for NoRPs (RPs). There is a 
bi-modal distribution, separated at $t_{\rm c,\,max}\sim 20 t_{\rm ff}$ 
as marked by the vertical dashed line, which distinguishes short-lived 
(SL) from long-lived (LL) clumps. The RPs have $\sim 54\%$ as many clumps 
as the NoRPs, but only $\sim 19\%$ as many LLCs. 
}
\label{fig:lifetime} 
\end{figure} 

\smallskip
\Fig{lifetime} shows the distribution of clump lifetimes in the NoRPs 
(blue) and the RPs (red). The $y$ axis shows the average number of clumps 
per snapshot per bin of ${\rm log}(\tcm)$, shown on the $x$ axis. The clump 
lifetime has been normalized by the mass-weighted average free-fall time 
during its lifetime 
\be 
t_{\rm ff}=\sqrt{\frac{3\pi}{32G\rho}}\simeq \frac{0.54}{\sqrt{G\rho}}
\ee

\smallskip
There is a bi-modal distribution with a minimum near $\tcm\simeq 20t_{\rm ff}$. 
This is in the ball park of the output time step in the simulations, and 
it can therefore separate clumps that were identified in one or multiple 
snapshots. We also note that $\tcm\simeq 20t_{\rm ff}$ is the maximal clump 
lifetime observed in isolated galaxy simulations with much stronger radiative 
feedback \citep{Hopkins12a}. We therefore use this threshold to distinguish 
short-lived clumps (SLCs) from long-lived clumps (LLCs). In both sets of 
simulations, only $\sim 6\%$ of clumps that can be tracked for more than 
one timestep fall into the SLC category, and these make up less than $1\%$ 
of the SLCs themselves. 

\smallskip
Ignoring ZLCs, the RPs have $\sim 54\%$ as many clumps as the NoRPs, 
and $\sim 19\%$ as many LLCs. In the NoRPs, $\sim 49\%$ of clumps are 
LLCs compared to only $\sim 17\%$ in the RPs. We conclude that RP has 
a minor effect on clump formation, but it dramatically reduces the number 
of long-lived clumps. 

\begin{figure*}
\centering
\subfloat{\includegraphics[width =0.344 \textwidth]{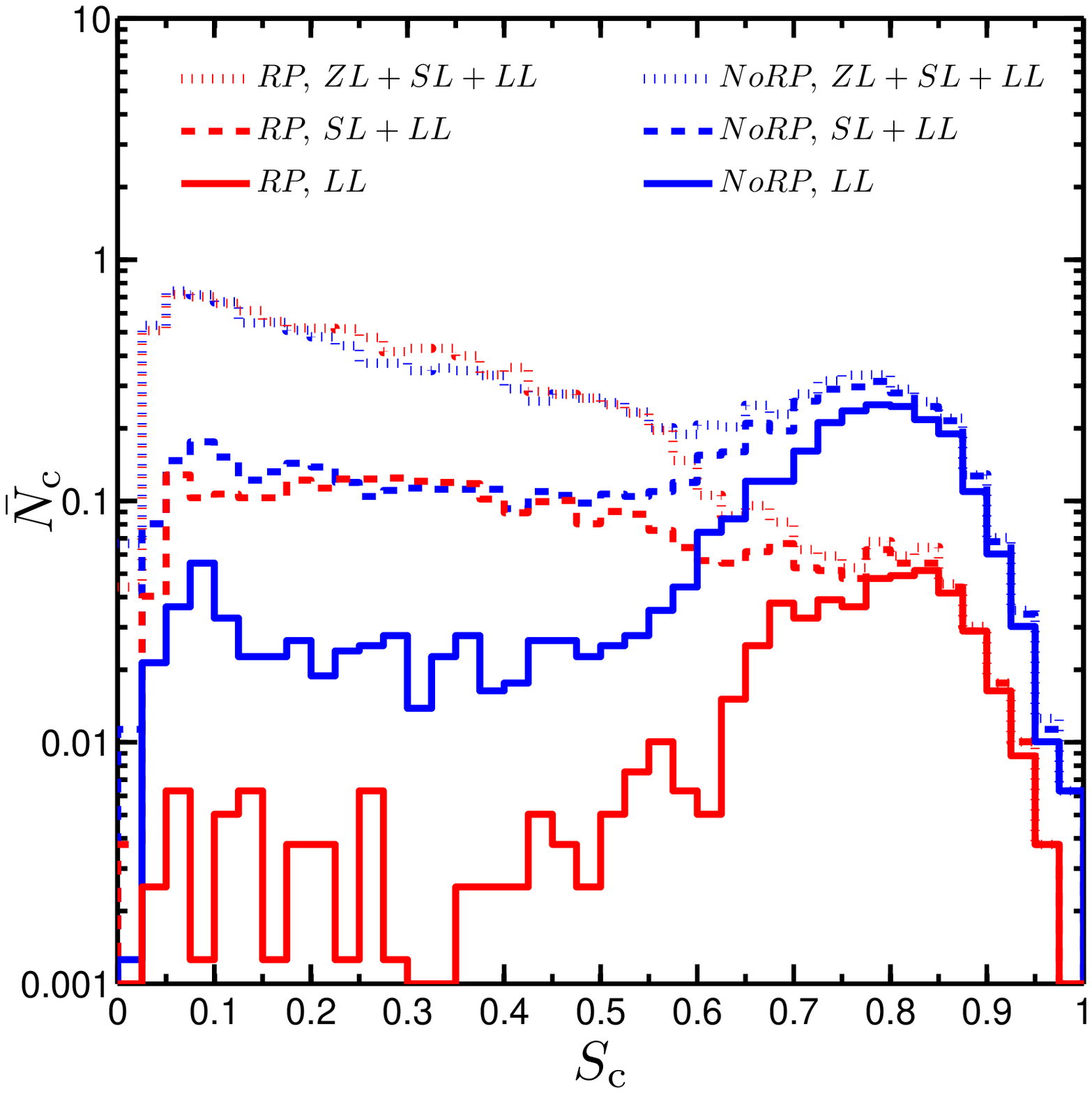}}
\hspace{-2.30mm}
\subfloat{\includegraphics[trim={1.76cm 0.10cm 0 0}, clip, width =0.302 \textwidth]{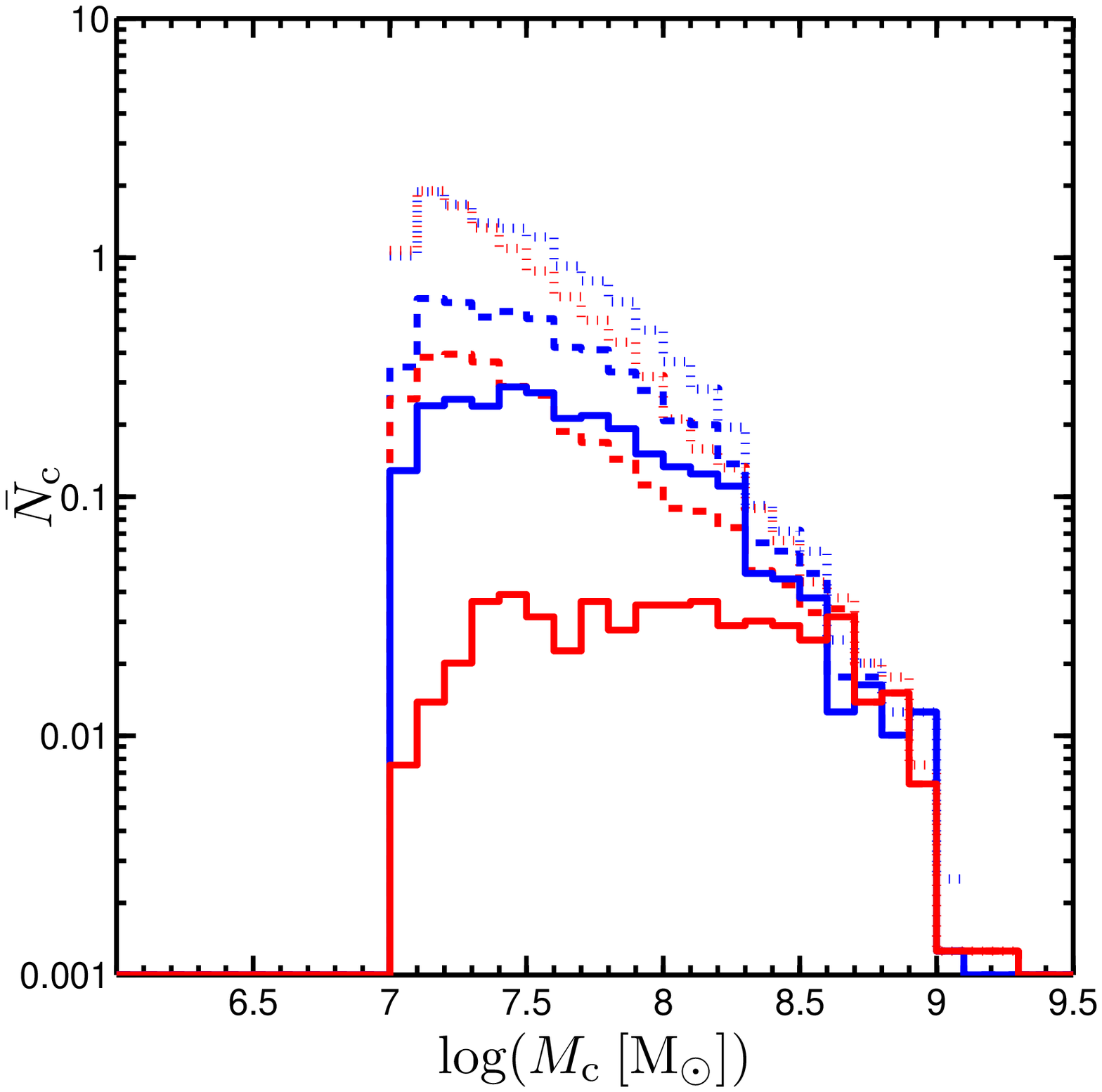}}
\hspace{-0.70mm}
\subfloat{\includegraphics[trim={1.76cm 0.10cm 0 0}, clip, width =0.301 \textwidth]{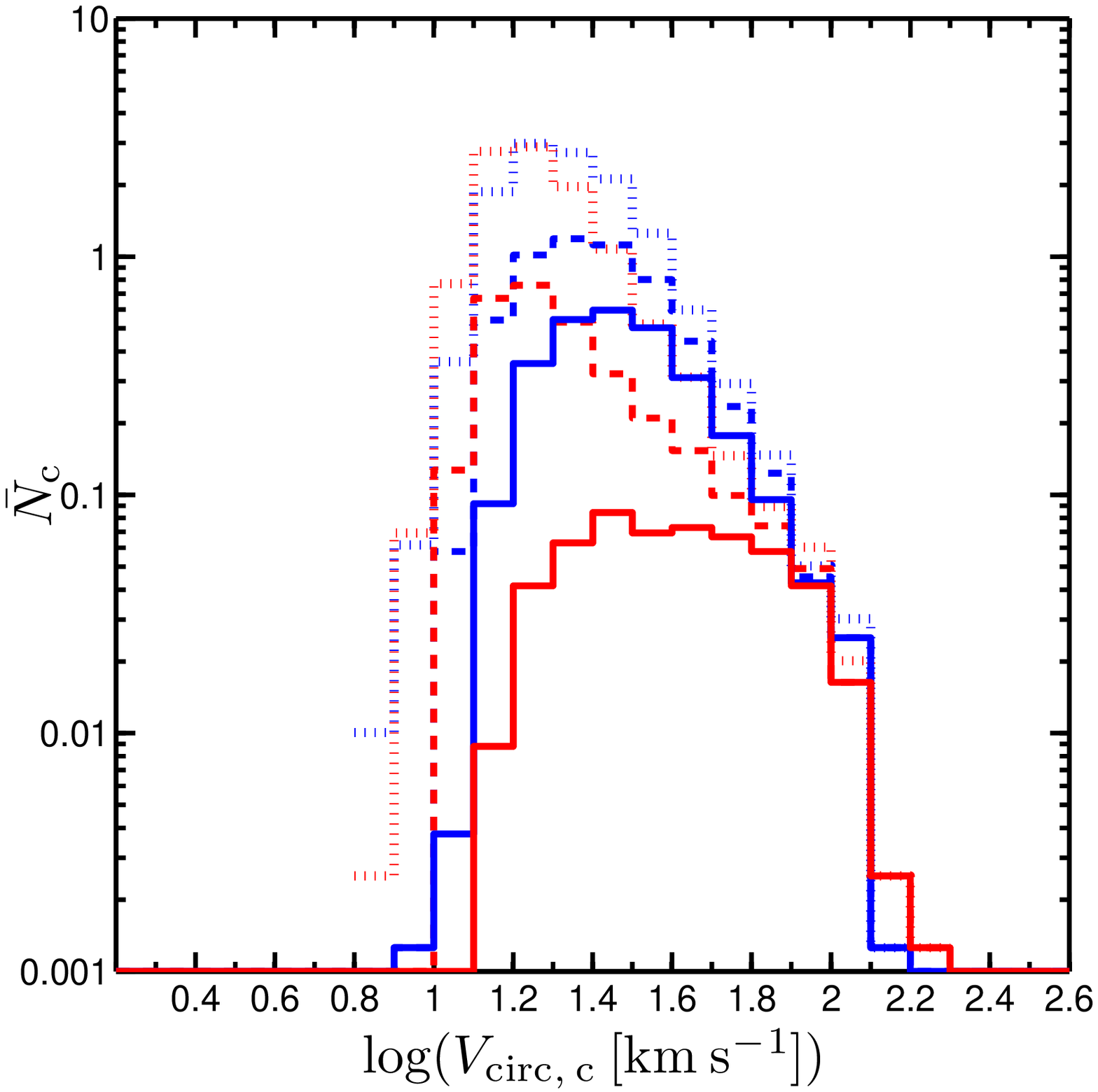}}\\
%\subfloat{\includegraphics[width =0.33 \textwidth]{residual_hist.eps}}
\subfloat{\includegraphics[width =0.332 \textwidth]{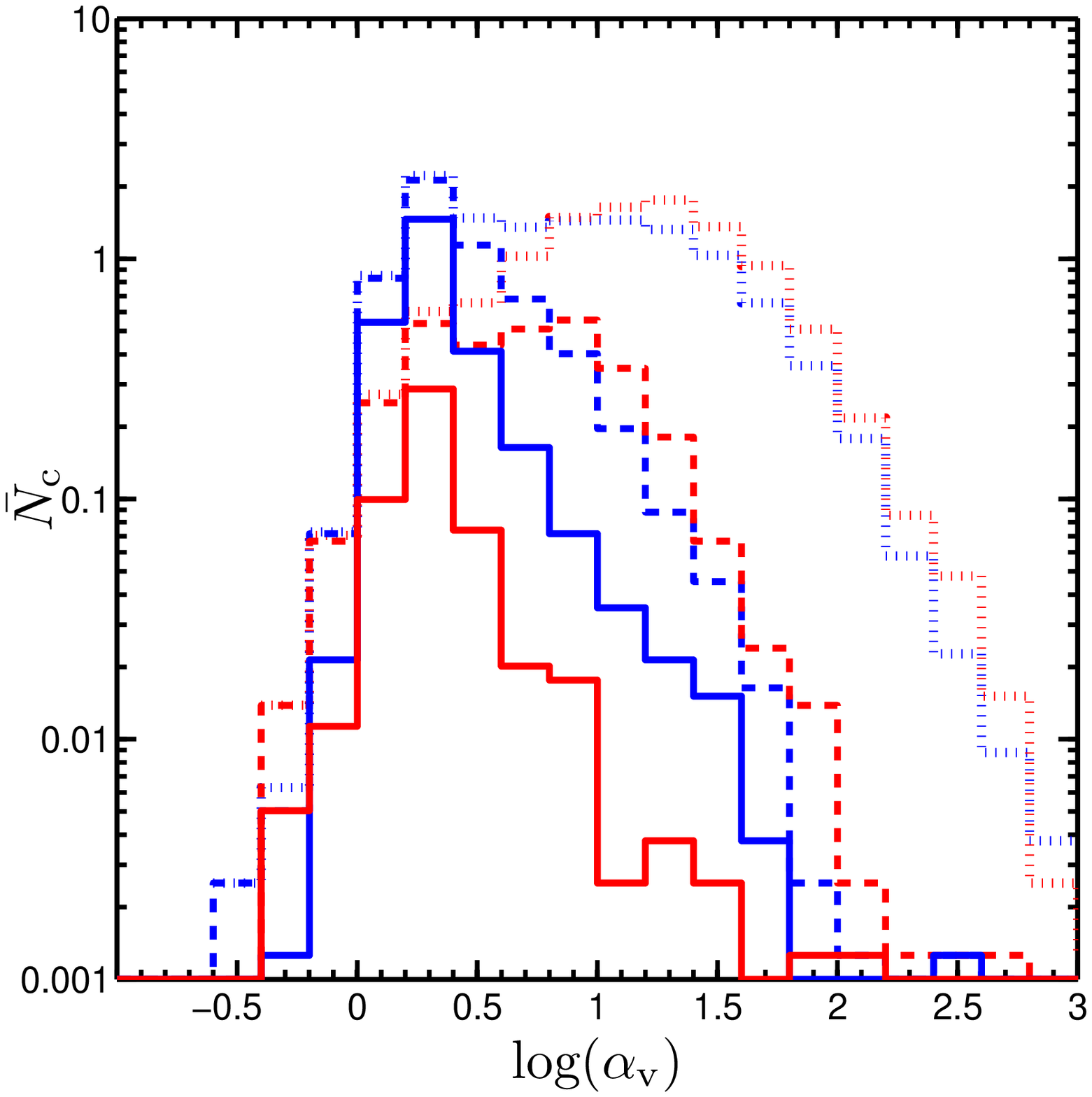}}
\hspace{+0.03mm}
\subfloat{\includegraphics[trim={1.76cm 0 0 0}, clip, width =0.302 \textwidth]{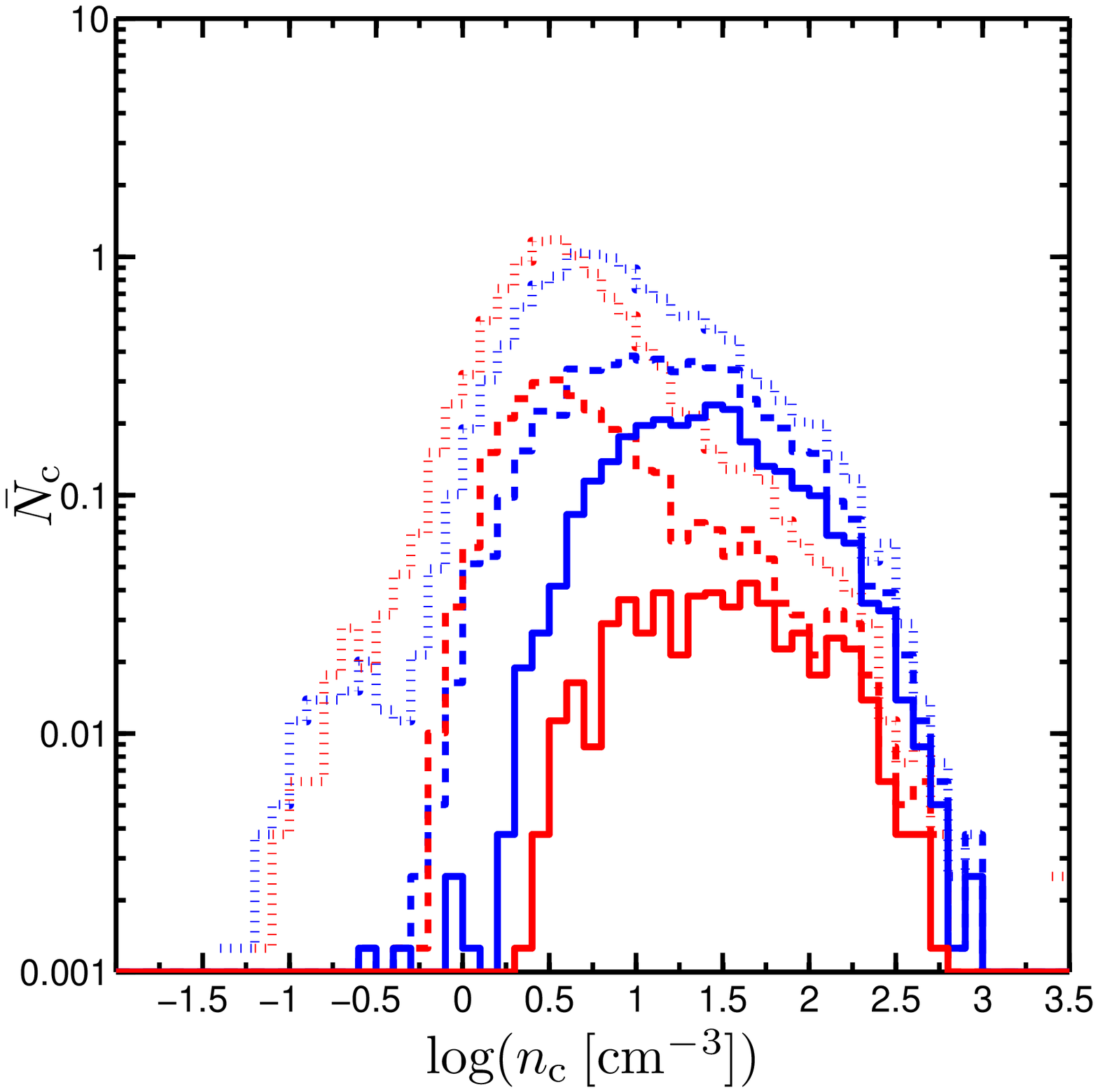}}
\hspace{-0.70mm}
\subfloat{\includegraphics[trim={1.76cm 0 0 0}, clip, width =0.2975 \textwidth]{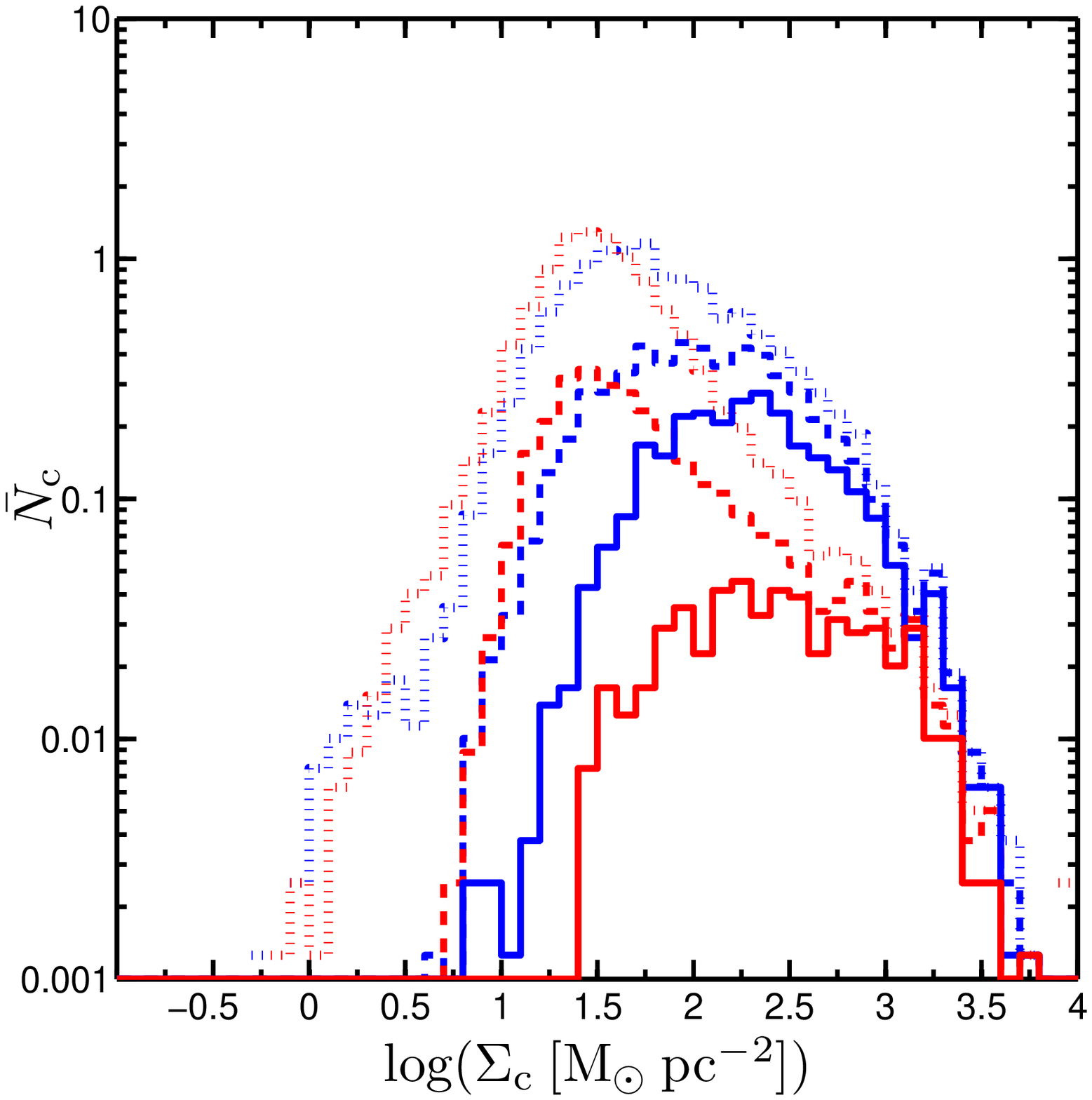}}\\
\caption{Properties of clumps in NoRP vs. RPs. In each 
panel we show the distribution of a different property, $x$, among 
in situ clumps in the simulations. The $y$ axis is the average number 
of clumps per snapshot per bin of $x$. In all panels, the blue/red 
lines show the distributions for NoRPs/RPs respectively. The dotted 
lines refer to all clumps including ZLCs, the dashed lines refer to 
non-ZLCs with $\tcm>0$, and the solid lines refer to LLCs only. 
\textit{Top left:} Clump shape, $\Sc$. \textit{Top centre:} Clump 
mass, $\Mc$. \textit{Top right:} Clump circular velocity, $V_{\rm circ,\,c}=\sqrt{G\Mc/\Rc}$. 
\textit{Bottom left:} Clump virial parameter $\alpha_{\rm v}=\sigma^2\Rc/(G\Mc)$ 
\textit{Bottom centre:} Clump baryonic volume density, $\nc=3\Mc/(4\pi \Rc ^3)$.
\textit{Bottom right:} Clump baryonic surface density, $\Sigma_{\rm c}=\Mc/(\pi \Rc ^2)$. 
The ZLCs have low densities and elongated shapes and are clearly unbound with 
$\alpha_{\rm v}\sim 10-1000$. LLCs typically have densities $\nc \gsim 10-30$ and 
are oblate or close to spherical, in agreement with \fig{M_R}. Most of them are 
close to virial equilibrium, with $\alpha_{\rm v}\sim 1-3$. RP feedback rapidly unbinds 
clumps with surface densities below $\sim 30 \msun \pc^{-2}$, and these SLCs have 
virial parameters $\alpha_{\rm v}\sim 3-30$. Clumps with 
$\Sigma_{\rm c}\gsim 300 \msun\:\pc^{-2}$, more massive than $\sim 10^{8.5} \msun$, 
appear unaffected by RP, while clumps with $V_{\rm circ,\,c}\lsim 20\kms$ are rapidly 
disrupted even in NoRPs. 
}
\label{fig:clump_compare} 
\end{figure*} 

\subsubsection{Clump structural and virial properties}
\smallskip
In \fig{clump_compare}, we compare structural and virial properties 
of clumps in the RPs and NoRPs, to try and discern what causes clump disruption. 
In each panel we show the distribution of a different property, represented 
on the $x$ axis, where the $y$ axis shows the average number of clumps per 
snapshot per bin of $x$, as in \fig{lifetime}. In clockwise order from the 
top left the properties are clump shape, $\Sc$; mass, $\Mc$; circular velocity, 
$V_{\rm circ,\:c}=\sqrt{G\Mc/\Rc}$; surface density, $\Sigma_{\rm c}=\Mc/(\pi\Rc^2)$; 
volume density, $\nc = 3\Mc/(4\pi\Rc^3)$; and virial parameter 
$\alpha_{\rm v}=2E_{\rm K}/|U|$, i.e. twice the ratio of kinetic to potential energies 
within the clump. For each property, we show the distributions including ZLCs, the 
distributions for non-ZLCs, and the distributions for LLCs only, as marked in the legend. 

\smallskip
The virial parameter is measured in a sphere of radius $\Rc$ centered on the 
peak baryonic density within the clump. This is a valid approximation for nearly 
spherical clumps and can serve as an order of magnitude estimate for elongated clumps 
as well. In the rest-frame defined by the center of mass velocity of baryons 
within the sphere, we measure the total baryonic kinetic energy, which gives us the 
mass-weighted 3D velocity dispersion, $\sigma^2=2E_{\rm K}/\Mc$. The virial 
parameter is
\be  
\label{eq:virial1}
\alpha_{\rm v} = a~\frac{\sigma^2\Rc}{G\Mc},
\ee
{\no}where $a$ is a constant of order unity that depends on the density profile 
within the clump. If the density profile is a power law in the clump-centric radius, 
$\rho\propto r^{-\beta}$, then 
\be 
\label{eq:virial2}
a = \frac{5-2\beta}{3-\beta}.
\ee
{\no}For a constant density sphere, $\beta=0$ and $a=5/3$. Previous studies of clumps in 
similar simulations to the ones used here as well as high-resolution simulations of isolated 
galaxies suggest that most clumps can be well approximated as having $\beta=2$ \citep{Ceverino12}. 
We adopt this approximation, which yields $a=1$ and $\alpha_{\rm v}= \sigma^2\Rc/(G\Mc)$, i.e. 
0.6 times the value for a constant density sphere.

\smallskip
The ZLCs comprise the majority of the sample, $\sim 84\%$ and $\sim 74\%$ in the 
RPs and NoRPs respectively. However, they are characterised by low masses and densities, 
filamentary shapes, and very large virial parameters, $\alpha_{\rm v}\sim 10-1000$, 
well above the SLCs which have $\alpha_{\rm v}\sim 3-30$. We conclude that the majority 
of ZLCs are slightly overdense patches of the disc, expected in any turbulent medium, and 
were not at any point gravitationally bound clumps. We hereafter ignore them, and use 
``clumps" to refer only to SLCs and LLCs, with masses $\Mc>10^7\msun$. 

\smallskip
LLCs appear to be close to virial equilibrium, with $\alpha_{\rm v}\sim 1-3$. 
Had we assumed a slope of $\beta=-2.3$ rather than $-2$ for the density profile, 
the peak of the distribution would have been at $\alpha_{\rm v}=1$. Other studies 
which have claimed that all clumps are short-lived \citep{Hopkins12a,Oklopcic16} 
measured the virial parameters of clumps in there simulations assuming constant 
density, i.e. $\beta=0$. They found that all clumps have $\alpha_{\rm v}>1$, while 
many have $\alpha_{\rm v}\lsim 3$, and claim this shows all clumps to be unbound. 
In our case, assuming $\beta=0$ shifts the peak of the distribution for LLC to 
$\alpha_{\rm v}=3$, consistent with these other studies. This shows that it is 
important to account for the density profile in the clump when using the virial 
parameter to asses whether or not the clump is bound. 

\smallskip
LLCs have a strong tendancy to be oblate or spherical, as we saw as 
well in \fig{M_R}. In both RPs and NoRPs, clumps with $\Sc\gsim 0.8$ 
are almost all long-lived, while clumps with $\Sc \lsim 0.6$ are almost 
all short-lived. Since clump shape correlates with density so that rounder 
clumps are denser, the LLCs are less prone to disruption by shear and tidal 
forces within the disc.

\smallskip
Very massive clumps, with $\Mc\gsim 10^{8.5}\msun$, are unaffected by the 
inclusion of RP. The number of such clumps is the same in the NoRPs and RPs, 
and $\sim 76\%$ of them are long-lived. Below this mass, there are fewer clumps 
in the RPs, and the slope of the mass function is slightly shallower. 
This is in qualitative agreement with the shallower slope in the high density 
power-law tail of the ISM mass-weighted density PDF (\fig{dens_PDF}). Below 
${\rm log}(\Mc)\sim 7.3$, the SLC mass function flattens and the number of LLCs 
begins to decline. This must be due in part to incompleteness, though there may 
be an additional physical effect disrupting clumps below this mass scale even without 
RP, as discussed below.

\smallskip
Due to an overall decrease in densities, clumps in RPs have shallower potential 
wells, making them more susceptible to ejective feedback from supernova. The number 
of LLCs begins to decline at $V_{\rm circ}\lsim 25\kms$, suggesting a typical scale 
for clump disruption by SN feedback. In the range $40 \lsim V_{\rm circ} \lsim 65 \kms$, 
nearly all clumps in the NoRPs are long-lived while there is a non-negligible fraction 
of SLCs in the RPs. This suggests that RP feedback boosts the efficiency of supernova 
feedback, by lowering the typical densities.

\smallskip
When RP is included, there are fewer very dense clumps, consistent with the 
overall decrease in ISM densities. Most LLCs have densities $\nc \sim 10-30 \cmc$, 
as expected from the power-law portion of the ISM density PDF (\fig{dens_PDF}), 
though densities in the range $\nc \sim 3 - 200 \cmc$ are common. Nearly all 
clumps with densities greater than $\sim 50\cmc$ or surface densities greater 
than $\sim 300 \msun \pc^{-2}$ are LLCs, in both RP and NoRPs. On the other hand, 
in the RPs there is a sharp decrease in the number of LLCs with surface densities 
below $\sim 30  \msun \pc^{-2}$.

\subsubsection{Clump disruption scale}
\smallskip
The grey lines in each panel of \fig{M_R} represent a constant circular 
velocity of $V_{\rm circ}=20\kms$. This seems to mark a transition between 
ZLCs and non-ZLCs in the NoRPs (bottom left panel), seen as well in the 
top-right panel of \fig{clump_compare}. Based on \citet{Dekel86}, 
\citet{DSC} showed that if the star-formtation efficiency per free-fall 
time in the clump is $\epsilon_{\rm ff}\sim 10\%$, then clumps with circular 
velocities $V_{\rm circ}\lsim 30\kms$ can be disrupted by supernova feedback. 
In our simulations, the star formation efficiency averaged over clumps is 
typically $\epsilon_{\rm ff}\sim 0.03-0.04$, though values as high as 
$10\%$ are not uncommon. This yields a minimum potential depth of 
$V_{\rm circ}\sim 20-30\kms$ below which clumps should be susceptible 
to ejective feedback from supernova, which may explain why the number 
of LLCs decreases below this value even in NoRPs. 

\smallskip
This may also be related to the decrease in the number of LLCs 
below $\Mc\sim 10^{7.3}\msun$ (\fig{clump_compare}). On the one hand, 
clumps must have $V_{\rm circ,\:c}=\sqrt{G\Mc/\Rc}\gsim 20\kms$ to survive. 
On the other hand, the self-gravitating regime of the ISM is at densities 
$\nc=3\Mc/(4\pi\Rc^3)\gsim 10\cmc$ and the number of LLCs declines towards 
lower densities (\fig{clump_compare}). These two scales meet at a unique mass, 
which is $\Mc\simeq 10^{7.4}\msun$. At lower masses, clump survival is limited 
by the potential-well $\propto \Mc/\Rc$, while at higher masses the clumps are 
limited by density $\propto \Mc/\Rc^3$, resulting in a kink in the mass function. 
Unfortunately, as this is very close to our incompleteness limit, it is 
difficult to draw firm conclusions from the simulations.

\smallskip
When RP is included, clump survival becomes limited by surface density. 
This is apparent in \fig{M_R}, where the white lines mark a constant 
surface density of $\Sigma_{\rm c}=200 \msun\:\pc^{-2}$, and the 
number of non-ZLCs in the RPs sharply declines to the left of this line 
(bottom right panel). \Fig{clump_compare} shows that above 
$\Sigma_{\rm c}=300 \msun\:\pc^{-2}$, nearly all clumps are long lived 
while there are hardly any LLCs with $\Sigma_{\rm c}<30 \msun\:\pc^{-2}$. 
A clump with $\Sigma_{\rm c}=300 \msun\:\pc^{-2}$ and $\nc=10\cmc$ has a 
mass of $\Mc\sim 10^{8.5}\msun$, explaining why above this mass, nearly 
all clumps are long-lived. 

\smallskip
To get a rough understanding for where this scale comes from, we consider 
the very simple case of a clump of mass $M$ and radius $R$ that forms stars 
in a single burst in a small region at its centre, and ask when the radiation 
pressure from this burst is enough to unbind the remaining clump. Equating 
the inward gravitational force acting on a spherical shell at radius $R$ to 
the outward force of the radiation pressure, assuming a single scattering per 
photon, yields the equation
\be 
\label{eq:unbind}
\frac{GM}{R^2} = \frac{L}{4\pi R^2 c {\tilde{\Sigma}}}
\ee
{\no}where ${\tilde{\Sigma}} = M/(4\pi R^2)$ is the mass per 
unit area which the radiation is pushing. This yields a critical 
surface density of 
\be 
\label{eq:Sig_crit}
\Sigma = 4{\tilde{\Sigma}} = \frac{L}{M \pi Gc} = \frac{\Gamma f_*}{\pi Gc} \simeq 380 f_* \msun\,\pc^{-2}
\ee
{\no}where $\Gamma=10^{36}~{\rm erg~s^{-1}~\msun^{-1}}$ is the parameter used in our 
simulations (\equ{RP}), and $f_* = M_*/M$ is the fraction of the clump mass turned into 
stars. Clumps with $\Sigma\sim 30 \msun\,\pc^{-2}$ become unbound after turning only 
$\sim 10\%$ of their mass into stars, while clumps with $\Sigma\sim 200 \msun\,\pc^{-2}$ 
are able to turn half their mass into stars before blowing out the remaining gas, allowing 
them to remain bound.

\smallskip
This model is too simplistic because the radiation pressure in the simulations only affects 
the cells neighbouring star particles younger than $5\Myr$ old, and not all the stars form 
in a single cell in a single timestep. In addition, the clump is non-spherical, the medium 
is non-uniform and supersonic turbulence also adds effective pressure as does the net 
effect of supernova feedback. Nevertheless, this crude estimate gives a sense for the 
range of surface densities observed in LLCs. We also note that more realistic models 
which estimate the scale for rapid disruption of giant clumps assuming weak radiation 
trapping \citep[e.g.][and references therein]{KrumholzDekel}, also predict that clumps 
with $\Sigma \gsim 100 \msun\:\pc^{-2}$ and $\Mc\gsim 10^{7.5}$ with $\epsilon_{\rm ff}\sim 0.03$ 
are rapidly disrupted, while clumps with $\Sigma \gsim 300 \msun\:\pc^{-2}$ and 
$\Mc\gsim 10^{8.5}$ are not.

\smallskip
To summarise, bound star-forming clumps only form at densities $\nc\gsim 10 \cmc$ 
where the ISM is self-gravitating. Clumps with $V_{\rm circ,\:c}\lsim 20\kms$ are 
rapidly disrupted by ejective feedback from supernovae, and in the RPs, clump 
disruption can occur even $V_{\rm circ,\:c}\lsim 40\kms$. Radiation pressure 
introduces a critical surface density for clump survival, where clumps with 
$\Sigma_{\rm c}<30 \msun\:\pc^{-2}$ are rapidly disrupted, while clumps with 
$\Sigma_{\rm c}>300 \msun\:\pc^{-2}$ are all long lived. The interplay between 
these three scales results in a kink in the clump mass function at 
$\Mc \sim 10^{7.3}\msun$ and in massive clumps with $\Mc\gsim 10^{8.5}$ being 
long-lived. However, we note that the number of LLCs in the NoRPs decreases 
below $\Sigma_{\rm c} \sim 30 \msun\:\pc^{-2}$ as well, albeit less sharply than 
in the RPs, since these clumps also have circular velocities close to the threshold. 
It is therefore difficult to quantitatively asses to what extent RP feedback alters 
the fundamental physical property responsible for clump disruption.

\subsection{Bulge and Ex Situ Clumps}
\smallskip
RP has little effect on the central bulge clumps. A central clump is 
identified in $\sim 83\%\:(96\%)$ of all galaxies in the RPs (NoRPs) 
respectively, and in $\sim 97\%\:(100\%)$ of the galaxies 
at $z<4$. Nearly all the bulge clumps have high densities and masses, 
$n>10\cmc$ and $\Mc>10^{8.5}$, and are thus largely unaffected by the 
RP feedback, showing similar distributions in both sets of simulations. 
Their typical masses are $\sim 10^{9.5}\msun$, or $\sim 0.3-0.5\Md$, 
and their typical SFRs are $\sim 0.5{\rm SFR_d}$.

\smallskip
On the other hand, RP feedback greatly reduces the number of ex situ 
clumps. Ex situ clumps are merging galaxies, and we may have expected 
similar numbers in the two sets of simulations, since the cosmic evolution 
is meant to be the same. However, even among very massive clumps with 
$\Mc>10^{8.5}\msun$ in baryons, which based on the fate of the in-situ 
clumps with such high masses are not expected to be strongly affected 
by RP, there are $\sim 3.5$ times as many ex situ clumps in the NoRPs. 
The reduction in the number of ex situ clumps likely results from RP 
acting on these merging dwarf galaxies at early times, when they were 
less massive, and puffing them up. The increased feedback causes more 
intense outflows and lowers the central densities of the merging galaxies, 
making them more vulnerable to tidal stripping as they merge with the main 
host. A detailed study of the evolution of these ex situ clumps is beyond 
the scope of this paper. 

\smallskip
Ex situ clumps are $\sim 10\%$ and $\sim 5\%$ of the off centre clumps 
in number, $\sim 35\%$ and $\sim 37\%$ in mass and $\sim 15\%$ and $\sim 29\%$ in 
SFR in the NoRPs and RPs, respectively. Among the most massive clumps with $\Mc>10^{8.5}\msun$, 
the ex situ contribution increases to $\sim 58\%$ and $\sim 36\%$ in number, 
$\sim 75\%$ and $\sim 60\%$ in mass, and $\sim 51\%$ and $\sim 56\%$ in SFR. 
This is comparable to the ex situ contribution found by M14, where the simulated 
galaxies were more massive by a factor of $\sim 5-10$, and the typical clump mass 
was $\Mc \gsim 10^8 \msun$.

\begin{figure*}
\centering
\subfloat{\includegraphics[width =0.45 \textwidth]{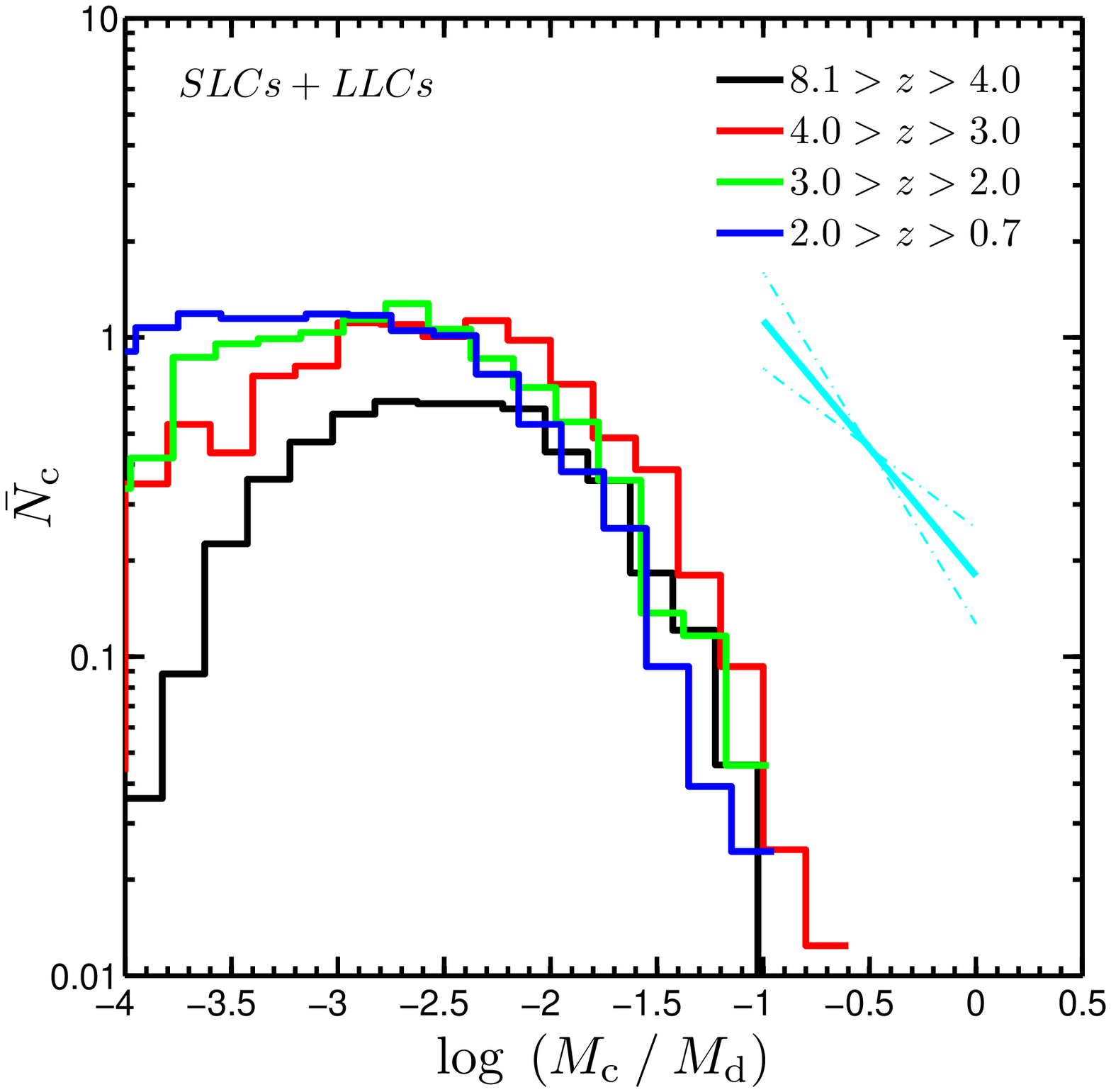}}
\hspace{-0.5mm}
\subfloat{\includegraphics[trim={1.95cm -0.06cm 0 0}, clip, width =0.399 \textwidth]{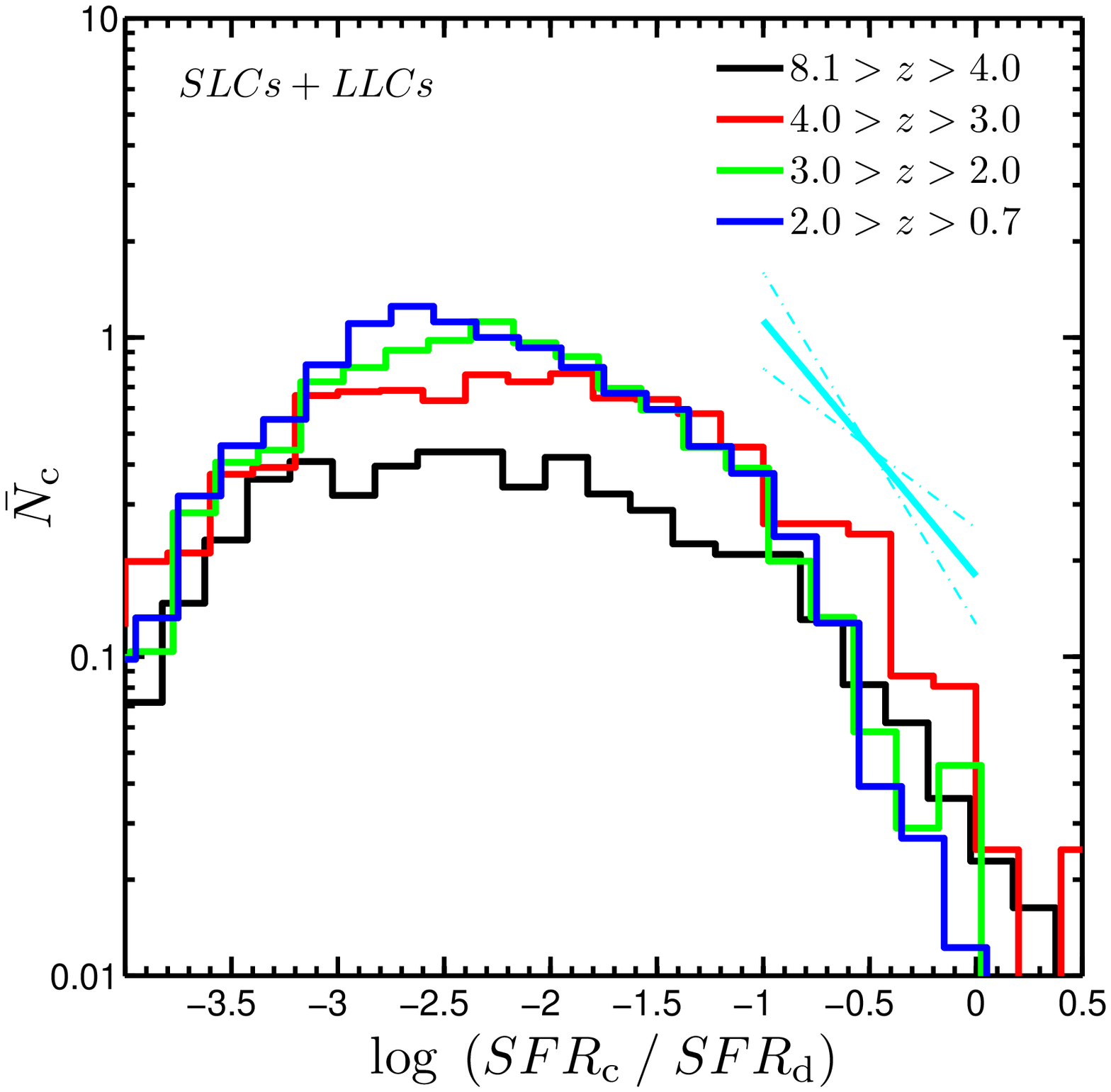}}
\caption{Mass function and SFR function for in situ clumps in the 
RPs. The $x$ axes show the clump mass normalized to the disc mass, $m=\Mc/\Md$ 
(left), and the clump SFR normalized to the disc SFR, $s=SFR_{\rm c}/SFR_{\rm d}$ 
(right). The $y$ axis shows the average number of clumps per snapshot per 
logarithmic interval of $x$. Different colours represent different redshift 
bins, as marked in the legend. The cyan lines mark slopes of $\alpha=1.5,1.8,2.1$ 
where $dN/dx \propto x^{-\alpha}$. The peak of the mass function decreases 
from ${\rm log}(m)\sim -2.5$ at $z>3$ to $\sim -3.5$ at $z<2$, likely due to the 
fact that our disc masses grow monotonically in time, while the minimal resolved 
clump mass remains fixed. In the range $-2\lsim {\rm log}(m) \lsim -1$ the mass 
function can be described by a power-law with a slope slightly shallower than $-2$, 
and independent of redshift. The overall behaviour of the SFR function appears similar 
at all redshifts, and is nearly identical in the two latest redshift bins, at $2<z<3$ 
and $z<2$ (green and blue lines respectively). The peak is shifted slightly from 
${\rm log}(m)\sim -2$ at $z>3$ to $\sim -2.5$ at $z<2$, though as with the mass 
function, this peak may be due to our limited resolution for clump detection while 
our galaxies grow monotonically with time. At values of $-2\lsim {\rm log}(s) \lsim -1$, 
the distribution can be described as a power law with a slope slightly shallower than 
$-2$, while the relation steepens at $-1\lsim {\rm log}(s)$. 
}
\label{fig:functions} 
\end{figure*} 

%%%%%%%%%%%%%%%%%%%%%%%%%%%%%%%%%%%%%%%%%%%%%%%%  
\section{Properties of Clumps in RP Simulations} 
\label{sec:obs} 
\smallskip
In this section we study physical properties of clumps 
that may be related to observations. We focus on 
the RPs as a better representation of the real Universe, 
and therefore use the full RP sample which consists of 
1130 snapshots: 320 at $4 \le z$, 161 at $3 \le z < 4$, 
241 at $2 \le z < 3$, 408 at $0.75 \le z < 2$. 

%--------------------------
\subsection{Mass Functions and SFR Functions} 
\label{sec:func} 

\smallskip
\Fig{functions} shows the distribution of masses and SFRs for in situ 
clumps normalized to the corresponding values in their host discs, 
$m\equiv\Mc/\Md$ and $s\equiv SFR_{\rm c}/SFR_{\rm d}$, for different 
redshift bins. While we refer here to the clump baryonic mass, we note 
that the stellar mass functions are rather similar, especially at $z<3$.

\smallskip
The peak of the mass function decreases from ${\rm log}(m)\sim -2.5$ at 
$z>3$ to $\sim -3.5$ at $z<2$. While this roughly coincides with the predicted 
fragmentation mass relative to the disc mass of $\delta_{\rm cold}^2$ \citep{DSC}, 
it seems likely that the peak in the mass function is due to the fact that our 
disc masses grow monotonically in time, while the minimal resolved clump mass 
remains fixed. Thus the characteristic value of $m$ becomes smaller at later 
times. In the range $-2\lsim {\rm log}(m) \lsim -1$, where we are complete, 
the mass function can be described by a power-law, $dN/dm \propto m^{-\alpha}$ 
with $\alpha \sim 1.7-1.8$, roughly independent of redshift. This is similar 
to observed slopes of GMC mass functions in the local galaxies 
\citep[e.g.][]{Williams97,Rosolowski05}. Note that a mass function with a slope 
of $\alpha=-2$ has equal mass on all scales, and tends to emerge generically in 
systems dominated by gravitational collapse \citep[e.g.][]{Press74,Bond91,Hopkins13}.

\smallskip
The distribution of SFR is broader, though it exhibits the same 
general shape as the mass function. The overall behaviour 
appears similar at all redshifts, and is nearly identical in the 
two latest redshift bins, $2<z<3$ and $z<2$ (green and blue lines 
respectively). The peak decreases slightly with time, from 
${\rm log}(m)\sim -2$ at $z>3$ to $\sim -2.5$ at $z<2$, though as 
with the mass function, this may be due to our limited resolution 
for clump detection while our galaxies grow monotonically with time. 
At values of $-2\lsim {\rm log}(s) \lsim -1$ the distribution can 
be described as a power law with a slope of $\sim -1.6$, slightly 
shallower than the mass function, though this steepens towards higher 
values. The maximal value of $s$ seems to decrease by $\sim 0.5~{\rm dex}$ 
from $4<z$ to $z<2$, indicating that individual giant clumps contribute 
less to the total SFR at later times. Values of ${\rm log}(s)>0$ can 
occur if clumps are not co-rotating with the disc, so that their stars 
are not considered ``disc stars" with $j_{\rm z}/j_{\rm max}>0.7$ and 
therefore are not counted towards the disc SFR. Associating SFR with UV 
luminosity, our SFR functions can be compared with observed UV luminosity 
functions of clumps. At $z<3$, more than $60\%$ of the clumpy galaxies 
in our sample have $9.8<{\rm log}(\Md)<10.6$, and our SFR function is 
similar to the UV luminosity function of clumpy galaxies in the same mass 
and redshift ranges \citep[][figures 7-9]{Guo15}.

%--------------------------
\subsection{Migration of Long-Lived Clumps}
\label{sec:grad} 

\smallskip
M14 predicted gradients in clump properties such as mass, stellar 
age and sSFR, with galactocentric distance due to clump migration. 
Observations of such gradients by \citet{Forster11b} and \citet{Guo12} 
have been interpreted as evidence of clump survival and migration in 
real galaxies. However, M14 did not track clumps in time and did not 
directly address gradients that may be present in the properties of 
SLCs at birth, due to gradients in the underlying disc properties. 
Here, we are able to directly compare LLCs and SLCs, and determine 
stronger observational constraints for the clump migration scenario. 
We also examine properties of the ex situ clump population and address 
differences between in situ and ex situ clumps. 

\begin{figure}
\subfloat{\includegraphics[width =0.475 \textwidth]{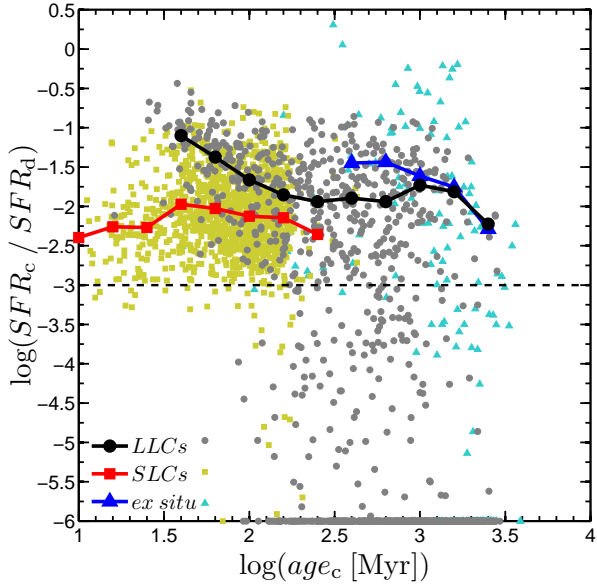}}
\caption{Clump SFR normalized to disc SFR, $s=SFR_{\rm c}/SFR_{\rm d}$, 
as a function of clump stellar age. Shown are LLCs (grey circles), SLCs 
(orange squares) and ex situ clumps (blue triangles), with masses $\Mc>10^7\msun$ 
in discs with masses $\Md>10^{10}\msun$ at redshifts $z<2.5$. Clumps with 
$s=0$ have been artificially placed at ${\rm log}(s)=-6$ so they show up 
on the plot. The horizontal dashed line at ${\rm log}(s)=-3$ marks our 
distinction between star-forming and quenched clumps.
%The dashed line at $s=-1.5$ represents the threshold used when 
%attempting to mimic observations of UV bright clumps (\se{gal_clump}). 
The thick black, red, and blue lines show the median values of ${\rm log}(s)$ 
in bins of stellar age for the LLCs, SLCs, and ex situ clumps respectively. 
Only star-forming clumps with ${\rm log}(s)>-3$ were considered when computing 
the medians. The bins have $\Delta {\rm log}(age_{\rm c})=0.2$ and the 
centres have been marked with corresponding symbols. We only include bins 
that contain at least 7 clumps. SLCs typically contribute $\sim 1\%$ of 
the disc SFR, with no significant age dependence, but with a large scatter 
of $\pm 0.5~{\rm dex}$. Young LLCs outshine SLCs, each contributing $\gsim 10\%$ 
of the disc SFR, but their SFR declines roughly as $age^{-1.5}$ and after 
$\sim 100\Myr$ they saturate at $\sim 1.5\%$, slightly higher than the 
median SLC. Ex situ clumps have ages of $\sim 1\Gyr$ with a very wide 
range of SFR.
}
\label{fig:age_sfr} 
\end{figure} 

\subsubsection{Sample selection}

\smallskip
We restrict our analysis in this section to the mass limited sample of 
clumps with $\Mc>10^7\msun$. This selection leaves $\sim 92\%$ of LLCs, 
but only $\sim 30\%$ of SLCs and $\sim 32\%$ of ex situ clumps. 

\smallskip
Furthermore, since the time between consecutive snapshots, $\Delta a=0.01$, 
is a larger fraction of the disc dynamical time at earlier times, our 
ability to track clumps for multiple timesteps and properly distinguish 
LLCs from SLCs becomes compromised at high redshifts. To avoid this issue, 
we limit our analysis to $z<2.5$, where the time between snapshots is roughly 
half an orbital time at the disc edge, and LLCs can be followed for an average 
of $3-4$ timesteps. This reduces the mass limited sample of LLCs by only $\sim 6\%$, 
consistent with our concern that properly identifying LLCs is more difficult at 
high redshift. The mass limited samples of SLCs and ex situ clumps are reduced 
by a further $\sim 34\%$ and $\sim 23\%$ respectively.

\smallskip
We reduce the sample further by only examining clumps in massive discs with 
$\Md>10^{10}\msun$. 
%There are two reasons for this additional cut. First of all, the normalized clump mass function (\fig{functions}) may be incomplete below $\Mc\sim 10^{-3} \Md$. Given our clump mass cut of $10^{7}\msun$, we only resolve the full clump mass spectrum for discs with $\Md>10^{10}\msun$. Secondly, 
This disc mass roughly corresponds to the transition between pre-compaction and 
post-compaction galaxies. Each galaxy in these simulations undergoes at some 
point one or more compaction events, where large amounts of gas flow towards 
the disc centre and lead to a compact starburst (a ``blue" nugget). This is 
followed by gas depletion from the central galaxy and the formation of an extended, 
massive star-forming ring \citep{Zolotov15,Tacchella16a,Tacchella16b}. This event 
marks major changes in galaxy properties and disc properties, and may well lead 
to systematic differences in clump properties. A detailed comparison of pre-compaction 
and post-compaction VDI is beyond the scope of this paper and is left for future 
work. To avoid this distinction, we limit our analysis to post-compaction discs, 
approximated here as $\Md>10^{10}\msun$. This has a small affect on the LLC sample, 
reducing it only by an additional $\sim 4\%$. The SLC and ex situ samples are reduced 
by a further $\sim 37\%$ and $\sim 34\%$ respectively. 

\smallskip
Finally, we further reduce the sample to account only for star-forming clumps. 
From \fig{functions} we see that many clumps have very low SFRs, well below the 
total disc SFR. Since observations target star-forming clumps, we wish to remove 
quenched clumps from the sample. In \fig{age_sfr} we show the SFR in clumps 
normalized to the disc SFR, $s=SFR_{\rm c}/SFR_{\rm d}$, as a 
function of the mean stellar age in the clump. We separate LLCs, SLCs and ex 
situ clumps according to the legend. Focusing on the LLCs, it is evident that 
there is a star-forming sequence and a population of quenched clumps with 
ages greater than $100-200\Myr$. The row of points at ${\rm log}(s)=-6$ represents 
upper limits for clumps with even lower SFRs. Most quenched clumps consist 
of relatively low mass LLCs that have very perturbed orbits taking them 
to large distances from the disc plane. These clumps often have average 
gas densities below the threshold density for star-formation, 
$n_{\rm gas,\:c}\lsim 1\cmc$, which artificially results in $s=0$.  
These clumps can survive for very long timescales of several $\Gyr$ and many 
of them have not yet migrated to the disc centre at $z=1$. It is possible 
that these clumps may evolve to become globular clusters at $z\sim 0$ as 
proposed by \citet{Shapiro10}. However, the study of quenched clumps is beyond 
the scope of this paper, which focuses on the star-forming clumps that are 
observed at high-$z$. We defer the study of quenched clumps to future work, 
selecting here only star-forming clumps with ${\rm log}(s)>-3$. 
Selecting clumps with gas mass $M_{\rm gas,\:c} > 10^6\msun$ or gas density 
$n_{\rm gas,\:c} > 1\cmc$ both result in very similar populations. This removes 
an additional $\sim 43\%$ of the LLCs and $\sim 56\%$ of the ex situ clumps, but 
only $\sim 4\%$ of the SLCs. 

\smallskip
To summarize, we select star-forming clumps with $SFR_{\rm c}>10^{-3}SFR_{\rm d}$, 
in a mass limited sample with $\Mc>10^7\msun$, in massive post-compaction 
discs with $\Md>10^{10}\msun$ at redshifts $z<2.5$. Counting each appearance 
of every clump as an independent measurement, separated by roughly half a disc 
orbital time, our final sample (hereafter referred to as our ``clean sample") 
consists of 85 ex situ clumps, 1209 SLCs, and 461 LLCs. However, the number 
of unique clumps in the sample is 56 ex situ clumps, 1086 SLCs and 150 LLCs.

\smallskip
We note that since most observations tend to target UV bright clumps, this 
can introduce a bias in their sample of clumps. We discuss this in \se{bias}.

\begin{figure}
\subfloat{\includegraphics[width =0.48 \textwidth]{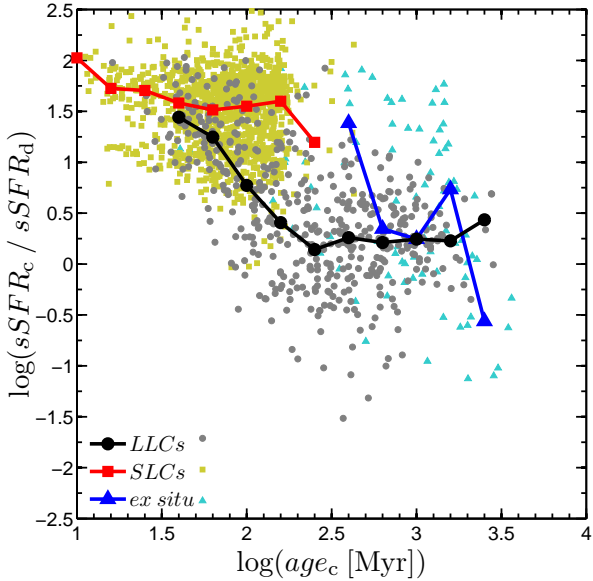}}
\caption{sSFR in clumps normalized to the disc value for our clean sample. 
Symbols and lines are the same as in \fig{age_sfr}. SLCs have very high 
sSFR, $\sim 10-100$ times higher than their host disc, with a median ratio 
of $\sim 30$, and no dependence on stellar age. The youngest LLCs have similar 
values, but their sSFR decreases rapidly, scaling roughly as $age^{-2}$, and 
is $\lsim 2$ times the disc average after $\sim 100-200\Myr$. Ex situ 
clumps have old ages of $\sim 0.5-1\Gyr$ and a wide range of sSFR values, 
though the median is similar to LL in situ clumps.
}
\label{fig:age_ssfr} 
\end{figure} 

\begin{figure}
\subfloat{\includegraphics[width =0.475 \textwidth]{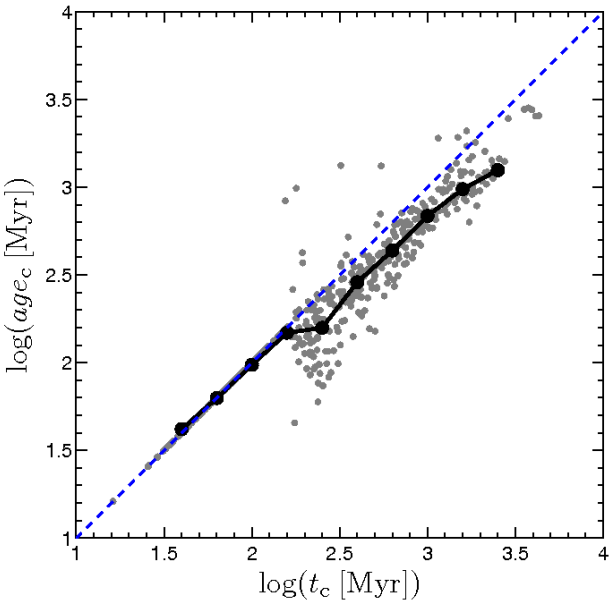}}
\caption{Mass-weighted mean stellar age in clumps as a function of clump 
time, for the LLCs in our clean sample. Grey points and the black line and 
symbols are the same as in \fig{age_sfr}. The blue dashed line corresponds 
to $age_{\rm c}=\tc$. In the first timestep when the clump is identified, 
$\tc$ is set to the stellar age by construction. In later timesteps, the 
stellar age is typically $\sim 0.1-0.2~{\rm dex}$ lower than the actual time 
since clump formation, due to continued star-formation and stripping of 
old stars.
}
\label{fig:age_time} 
\end{figure} 

\subsubsection{Clump SFR and stellar age}

\smallskip
From \fig{age_sfr}, we see that the three species of clumps, SLCs, LLCs, 
and ex situ clumps, can be distinguished by their stellar ages. SLCs 
typically have stellar ages of $\sim 50-100\Myr$ with a median age of 
$\sim 85\Myr$, and the very oldest have ages of $\sim 150 \Myr$. LLCs 
have median stellar ages of $\sim 240 \Myr$ and the oldest have ages of 
up to $\sim 1\Gyr$, which is the median age of ex situ clumps, that can 
be up to $\sim 3\Gyr$ old.

\smallskip
We have marked in the figure the median value of ${\rm log}(s)$ in bins of stellar 
age, for each of the three clump species. During their short lifetime, 
the SFR of SLCs does not vary systematically, and the median is $\sim 1\%$ 
with a scatter of $\sim \pm 0.5~{\rm dex}$. The evolution of LLCs, on the other hand, 
is more well defined. The youngest clumps contribute $\lsim 20\%$ to the disc SFR, 
and this value steadily decreases as $s\propto age^{-\alpha}$ with $\alpha \sim 1.5$, 
until it saturates at $\sim 2\%$ after $\sim 100\Myr$, slightly higher than SLCs. 
Ex situ clumps have a very broad range of SFRs that are not correlated with the 
disc. 

\begin{figure}
\subfloat{\includegraphics[width =0.48 \textwidth]{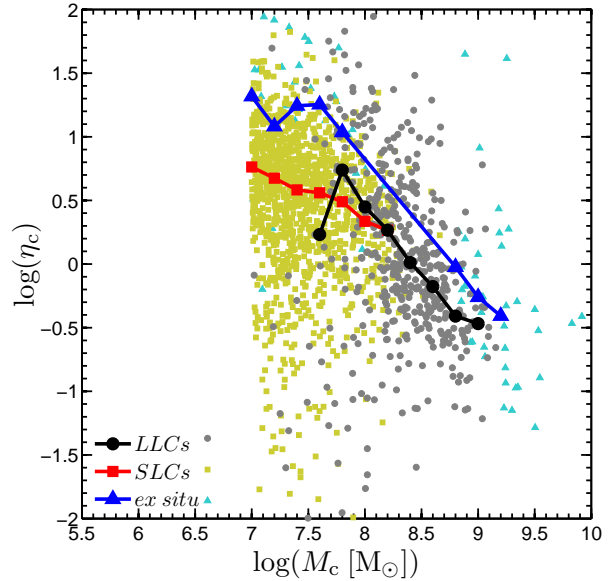}}
\caption{Outflows from clumps. We show the mass loading factor, 
$\eta={\dot {M}}_{\rm gas,\,out}/{\rm SFR}$, for clumps as a 
function of clump mass, $\Mc$. Symbols are the same as in \fig{age_sfr} 
and the lines show the medians in bins of ${\rm log}(\Mc)$. Most 
clumps with $\Mc\gsim 10^8\msun$ are long-lived. The mass loading 
factor tends to decrease with mass, though the scatter is extremely 
large, especially for the SLCs. Massive LLCs, $\Mc\lsim 10^{8.5}\msun$, 
have mass loading factors $\eta\sim 0.3-3$, with a median value of $\sim 1$. 
The most massive clumps with $\Mc\sim 10^9 \msun$ exhibit smaller 
values, $\eta\sim 0.2-2$, with a median of $\sim 0.3$.
}
\label{fig:outflow} 
\end{figure} 

\begin{figure}
\subfloat{\includegraphics[width =0.48 \textwidth]{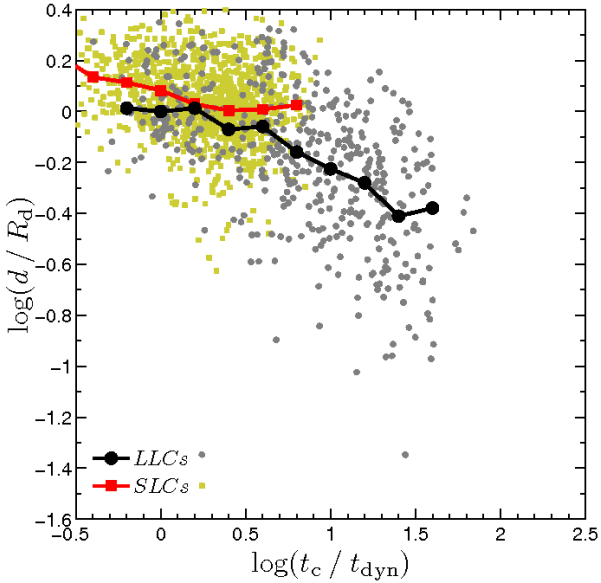}}
\caption{Clump migration. We show the clump galactocentric distance 
normalized by the disc radius as a function of the clump time normalized 
by the disc dynamical time. SLCs are typically at the disc edge, where 
most new clumps are formed, but can be found at distances as small as 
$\sim 0.5\Rd$. LLCs clearly migrate closer to the disc centre as time 
goes by. After $\sim 30 \td$, at the end of migration, most 
LLCs are at $\sim 0.3\Rd$ and the innermost clumps are at $\sim 0.1\Rd$.
}
\label{fig:dist_time} 
\end{figure} 

\smallskip
In \fig{age_ssfr} we show clump sSFR normalized by the sSFR in the host 
disc for our clean sample, as a function of clump stellar age. SLCs form a cloud, 
with ages uniformly distributed between $\sim 30-150\Myr$ and a tail 
down to $10\Myr$. Their sSFR values are $\sim 10-100$ times greater 
than the disc, with a median ratio of 30, independent of age. The youngest 
LLCs have comparable sSFR values, but these very quickly decline, scaling 
as $sSFR\propto age^{-\alpha}$ with $\alpha \sim 2$ in the age range 
$40-250\Myr$, after which it saturates at only slightly higher values 
than the disc average. The ex situ clumps have comparable ages and sSFR 
ratios to the oldest in situ clumps.

\smallskip
It is important to realize that the stellar age of LLCs is typically 
younger than the true time that has elapsed since the formation of the 
clump. This is due to ongoing star-formation in the clump as well as 
stripping of old stars, as was pointed out by \citet{Bournaud14}. In 
\fig{age_time} we show the stellar age as a function of clump time for 
the LLCs in our clean sample, together with the median age in bins of 
time. For the first $\sim 150\Myr$, the two agree by construction, 
since this is how we initialize the clump time when it is first identified. 
However, at later times, the stellar ages are $\sim 0.2~{\rm dex}$ below 
$\tc$. This makes the scaling of sSFR with time for LLCs slightly 
shallower than seen in \fig{age_ssfr}.

\subsubsection{Outflows from clumps}
\smallskip
In \fig{outflow} we show the mass loading factor, defined as 
$\eta={\dot {M}}_{\rm gas,\,out}/{\rm SFR}$, for the clumps 
in our clean sample as a function of clump mass, $\Mc$. We 
computed the outflow rate of gas through thin spherical shells 
with width $\Delta r=140\pc$ centred at radii $r=\Rc$, $1.5\Rc$ 
and $2\Rc$, only considering gas that is outflowing, $V_{\rm r}>0$, 
with velocity magnitude larger than the escape velocity from the 
clump at each shell, $V>V_{\rm esc}=(2G\Mc/r)^{1/2}$. Since the 
clump is not isolated but located within a dense disc, this is not 
the true escape velocity from the system, but it does give a rough 
approximation for the outflows generated at the clump base. The 
value used in $\eta$ is the average of the value obtained from the 
three shells, which have a typical scatter of $\sim 0.1~{\rm dex}$ 
between them. The mass loading factor decreases with mass, so that 
LLCs tend to have lower values than SLCs, though the scatter is very 
large. At masses $\Mc\lsim 10^{8.5}\msun$, the median mass loading 
factor is of order unity, $\eta\sim 1$, and values as high as $\eta\gsim 3$ 
are common. These clumps are the launching sites of galaxy-scale outflows, 
whose extent is correlated with the distribution of clumps \citep{Ceverino16b}. 
The most massive clumps with $\Mc\sim 10^{9}\msun$ have median values of 
$\eta\sim 0.3$, though values of $\eta\gsim 1$ are not uncommon. These 
may correspond to observations of very massive clumps with strong outflows 
\citep{Genzel11,Newman12}.

\subsubsection{Clump migration and gradients of clump properties}

\begin{figure*}
\centering
\subfloat{\includegraphics[trim={0 1.75cm 0 0}, clip, width =0.333 \textwidth]{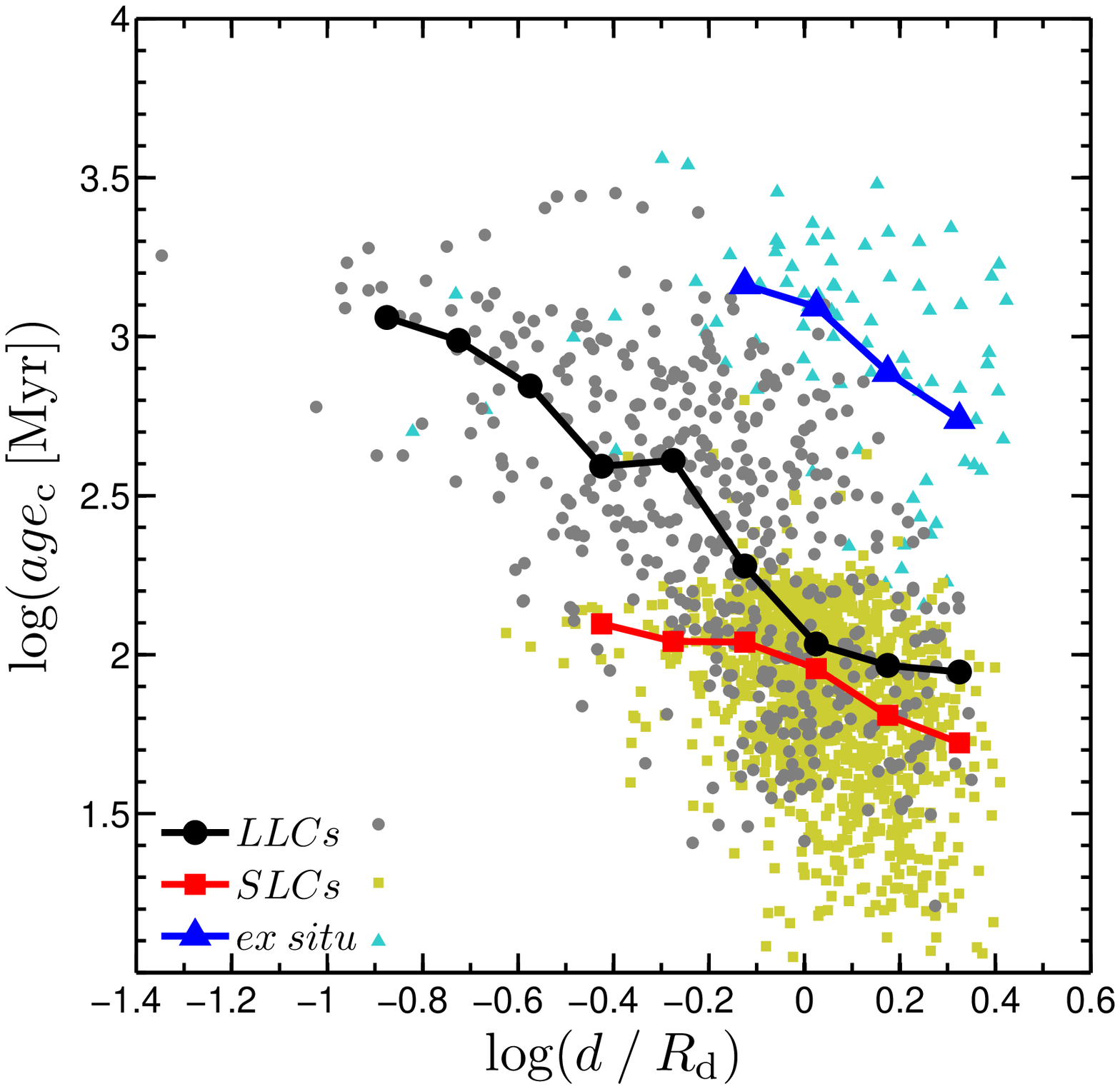}}
\subfloat{\includegraphics[trim={0 1.75cm 0 0}, clip, width =0.333 \textwidth]{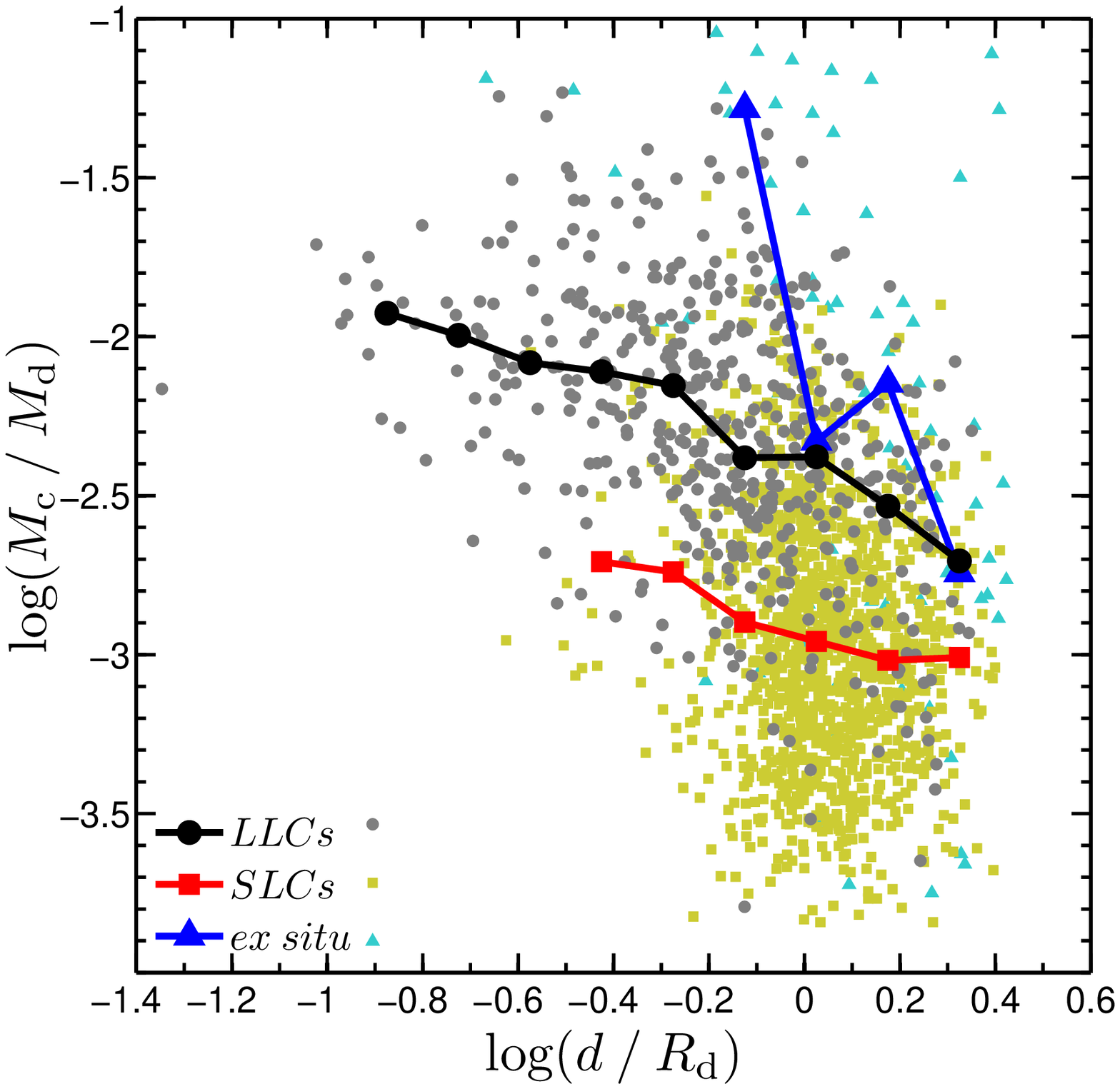}}
\subfloat{\includegraphics[trim={0 1.75cm 0 0}, clip, width =0.333 \textwidth]{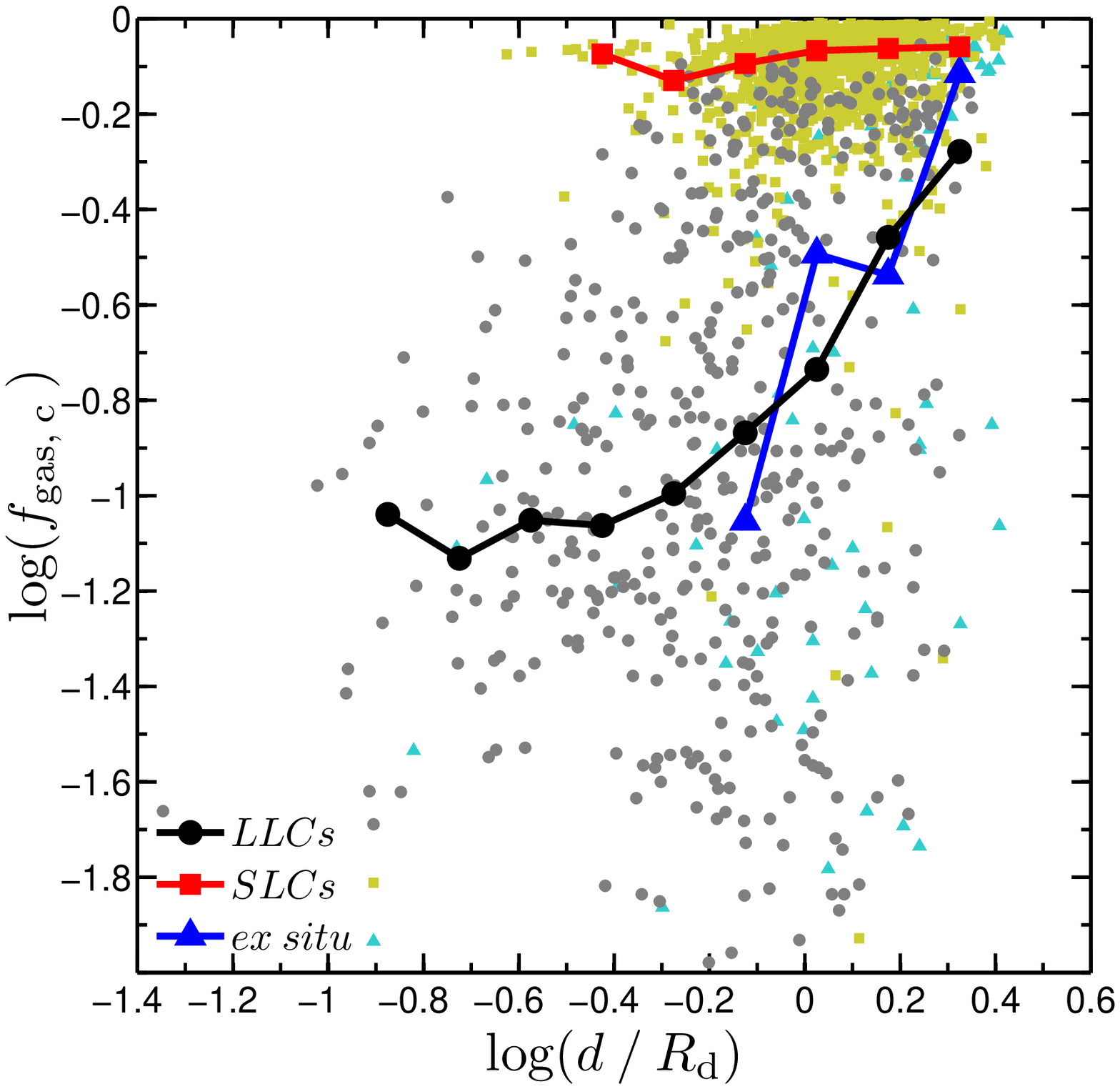}}\\
\vspace{-4.5mm}
\subfloat{\includegraphics[width =0.333 \textwidth]{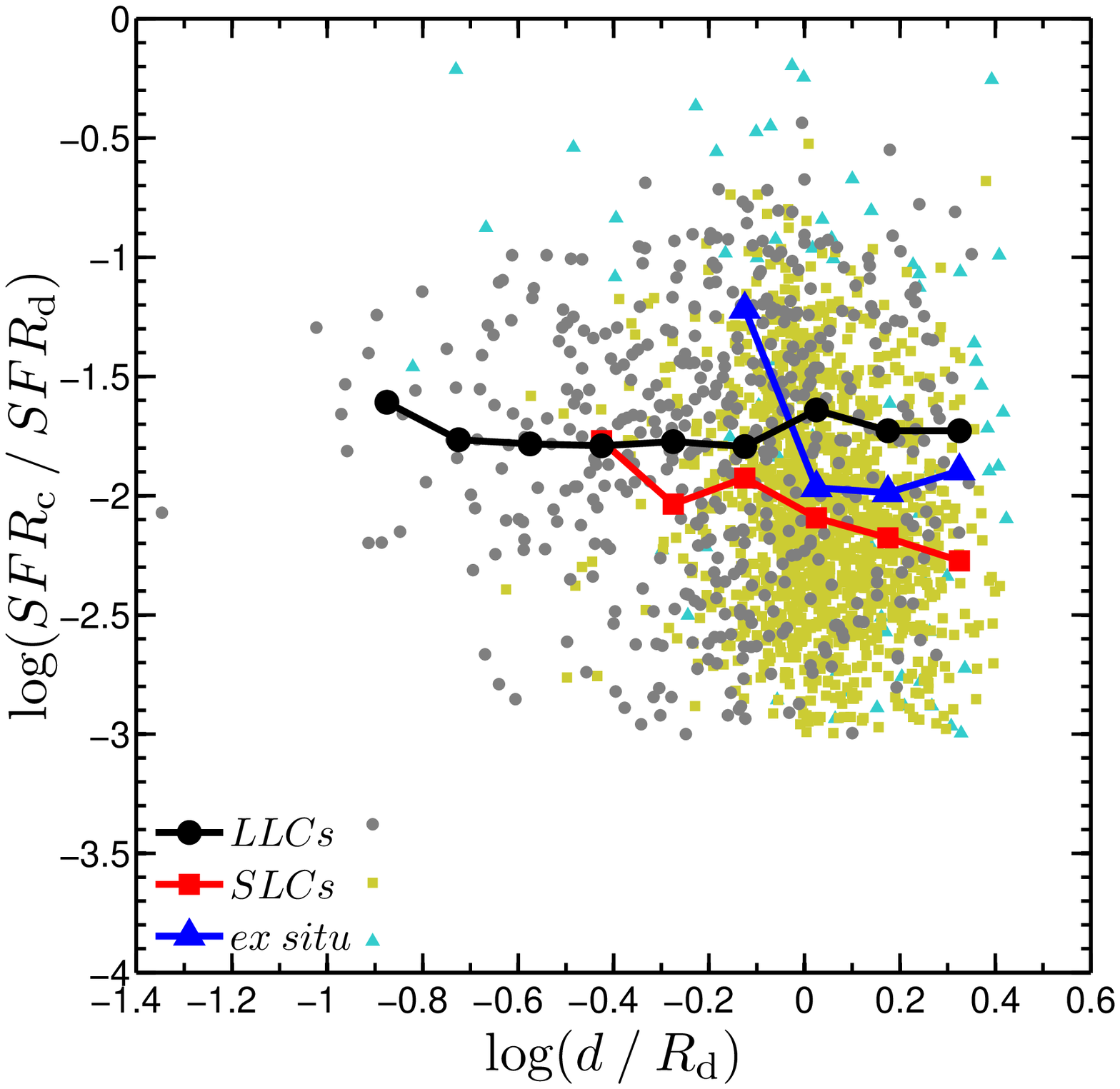}}
\subfloat{\includegraphics[width =0.333 \textwidth]{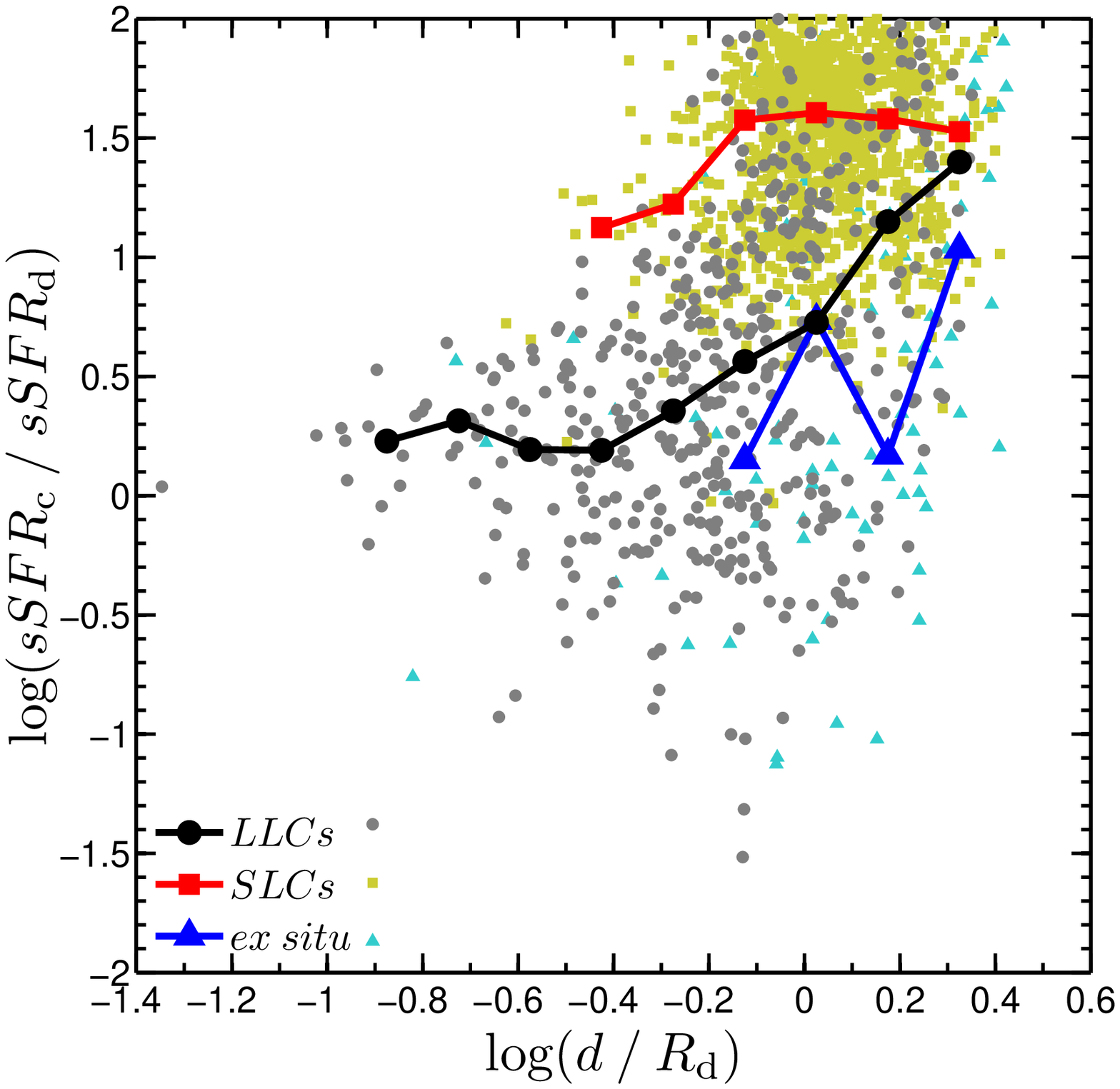}}
\subfloat{\includegraphics[width =0.333 \textwidth]{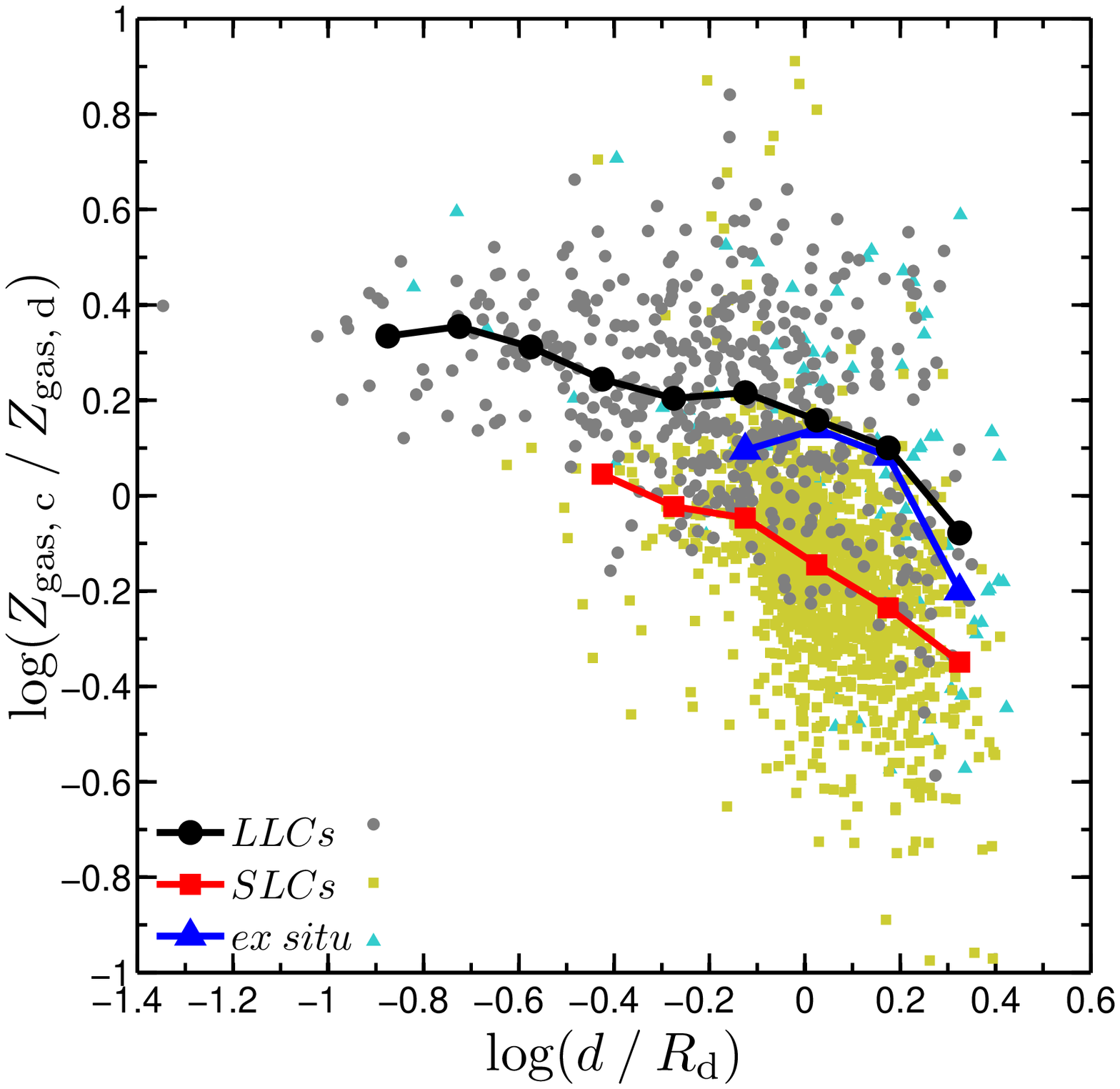}}\\
\caption{Gradients of clump properties with galactocentric distance. 
In each panel, the $x$ axis is the distance from the galaxy centre 
normalized by the disc radius, $d/\Rd$ and the $y$ axis refers to a 
different clump property. \textit{Top left:} Mass-weighted mean stellar 
age; \textit{top centre:} Baryonic clump mass normalized to the disc mass; 
\textit{top right:} Gas fraction; \textit{bottom left:} Clump SFR normalized 
to the disc SFR; \textit{bottom centre:} Clump sSFR normalized to the disc 
sSFR; \textit{bottom right:} Clump metallicity normalized to the mean disc 
metallicity. Lines and symbols are the same as in \fig{age_sfr}. The medians 
were calculated in equally spaced bins of $0.15~{\rm dex}$ in ${\rm log}(d/\Rd)$. 
There are hardly any SLCs within $\sim 0.5\Rd$ while LLCs can be found at 
$\sim 0.1\Rd$ at the end of migration. LLCs have a strong mass gradient, 
gaining a factor of $\sim 2-3$ in mass during migration, while SLCs show 
no such gradient. LLCs also exhibit a steep age gradient, where clumps closer 
to the disc centre have older stellar ages, while the corresponding gradient 
for SLCs is much shallower. SLCs have much higher gas fractions than LLCs which 
rapidly turn an order unity fraction of their mass into stars. While there is 
no strong trend of SFR with distance, the sSFR of LLCs steadily declines during 
migration while the SLCs have a roughly constant value. LLCs have a shallower 
metallicity gradient and higher metallicity values than SLCs. Ex situ clumps are 
much older, more massive and appear concentrated near the outer disc, though they 
are similar to LLCs in other properties.
}
\label{fig:gradients} 
\end{figure*} 

\smallskip
\Fig{dist_time} examines the migration of in situ clumps, LLCs and SLCs, 
in our clean sample. We show the clump galactocentric distance normalized 
by the disc radius, $d/\Rd$, as a function of the clump time normalized by 
the disc dynamical time, $\tc/\td$. The dynamical time is defined as 
$\td=R_{\rm rot}/V_{\rm rot}$, where $R_{\rm rot}$ and $V_{\rm rot}$ 
are the mass-weighted mean radius and rotational velocity of cold gas in 
a cylindrical ring $1\kpc$ thick in the vertical direction and extending 
from $0.5\Rd$ to $\Rd$ in the radial direction\footnote{For a uniform disc, 
$R_{\rm rot}\simeq 0.78\Rd$. For an exponential disc where $\Rd=2R_{\rm 1/2}$, 
which is the radius that contains $\sim 85\%$ of the mass, 
$R_{\rm rot}\simeq 0.71\Rd$. Assuming a flat rotation curve, this means that 
the dynamical time used here is roughly $75\%$ of the dynamical time at $\Rd$, 
where most new clumps are formed.}. We also show the median distance in bins 
of clump time. The median lifetime of a SLC is $\sim 2\td$ and the oldest of 
them live for $\sim 4\td$. They are mostly located at $d\simeq\Rd$, though many 
of them can be found at $d \sim 0.5\Rd$. There is no trend of distance 
with time, so any radial migration of SLCs during their lifetime is 
negligible. The youngest LLCs have a similar radial distribution to 
the SLCs, reflecting the star-forming ring where clumps are formed. 
After $\sim 4\td$ there is clear radial migration, and the median 
distance scales as $d\propto t^{-\alpha}$ with $\alpha \sim 0.5$. We 
do not identify any LLCs older than $\sim 30\td\sim 4-5 t_{\rm orb}$, 
comparable to the expected migration time, $t_{\rm mig}\simeq \delta^{-2}\td$ 
\citep{DSC}, where $\delta\sim 0.15$ in our simulations. At this time, 
LLCs are typically found at $\sim 0.3-0.4\Rd$, but can be found as close 
to the centre as $\sim 0.1\Rd$. We do not find any SLCs at such small 
distances, showing that clump formation is limited to the external disc, and 
the existence of clumps at such small radii is a clear sign of migration. 
This is a consequence of limiting our analysis to massive post-compaction 
discs. These have developed a central bulge and begun to stabilize in their 
centres, while VDI and clump formation is limited to a star-forming ring at 
radii $0.5-1\times\Rd$. At higher redshifts, in pre-compaction discs that 
lack a central mass concentration, our simulations do show clump formation 
at smaller radii. It is possible that these clumps may behave differently 
to the ones studied here, but as stated previously this is beyond the scope 
of the current analysis. Observations of clumpy unstable discs at $z\sim 2$ 
show similar unstable rings around stable centres \citep{Genzel14}, and our 
predictions should be compared to such systems.

\smallskip
\Fig{gradients} shows gradients of clump properties for all clumps in our 
clean sample. In each panel, the $x$ axis shows the galactocentric distance 
normalized by the disc radius, $X \equiv d/\Rd$, and the $y$ axis shows 
different clump properties. We show points for individual clumps as well as the 
median value in equally spaced bins of distance. For each property discussed below, 
we quote the logarithmic slope of the radial dependence of the median values, i.e. 
$\alpha$ where $y\propto X^{\alpha}$.

\smallskip
\textbf{Stellar Age:} 
The top left panel shows stellar age. For SLCs, the clump time and age are by 
construction the same, so this can be easily compared with \fig{dist_time}. 
Exterior to $\sim 0.8\Rd$, the median SLC ages have a slope of $\alpha \sim -0.6$, 
though the scatter about this relation is very large, and interior to $0.8\Rd$ 
there is no radial trend. The LLC profile is steeper, the median ages exhibiting 
a slope of $\alpha \sim -1$, very similar to the slope found by M14 in simulations 
with weaker feedback. We conclude that an age gradient in the outer disc is not by 
itself evidence of clump migration, but migration causes a steeper age gradient 
which extends to smaller radii in the disc. Ex situ clumps have much older ages 
than in situ clumps, comparable to the mean stellar age of the disc, and show no 
systematic gradient. This is consistent with them being mergers, and consistent 
with the results of M14.

\smallskip
\textbf{Mass:} 
The top centre panel shows clump baryonic mass normalized by the disc mass. 
The mass range of SLCs is roughly $1~{\rm dex}$ with a median at $\sim 0.1\%$ of 
the disc mass and no appreciable trend with distance, except for the innermost 
SLCs, which can in fact be identified for 2 snapshots. On the other hand, the 
LLCs grow in mass by $\sim 0.7~{\rm dex}$ during migration. In the outer disc, $d>0.5\Rd$, 
the slope is $\alpha \sim -0.9$, while in the inner disc it flattens to $\alpha \sim -0.4$, 
very similar to the mass profile found by M14. Fitting a single slope from 
$\Rd$ to $0.1\Rd$ results in $\alpha\sim -0.6$. LLCs grow in mass during 
migration by accreting gas from the disc \citep[][M14]{DK13}. While it has been 
suggested that a mass gradient could be present at birth, due to higher velocity 
dispersions closer to the disc centre leading to a larger Jeans mass \citep[e.g.][]{Genel12a}, 
we find no evidence for this among our clean sample of SLCs (though see \se{bias} 
for how observationally selecting only the brightest clumps may introduce such a 
gradient). Ex situ clumps are more massive than in situ clumps at all radii, and 
they also exhibit a trend for more massive clumps to be located closer to the centre, 
as found by M14. This can be caused by mass growth of ex situ clumps that join the 
disc rotation and migrate in a similar manner to in situ clumps, or by more massive 
mergers sinking to the centre faster.

\smallskip
\textbf{Gas fraction:} 
The top right panel shows the gas fraction. SLCs have very high gas fractions. 
the median values decrease from near unity at the outer disc to $\sim 70\%$ 
for the innermost SLCs, wit a flat slope of $\alpha \sim 0.10$. The $67\%$ 
scatter about the median is $0.64-0.88$, so SLCs turn only $\lsim 30\%$ of 
their mass into stars before being disrupted. Since the oldest SLCs are 
$\sim 10 t_{\rm ff}$, this translates to a time averaged SFR efficiency 
per free fall time of $\sim 3\%$. Even the most gas rich LLCs rarely have 
gas fractions greater than $\sim 65\%$, and their median profile shows a 
dramatic decline very early on. By the end of migration, LLCs have median 
gas fractions of $\sim 8\%$, almost a factor of 10 lower than at the disc 
outskirts, and the $67\%$ scatter about the median is $\sim 4-24\%$. 
LLCs turn most of their mass into stars which is why they are able to remain 
bound. Ex situ clumps have similar gas fractions to LLCs.

\smallskip
\textbf{SFR:} 
The bottom left panel refers to the clump SFR normalized to the disc total. 
SLCs have a tendency for higher SFRs closer to the disc centre, with a slope 
of $\alpha \sim -0.65$, though the scatter is very large. The median SFR of 
LLCs is constant with distance at a value of $\sim 1-2\%$ of the disc total, 
though here too the scatter is large, $\sim 1~{\rm dex}$. The constancy of 
the SFR during migration reflects a steady state between inflows from the 
disc and SF+outflows, and is consistent with \fig{age_sfr}. 

\smallskip
\textbf{sSFR:} 
The bottom centre panel refers to sSFR, normalized to the disc mean, and can 
be compared to \fig{age_ssfr}. SLCs have very high sSFRs of $\sim 30$ times 
the disc average, except for the innermost clumps which actually survive for 
two timesteps. The LLCs have comparable sSFR to the SLCs at birth, but this 
value steadily declines during migration, with a typical slope of $\alpha \sim 1.6$. 
Interior to $\sim 0.4\Rd$, the sSFR saturates at roughly $1-2$ times the disc 
mean. The median sSFR for ex situ clumps is slightly lower than that of LLCs.

\smallskip
\textbf{metallicity:} 
The bottom right panel examines clump gas metallicity normalized by the mass weighted 
mean metallicity in the disc. The SLCs exhibit a slope of $\alpha \sim -0.6$ exterior 
to $0.8\Rd$, saturating at the average disc value at smaller radii. This is very similar 
to the metallicity gradient in the underlying disc, since the SLCs do not self-enrich much. 
In fact, SLCs at large radii acquire gas with very low metallicities from primordial gas 
inflow \citep{Ceverino16a}. Averaged over the whole disc, the LLCs have a shallower 
gradient of $\alpha \sim -0.3$, similar to the gradient found by M14, though exterior 
to $0.8\Rd$ their gradient appears similar to that of the SLCs. They have higher metallicity 
values than SLCs overall since they continually form stars and can self-enrich during their 
migration. Observations of gradients in clump metallicity that are shallower than the underlying 
disc may thus be evidence for clump migration. However, ex situ clumps have very similar behaviour 
to LLCs, with nearly identical distributions, presumably because ex situ clumps can join the 
disc rotation and migrate towards the disc centre while accreting fresh gas from the underlying disc, 
similar to migrating LLCs (M14).

\begin{figure*}
\centering
\hspace{-1.4mm}
\subfloat{\includegraphics[trim={0 -0.45cm 0 0}, clip, width =0.3367 \textwidth]{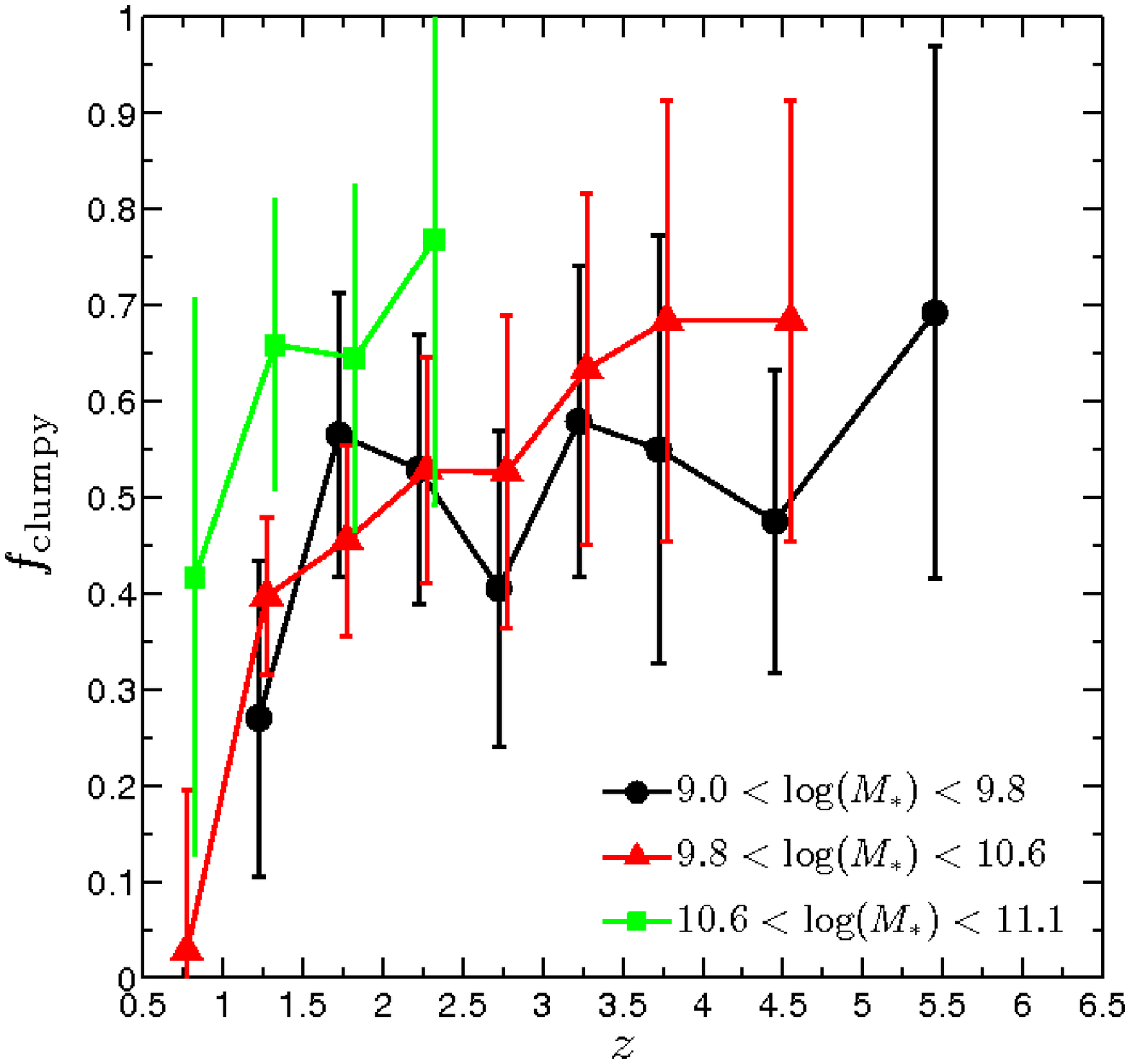}}
\subfloat{\includegraphics[width =0.329 \textwidth]{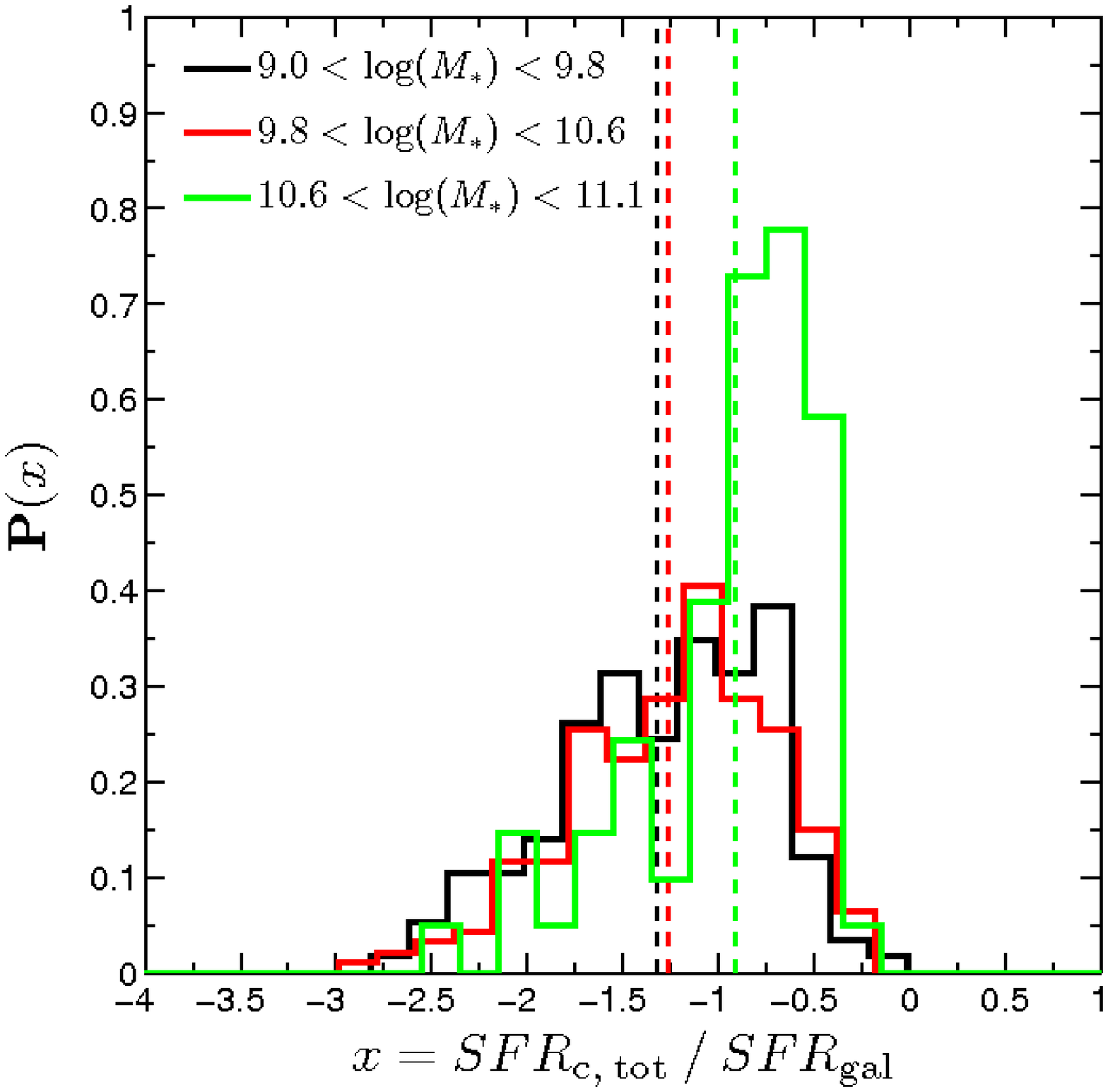}}
\subfloat{\includegraphics[width =0.329 \textwidth]{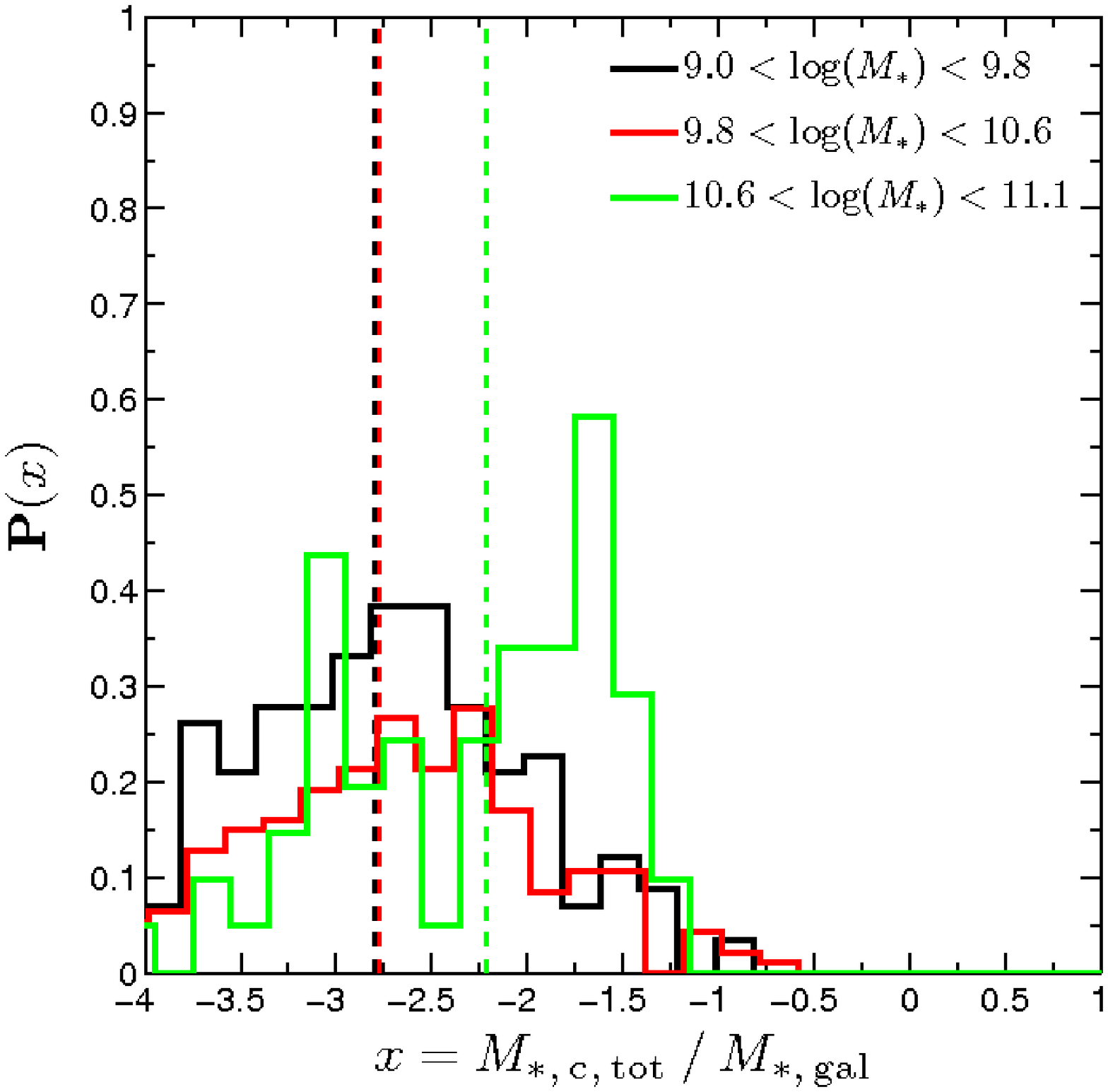}}\\
\vspace{-4.5mm}
\subfloat{\includegraphics[trim={0 -0cm 0 0}, clip, width =0.3374 \textwidth]{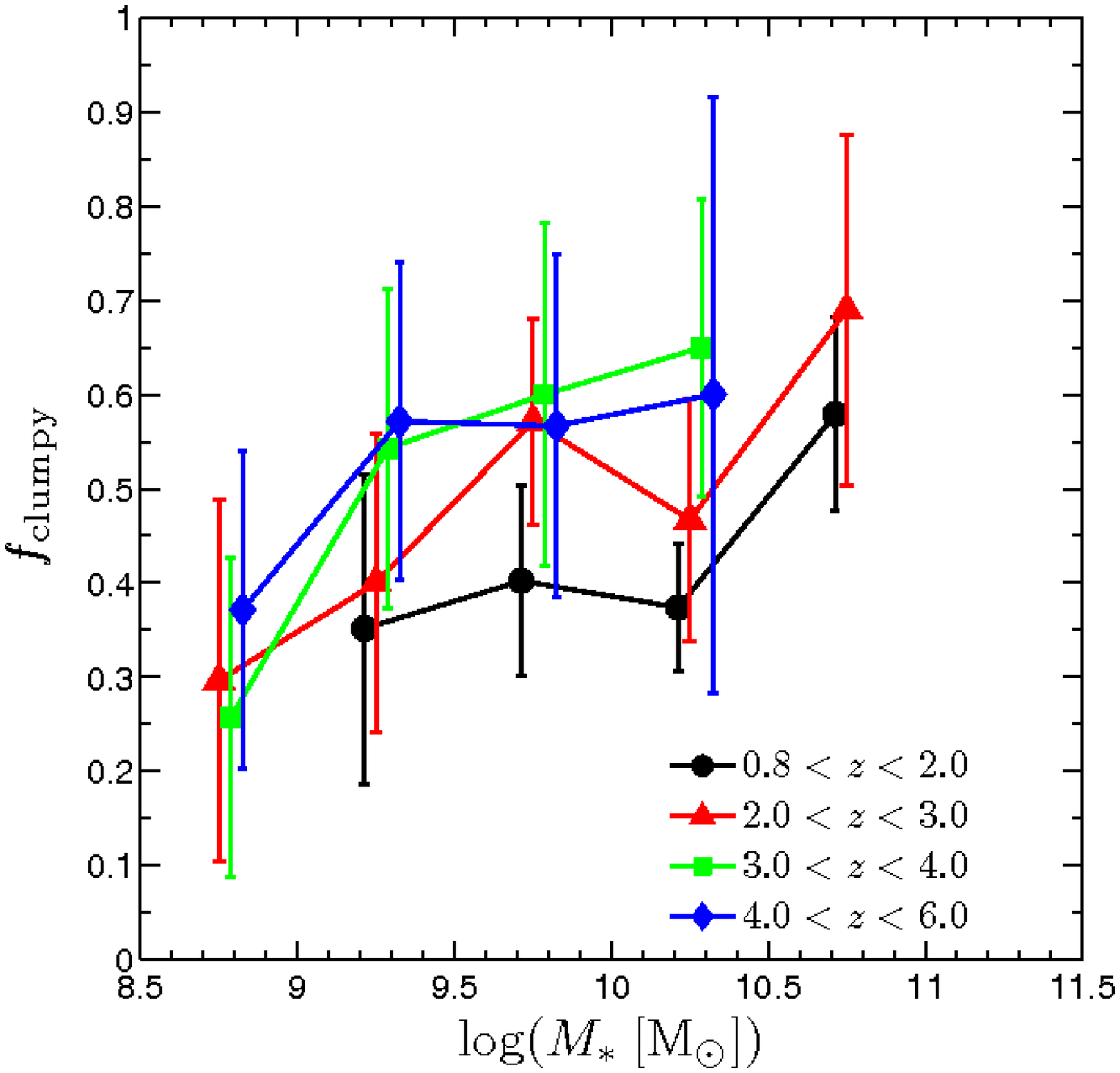}}
\subfloat{\includegraphics[width =0.329 \textwidth]{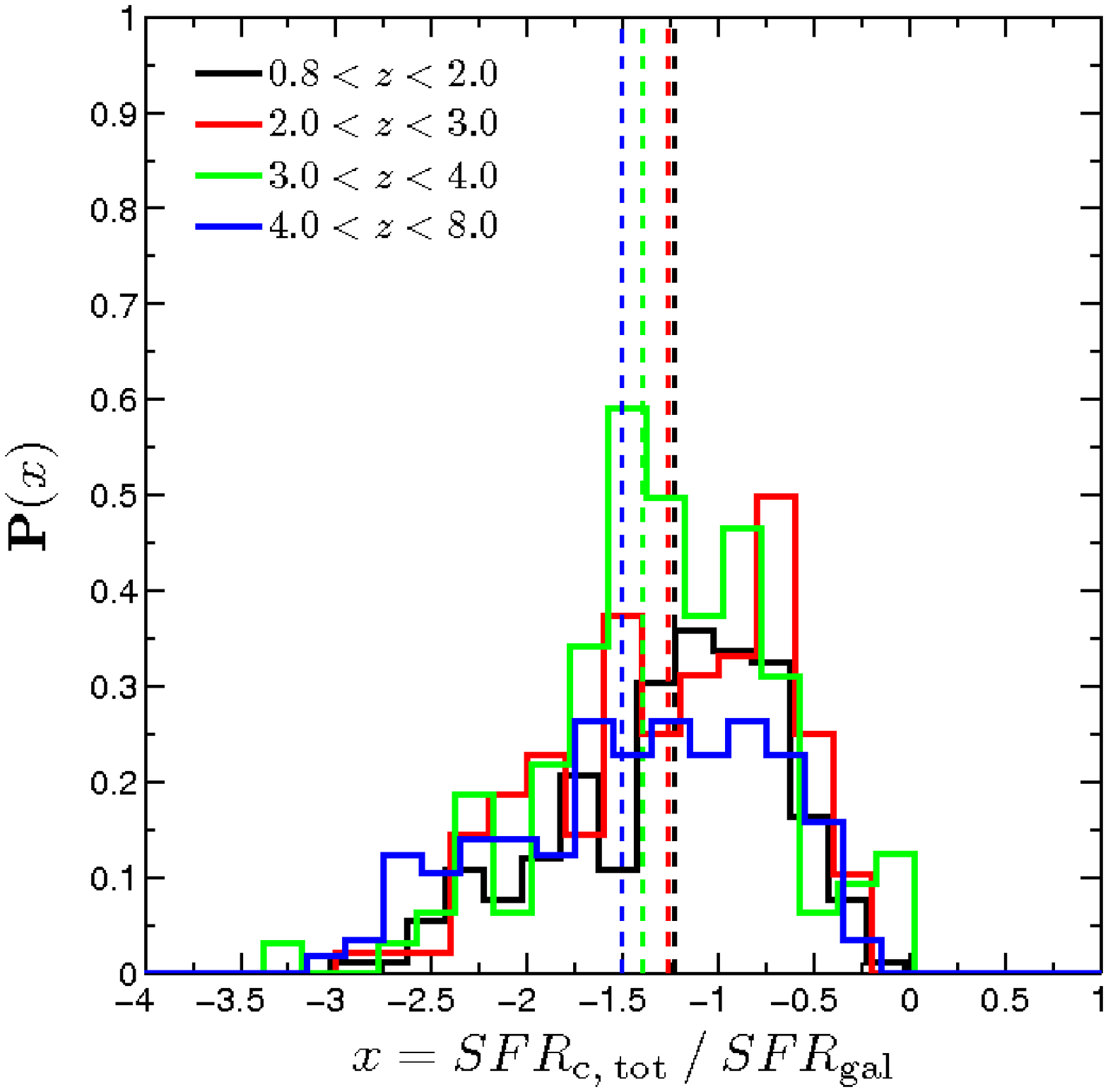}}
\subfloat{\includegraphics[width =0.329 \textwidth]{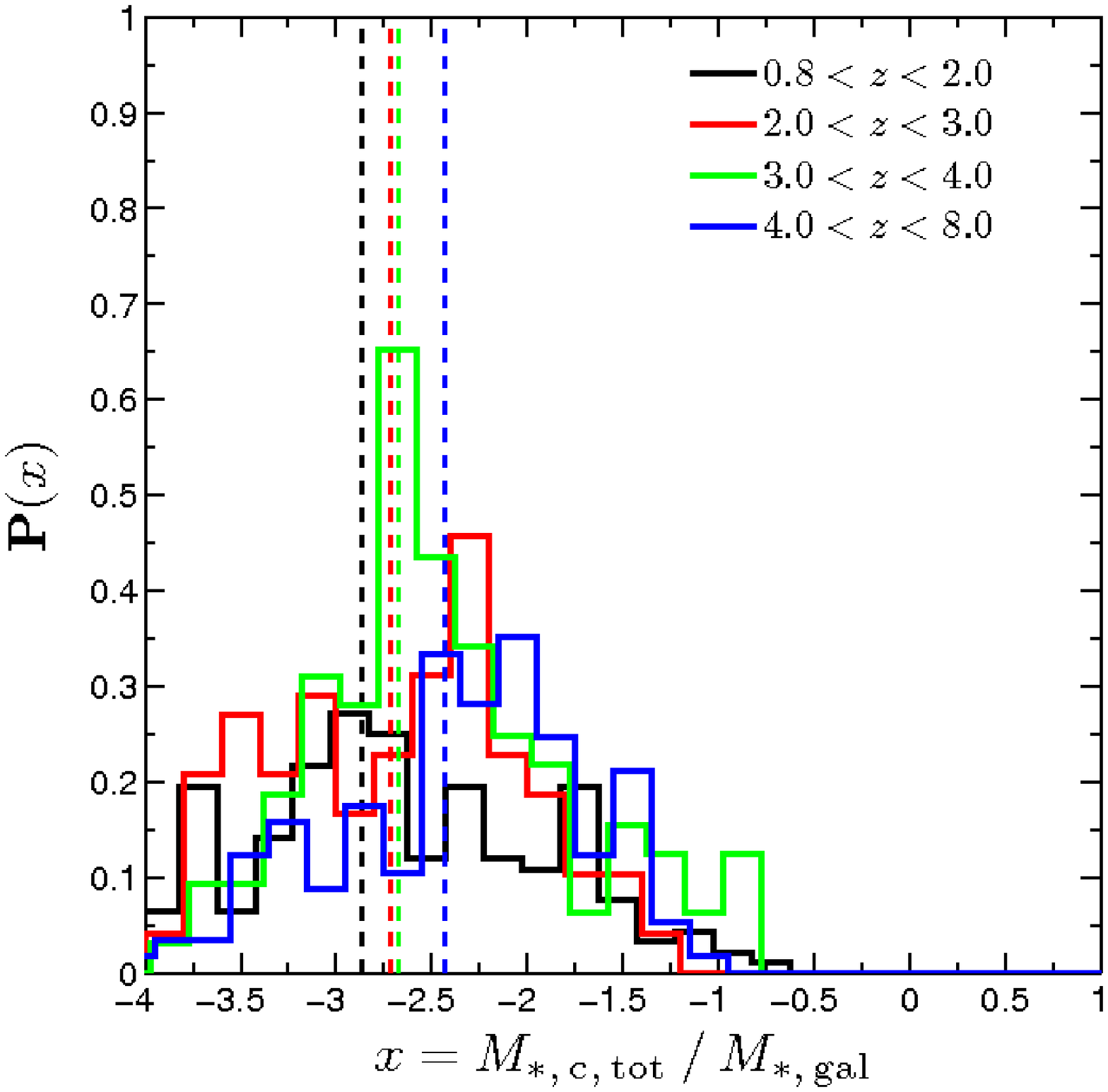}}\\
\caption{Clumpiness of simulated galaxies. We show various measures of the 
clump contribution to the galaxy population, using clumps within a sphere 
of radius $\Rd$ that have $\Mc>10^7\msun$ and $SFR_{\rm c}/SRF_{\rm d}>10^{-3}$. 
On the left we show the fraction of clumpy galaxies as a function of redshift 
in three bins of stellar mass (top) and as a function of stellar mass in 
four redshift bins (bottom). Galaxy stellar mass is computed within a sphere 
of radius $\Rd$. The error bars are Poissonian, and we only show bins where 
we have data from at least 5 snapshots among at least 3 different simulations. 
The clumpy fraction is $\sim 40-60\%$ over a broad range of 
stellar masses and redshifts, though it tends to decrease towards low redshifts 
and low mass galaxies. The middle column shows the combined contribution of clumps 
to the SFR of clumpy galaxies (computed in the same sphere). We show the distributions 
in the same bins of galaxy mass (top) and redshift (bottom) used previously. Each 
histogram is normalized to an integral equal to the clumpy fraction in the bin. 
Vertical dashed lines show the median values of each bin, according to the color 
legend. The distributions are very similar at all redshifts, with median contributions 
of $\sim 3-6\%$ and a slight trend for higher values at lower redshfits. High 
mass galaxies have a higher contribution from clumps, $\sim 12\%$ compared to 
$\sim 6\%$ in lower mass galaxies. On the right we show distributions for the 
combined contribution of clumps to the stellar mass of clumpy galaxies, in the 
same bins used previously. Vertical dashed lines show the median values of each 
bin. The distributions are overall similar at all redshifts, but the median value 
declines towards lower redshift by $\sim 0.4~{\rm dex}$ from $4<z$ to $z<2$. Again, the 
contribution of clumps to the high mass galaxies is larger than the low mass galaxies, 
by roughly $\sim 0.5~{\rm dex}$ in the median, though the low end of the distribution is 
very similar.
}
\label{fig:clumpiness} 
\end{figure*} 

\smallskip
To summarize, LLCs in our simulations exhibit very similar gradients to those predicted 
by M14 in most of their properties. SLCs can also show gradients in stellar age and gas 
fraction, though these are typically much shallower, while their metallicity gradient is 
steeper. SLCs do not show a gradient in mass (though see \se{bias}) or sSFR, contrary to 
LLCs. Furthermore, SLCs are not expected within $\sim 0.5\Rd$ in post-compaction discs, 
while LLCs exist at distances $d\gsim 0.1\Rd$ where they have higher masses, older ages, lower 
gas fractions and lower sSFRs. These can be used to observationally test the clump migration 
scenario. Ex situ clumps have comparable SFRs, gas fractions and metalicities to LLCs, but 
they are more massive and older, and are typically located at large distances similar to SLCs.

%--------------------------
\subsection{Clumpiness of Galaxies} 
\label{sec:gal_clump} 

\smallskip
An important observational constraint of cosmological models 
of VDI is the fraction of clumpy discs as a function of galaxy 
mass and redshift and the total contribution of clumps to the 
galaxy SFR and mass. We address this in \fig{clumpiness}. As in 
the previous section, we limit this analysis to the mass limited 
sample of star-forming clumps, i.e. clumps with $\Mc>10^7\msun$ 
and $s=SFR_{\rm c}/SFR_{\rm d}>10^{-3}$. In addition, we only 
account for clumps within a sphere of radius $\Rd$. However, unlike 
the previous section, we do not limit the disc mass or redshift range. 
The ex situ clumps comprise only $\sim 4\%$ of the clumps in the sample, 
though they contribute $\sim 38\%$ of the mass and $\sim 22\%$ of the SFR. 

\smallskip
The two left panels in \fig{clumpiness} show the fraction of clumpy 
galaxies in our sample, $f_{\rm clumpy}$, i.e. the fraction of galaxies 
that contain at least one clump obeying the above criteria. In the top 
panel we show this as a function of redshift in bins of stellar mass 
and on the bottom we show this as a function of stellar mass in bins 
of redshift. Stellar mass here refers to the total stellar mass within 
$\Rd$. The error bars refer to the Poisson error within each bin. We 
caution, however, that while we have many snapshots, they are from a 
limited number of independent galaxies, especially below $z\sim 2$. 
Therefore, the fraction of clumpy galaxies is a somewhat biased measure, 
because whether a particular galaxy is clumpy or not in a particular snapshot 
will influence the chances that it will be clumpy in the next snapshot. 
We try to mitigate this by only plotting bins where we have at least 5 
snapshots from at least 3 independent galaxies. Nevertheless, the errors 
on $f_{\rm clumpy}$ may be larger than indicated by the error bar.

\smallskip
There is a tendency for galaxies to become less clumpy with time, though 
the fraction of clumpy galaxies among low mass galaxies is fairly constant 
at $\sim 50\%$ from $z\sim 1.5-4$. Intermediate mass galaxies have 
comparable clumpy fractions at $z\sim 1-3$, and slightly higher fractions 
at higher redshifts, though this is within the scatter. The clumpy fraction 
is higher among massive galaxies, though this declines steeply from 
$z\sim 2.5-1$, and is also within the scatter. The clumpy fraction 
is also fairly constant at $f_{\rm clumpy}\sim 40-60\%$ over a large range 
in stellar mass, with a tendency for the most massive galaxies to be clumpier 
than the least massive galaxies. 

\smallskip
In the middle two panels, we examine the total contribution of clumps 
to the galaxy SFR, calculated within a sphere of radius $\Rd$. We show 
the distributions for the same bins of stellar mass (top) and redshift 
(bottom) used in the $f_{\rm clumpy}$ measurements. Each histogram is 
normalized to an integral equal to the fraction of clumpy galaxies in 
the respective bin. The distribution is only weakly dependent on redshift. 
The median contribution of clumps to the total SFR grows from $\sim 3-6\%$ 
between $4<z$ and $z<2$, while the FWHM of the distribution is $\lsim 1~{\rm dex}$ 
at all redhsifts. On the other hand, there is a tendency for the most massive 
galaxies to have a higher contribution to their SFR from clumps, $\gsim 12\%$ 
as opposed to $\sim 6\%$ for less massive galaxies. However, as before, the 
distributions are wide with a FWHM of $\sim 1~{\rm dex}$. 

\smallskip
In the right two panels we show the total contribution of clumps to the 
galaxy stellar mass, in the same mass and redshift bins as above. These 
are lower than the SFR contributions by $1-1.5$ orders of magnitude. The 
distributions are similar at all redshifts, though the median contribution 
increases towards higher redshift, from $\sim 0.15\%$ at $z<2$ to $\sim 0.35\%$ 
at $4<z$. The two low mass bins have a very similar distribution, as they 
did for the SFR contribution, while the highest mass bin is again skewed 
towards high values, with a median of $\sim 0.55\%$ compared to $\sim 0.2\%$. 

%%%%%%%%%%%%%%%%%%%%%%%%%%%%%%%%%%%%%%%%%%%%%%%%  
\section{Discussion}
\label{sec:disc} 

\subsection{Clump Survival vs Disruption}
\smallskip
There are two extreme scenarios concerning the life and fate 
of high-redshift giant clumps formed by VDI. In one scenario, 
the clumps, despite undergoing outflows, remain intact and even 
grow by accretion as they migrate into the disc centre on an 
orbital timescale, $\sim 250-500\Myr$ at $z\sim 2$. In the 
competing scenario, the clumps disrupt on a dynamical timescale, 
$\sim 50-100\Myr$ at $z \sim 2$, well before they complete their 
migration. Most previous numerical studies of VDI and high-$z$ 
clump formation have typically fallen into one of these two 
categories. In simulations that incorporated only supernova 
feedback \citep[e.g.][]{Bournaud07,Agertz09,CDB,Ceverino12,M14,Tamburello15}, 
all massive clumps were long lived and migrated towards the 
disc centre where they coalesced in the bulge. In simulations 
that incorporated strong momentum driven feedback, either using 
sub-grid models for galactic scale winds \citep{Genel12a} or 
models for RP feedback with very high trapping factors 
\citep{Hopkins12a, Oklopcic16}, all clumps were rapidly destroyed. The 
simulations used in this study fall into an intermediate 
category, where RP feedback is implemented but without the 
additional boost from photon trapping. Recent simulations of 
isolated galaxies by \citet{Bournaud14} which employed only 
a moderate boost factor fall into a similar category. This 
offers a compromise between the two extreme scenarios, where 
low mass clumps with low surface densities are disrupted in 
only a few free fall times, while massive, dense clumps are 
long lived.

\smallskip
Theoretically, the effect of radiation pressure and the strength 
of photon trapping in different environments is unclear. Idealized 
radiation hydrodynamics (RHD) simulations by \citet{KT12,KT13} 
showed that radiation trapping in clump-like environments is 
negligible because the wind destabilizes due to radiation 
Rayleigh-Taylor instability and radiation escapes through 
optically-thin holes in the wind. On the other hand, \citet{Davis14} 
simulated a similar setup to \citet{KT13}, but using a different 
method for closing the RHD equations. They found that in very dense 
regions with very large optical depths, the effective photon 
trapping may be a factor of a few higher than advocated by \citet{KT13}. 
However, it is unclear if the physical conditions they studied are 
relevant for high-$z$ clumps. Recently, \citet{Roshdal15} performed 
RHD simulations of isolated galaxies, using yet a third method for 
closing the RHD equations, and found that the net effect of RP is 
gentler and less effective than often assumed in subgrid models. 

\smallskip
Several analytic studies have attempted to estimate the effect of 
feedback on clumps. \citet{murray10} argued that momentum-driven 
radiative feedback could disrupt the clumps on a dynamical 
timescale, as it does in local giant molecular clouds. Then 
\citet{KrumholzDekel} showed that such an explosive disruption is 
not expected to occur in the high-redshift giant clumps unless the 
SFR efficiency in a free-fall time is $\epsilon_{\rm ff}\sim 0.1$, 
much larger than what is implied by the Kennicutt relation at $z=0$. 
Such a deviation from the local Kennicutt law is inconsistent with 
observations \citep{Tacconi10,Tacconi13,Freundlich13}. \citet{DK13} 
proposed instead that the observed outflows from high-redshift clumps 
\citep{Genzel11,Newman12}, with mass loading factors of order unity, 
are driven by steady momentum-driven outflows from stars over many 
tens of free-fall times. Their analysis was based on the findings 
of \citet{KT13} and assumed that each photon can contribute to the 
wind momentum only once, so the radiative force is limited to 
$\sim L/c$. When combining radiation, protostellar and main-sequence winds, 
and supernovae, \citet{DK13} estimated the total direct injection rate 
of momentum into the outflow to be $\sim 2.5 L/c$. The adiabatic phase 
of supernovae and main-sequence winds can double this rate. They conclude 
that most clumps are expected to complete their migration prior to gas 
depletion. 

\smallskip
Given the uncertainties in the theoretical analysis, it is worthwhile 
to simulate and compare different scenarios with varying feedback strengths 
and study the effect on clumps. In this respect, our simulations are extremely 
useful because they produce both SLCs and LLCs, and by studying the different 
properties of these two populations, we can observationally constrain the 
longevity of clumps in the real Universe. Detailed observations of clump 
properties such as age, mass, metallicity, gas fraction and SFR, their 
correlations with each other and their radial variation across the disc 
will be able test for the existence of long-lived migrating clumps, 
and thus place constraints on stellar feedback.

\subsection{Comparison With Observations}

\smallskip
Many of the results presented in this work can be directly compared to existing 
observations, and make predictions for upcoming observations. We review here 
some of the more recent observational studies targeting high-$z$ SFGs and compare 
our results with theirs.

\subsubsection{Clumpiness of galaxies}

\smallskip
Using a mass complete sample of 649 massive ($M_*>10^{10}\msun$) at 
$0.5<z<2.5$, \citet{Wuyts12} estimated the overall fraction of clumpy 
discs, finding it to range from $\sim 40-75\%$ at redshifts $z=1.5-2.5$ 
and $\sim 15-80\%$ at $z=0.5-1.5$, depending on the band used to identify 
clumps, with redder bands being less clumpy than bluer bands, and stellar 
mass maps being the least clumpy. However, in this study clumps are not 
detected individually as in our work, but are rather defined as off-centre 
pixels with heightened surface density in stacked, pixelated images 
of the galaxy population. Observations of several thousand galaxies in 
the CANDELS/GOODS-S and UDS fields in the redshift range $0.5-3$, where 
clumps are detected both visually and using an automated algorithm similar 
in spirit to the one implemented in this work \citep{Guo15}, reveal clumpy 
fractions of $\sim 40-60\%$ for galaxies in a similar mass and redshift 
range as studied in this work. They find that for low mass galaxies ($M_*<10^{9.8}$), 
this fraction is constant in time from $z=0.5-3$. Higher mass galaxies have 
comparable clumpy fractions at $z\gsim 2$, but these decline towards lower 
redshifts, reaching $20-40\%$ at $z \sim 0.5$. \citet{Shibuya16} used 
HST to study $\sim 17000$ galaxies at $z=0-8$ and found that the fraction of 
clumpy galaxies increases from $z=8$ to $z=1-3$ where it is $\sim 50-60\%$, 
and then declines at $z\lsim 1$ down to $\lsim 30-40\%$ by $z\sim 0.5$. Given 
the uncertainties in the observations and the simulations, the clumpy fractions 
predicted by our simulations (\fig{clumpiness}) are consistent with the data, 
though a larger sample of simulations covering a broader mass range is needed 
to place tighter constraints.

\subsubsection{Clump Masses and SFRs}

\smallskip
\citet{Wuyts12,Wuyts13} and \citet{Guo15} have measured the contribution of clumps
to the total SFR of SFGs, accounting for variations in dust extinction across 
the disc and between clump regions and off-clump regions. They find that in 
clumpy galaxies, the clumps contribute $\sim 5-15\%$ of the total SFR. 
\citet{Guo15} find a slight tendency for higher contributions at lower 
redshifts, but no systematic dependence on galaxy mass. These results are 
in good agreement with our simulations (\fig{clumpiness}).

\smallskip
\citet{Wuyts12} estimated a total contribution of clumps to the stellar mass 
of clumpy galaxies to be $\lsim 5\%$, when observing in blue wavelengths. This 
is quite a bit higher than the mass fractions in our simulations. However, this 
study did not attempt to resolve individual clumps, but rather studied stacked, 
pixelated images of galaxies, so a one-to-one comparison is difficult. Much of what 
we deem as off-clump material, as well as many of our ZLCs would likely be included 
in this mass estimate. We also note that we find a comparable contribution of clumps 
to the disc mass of clumpy galaxies, so the very low ratios of clump mass to total 
stellar mass may be influenced by our bulge to disc ratios being somewhat higher than 
observed galaxies, rather than by our clumps not being massive enough.

\smallskip
\citet{Guo15} examined the UV luminosity functions of clumps as a function of 
galaxy mass and redshifts. \citet{Livermore15} studied a sample of strongly 
lensed galaxies at $1<z<4$ and also measured the UV luminosity function of 
clumps, down to much smaller scales than available in other works, due to 
the magnification afforded by lensing. Their results are extremely similar 
to the shape of the SFR function for our simulated clumps. The rising slopes 
at the faint end, the declining slope at the bright end, and the increase of 
the high-end cutoff towards higher redshift are all very consistent with our 
results. A quantitative comparison requires detailed treatment of dust in the 
simulations, which is ongoing. However, the inescapable conclusion is that 
individual clumps contribute more to the galaxies UV light and SFR at higher 
redshifts. This is likely a result of the higher fragmentation scale at high 
redshifts due to the increased dust fractions \citep{Livermore15}. On the other 
hand, lower redshift galaxies contain more clumps, each of which is less bright, 
but which combined can contribute more to the total galaxy light.

\subsubsection{In Situ vs Ex Situ Clumps}

\smallskip
As already pointed out by M14, we find that in situ clumps and 
merging ex situ clumps are distinguishable. In order to get a clean sample 
of in situ clumps with which to test the predictions presented here, separating 
the two populations is desirable. As advocated by M14, clumps near the disc 
edge that are massive, $\lsim 0.1\Md$, with old stellar populations $\sim 1\Gyr$ 
old and low sSFRs of $\lsim 0.1 \Gyr^{-1}$ are likely to be ex situ. While 
in situ clumps can develop comparable properties late in their life, this only 
occurs deep within the disc.

% colors - SUNRISE
\smallskip
In many cases, young stellar ages and high sSFRs and gas fractions 
should translate to observed blue colors and vice versa. However, 
realistic luminosities and colors will depend on the effects of dust. 
In an ongoing work, dust and radiative transfer is incorporated into 
the simulated galaxies using the SUNRISE code \citep{Jonsson06}, 
thus creating realistic mock observations comparable to CANDELS 
data \citep{Moody14}. Prelimenary results show that in situ clumps 
are not affected much by dust near the disc edge in face-on images 
and they indeed tend to appear very blue, so the distinctions between 
in situ and ex situ discussed above should be observable. Similar 
conclusions were reached by \citet{Wuyts13} who found dust extinction 
to be weaker in UV selected clumps than in the interclump regions. 
However, in edge on views and toward the central parts of the disc, 
the in situ clumps are reddened, and the differences between the 
clump types become less pronounced. Observations find most off-centre 
clumps to be blue  \citep{Elmegreen05b,Forster11b,Guo12,Wuyts12,Guo15}. 
Based on our analysis of the simulations and the preliminary SUNRISE 
images, this suggests that the majority of the observed off-centre 
clumps are, indeed, in situ.
%also suggest that the off-centre clumps are peaks in the distribution of sSFR, similar to what we find for the in situ  clumps, which have sSFR values a factor of $\sim 5$ higher than the median of the smooth disc.

\smallskip
A few of the observed off-centre clumps are redder and more massive 
\citep{Genzel11,Forster11b}, and these may well be ex situ clumps. 
It is also possible that more ex situ clumps have been observed, but 
classified as mergers rather than clumps. In addition, central massive 
red clumps have been observed \citep{Elmegreen09,Guo12,Wuyts12} which 
seem to resemble our bulge clumps. 

\smallskip
Given the mounting theoretical and observational evidence that most of the 
high-$z$ SFGs are extended discs undergoing VDI and that external mergers are 
responsible for only a part of the clump population, it would not make sense 
to classify the high-z SFGs using the familiar classification schemes used at 
low redshifts. In particular, the high-$z$ VDI phase with giant clumps is 
unlikely and therefore unaccounted for at low redshift, where disc instability 
takes the form of a bar and spiral arms associated with secular evolution (though 
see \citet{Green14} for a sample of low-$z$ analogues to the high-$z$ clumpy discs). 
This calls for a new classification scheme for high redshift galaxies, which 
recognizes the dominance of VDI systems and explicitly differentiates between 
VDI galaxies and merging systems. Such a scheme will be devised using the SUNRISE 
images of the simulated galaxies together with the complete merger history of all 
the clumps. 

\subsubsection{Clump Migration}

\smallskip
The stellar ages of clumps and their predicted lifetimes are 
being estimated observationally, though with very large uncertainties.
\citet{Elmegreen05b} observed ten clump-cluster galaxies in 
the HUDF at $1.6<z<3.0$ with 5-10 clumps each, and estimated 
clump ages of $100-800 \Myr$, with an average of $340 \Myr$, 
hosted in older discs of $1.4-3 \Gyr$. Then \citet{Elmegreen09} 
found a very large range of ages for star-forming clumps, centred 
around $100\Myr$ but reaching values as high as $1\Gyr$. 
\citet{Genzel11}, who examined five $z\sim 2.2$ clumpy SFGs with 
SINFONI, estimated clumps to be between 10 and a few hundred 
$\Myr$ old, with typical clumps having stellar ages of $100-200\Myr$ 
and an upper envelope of $300\Myr$. Additional considerations 
led them to estimate the average lifetime of clumps to be $\sim 500 \Myr$. 
\citet{Forster11b} obtained ages for 7 clumps 
in one galaxy from the SINS survey, which ranged from a few 
tens to about $\sim 250\Myr$, centred on just below $100\Myr$. 
\citet{Wuyts12} pixelated and stacked galaxy images and defined 
clumps as off-centre pixels with elevated surface brightness above the 
background. At $z\sim 2$, the clump pixels have ages of $\sim 200 \Myr$, 
far younger than the off-clump pixels. At $z\sim 1$, both the clumps and 
the discs are older by about a factor of 2. \citet{Guo12} collected data 
on $\sim 40$ clumps from ten galaxies in the HUDF and found the distribution 
of clump ages to be roughly lognormal, centred on $\sim 300\Myr$, but 
covering a wide range from $10 \Myr$ to a few $\Gyr$, while the disc ages 
were concentrated in the range $0.3-1 \Gyr$. 

\smallskip
\citet{Zanella15} observed a massive, $\Mc \gsim 2.5\times 10^9\msun$, off centre clump at $z\sim 2$. 
They estimated its sSFR to be 30 times higher than the underlying disc, with a 
gas consumption timescale 10 times faster. They were also able to obtain robust 
age measurements, finding the stellar population to be $\sim 10\Myr$ old and 
suggesting that the clump is in the process of formation. Using this short time 
window for their discovery, together with the fact that they found one such case 
in a sample of 68 galaxies, they estimated the frequency of formation of such a 
massive clump to be $\sim 2.5 \Gyr ^{-1}$ per galaxy. Given the observations 
of $1-2$ clumps per galaxy with similar masses, they estimated a clump lifetime 
of $\sim 500\Myr$.

\smallskip
These studies use different methods for estimating the clump stellar ages, 
and all agree that their uncertainties are very large. Nevertheless, the 
fact that the clumps exhibit such a wide range of ages in any particular 
study and that even low estimates on clump ages are rarely far below $100 \Myr$ 
seem to favour a scenario where clumps are long lived. This is especially 
true when one considers that the observed stellar ages of clumps are likely 
underestimates of the true time since clump formation, due to continued star 
formation in the clump and tidal stripping of the older stars (\fig{age_time} 
and \citealp{Bournaud14}).

% age gradient
\smallskip
Some of these studies \citep{Forster11b,Guo12} have even attempted to measure 
radial variations of clump properties within the discs. They find evidence for 
older, redder and more massive clumps to be located closer to the disc centre. 
\citet{Forster11b} find a logarithmic slope for the ages of seven clumps in 
a single galaxy of $-2.06 \pm 0.63$, much steeper than the radial variation 
of the background disc. They note that even if the absolute values for clump 
ages are wrong, the relative trend should hold. In another galaxy from 
their sample, also containing seven clumps, they found clumps to become 
redder closer to the centre, with increased mass-to-light ratios. Clumps 
near the disc edge have masses of $\sim 10^8\msun$ while close to the 
centre the masses are $\gsim 6\times 10^8\msun$. \citet{Guo12} find that 
clumps closer to the disc centre have lower sSFRs, older ages, higher dust 
extinctions and higher stellar surface densities. Moreover, they find the 
radial variation in clump properties steeper than the global gradients in 
the background disc, and deduce that the clumps must have evolved separately 
from the disc in a state of quasi-equilibrium. They find clumps within 0.1 
times the disc radius to have sSFR values 5 times lower and stellar surface 
densities 25 times higher than clumps in the outer half of the disc, very 
similar to the LLCs produced in our simulations. The clumps near 
the disc centre have ages of roughly $700 \Myr$ as opposed to $\sim 100 \Myr$ 
in the outer disc. These studies, while crude and preliminary, are in good 
agreement with the predictions of clump migration, which seems to indicate 
that massive clumps survive for extended periods longer than an orbital time 
and evolve as they migrate inwards. 

\subsection{Caveats and Future Prospects}

\smallskip
As detailed in \se{limitations}, these simulations are still not 
doing the best job in terms of feedback and star-formation. The 
same simulations studied in this work in the NoRP and RP versions 
are currently being run in a third version which includes additional 
radiation trapping, with an effective boost factor of a few in dense 
regions, as well as photoheating from UV photons \citep{Ceverino14}. 
These simulations will be analyzed in a similar way to the NoRP and 
RP versions used here to determine the effect on clump survival. 
However, prelimenary tests reveal that this new feedback has only 
a minor effect on the stellar masses of the galaxies. Furthermore, 
\citet{Ceverino14} showed that the effect of these additional feedback 
mechanisms on the density and temperatue of ISM is less severe than 
the initial effect of RP as implemented here. Therefore, we do not 
expect the main conclusions of this work to change, namely that massive 
clumps will survive and migrate towards the centre.

\smallskip
However, the weak feedback may underestimate the importance of winds 
in LLCs. In work in preparation, we use an analytic bathtub toy 
model to predict the evolution of giant clumps subject to feedback 
driven winds, star formation and mass accretion from the surrounding 
disc. We find that for realistic mass loading factors, the clump mass 
should be constant during migration, with accretion balancing the outflows. 
However, in these simulations we find a strong mass gradient for LLCs, 
similar to the very weak feedback simulations studied in M14. This may 
affect the properties of LLCs, and must be tested in future studies of 
simulations with more realistic feedback.

\smallskip
Finally, our analysis has been performed in 3D, taking full advantage 
of the information available in the simulations. 
While this method has helped us develop a theoretical understanding 
of VDI, one should now worry about how to directly compare our 
results with realistic 2D observations that also suffer from dust 
effects, background and foreground contamination, and beam smearing. 
Previous studies \citep{Moody14,Snyder15} have used the radiative 
transfer code SUNRISE to ``observe" these simulated galaxies in a 
realistic way, e.g. for comparison with CANDELS data. Efforts are 
currently being made to use these mock-observations to compare the 
populations of clumps found in this work to observed clumps found 
using the method of \citet{Guo15}. This will greatly assist in relating 
the simulations to observations, and devising observational tests to 
distinguish between LLCs and SLCs, as well as between in situ clumps 
and mergers, allowing us to devise a new theory-motivated classification 
scheme for high-$z$ galaxies, which explicitly accounts for VDI.

%%%%%%%%%%%%%%%%%%%%%%%%%%%%%%%%%%%%%%%%%%%%%%%%  
\section{Conclusions}
\label{sec:conc} 

\smallskip
We have studied the lifetime, evolution and properties of 
giant clumps formed in high-$z$ disc galaxies using two suites 
of high-resolution cosmological simulations, run with and without 
radiative-pressure feedback from young stars. Our main goals were 
to address the effect of radiation pressure on the formation and 
survival of clumps, and to devise observable signatures of clump 
evolution during their inward migration. While the RP feedback 
used here is relatively weak compared to what is often assumed 
in other works, our predictions regarding observable differences 
between long-lived, migrating clumps and short-lived disrupting 
clumps, when combined with observations, can be used to place 
constraints on clump migration and the efficiency of feedback. 
Our main results can be summarized as follows:

\begin{enumerate}

\smallskip
\item RP feedback lowers the galactic stellar mass by a factor of $\sim 2$ 
at $z\sim 2$ and by up to a factor of $\sim 10$ in low mass galaxies at 
high redshift. While our stellar-to-halo mass ratios are still roughly 
a factor of $\sim 2$ above the \citet{Behroozi13} relation, they appear 
consistent with recent observations of SFGs at $z\lsim 2$ residing in halos 
of mass $\Mv\lsim 4\times 10^{11}\msun$ \citep[][figure 5]{Burkert15}.

\smallskip
\item RP has only a mild effect on the sizes of discs, increasing the 
thickness by $\sim 30\%$ and the radius by $\lsim 10\%$. The cold fraction 
within the disc radius increases when RP is included due to delayed star-formation 
and increased gas fractions. However, the Toomre $Q$ parameter is comparable 
in both feedback models, showing that one is not inherently more unstable than 
the other. In both feedback models, the ISM density transitions from a log-normal 
distribution, dominated by supersonic turbulence at low densities to a power 
law distribution, dominated by self-gravity, at high densities of $n\gsim 10\cmc$. 
In the RP simulations, the fraction of mass in dense self-gravitating gas is 
reduced by a factor of $\sim 2$. 

\smallskip
\item There is a bi-modality between short-lived clumps (SLCs) and long-lived clumps 
(LLCs), separated at a lifetime of $\sim 20$ free-fall times. While RP only mildly 
suppresses the total number of clumps that form, it greatly reduces the number of LLCs. 
There is also a large population of low-mass clumps with virial parameters 
$\alpha_{\rm v}\sim 10-1000$ that never formed significant stellar populations in them 
(zero-lifetime clumps, ZLCs). These were likely never bound structures and are not 
included in our analysis.

\smallskip
\item The majority of clumps have densities $\nc>10\cmc$, in the self-gravitating regime 
of the ISM density distribution, and circular velocities $V_{\rm circ,\:c}>20\kms$. Below 
this value, clumps are rapidly disrupted by supernova feedback even without RP. The clump 
mass function flattens when these two thresholds meet, below $\Mc\sim 10^{7.3}\msun$. 

\smallskip
\item When RP is included, it introduces a scale in surface density, with most clumps 
having surface densities $\Sigma_{\rm c}>200-300\msun\:\pc^{-2}$. This threshold meets 
the threshold in clump density of $\nc>10\cmc$ at a mass of $\Mc\sim  10^{8.2-8.5}\msun$. 
Above this mass, RP does not disrupt clumps and there are comparable numbers of LLCs in 
both simulation suites. 
%Since more massive galaxies than simulated here will tend to produce more massive clumps with higher surface densities, increasing the strength of our RP feedback by a factor of $\sim 3$ as advocated by, e.g., \citet{DK13}, will not cause disruption of the most massive clumps of $\Mc\sim 10^9 \msun$.

\smallskip
\item In addition to in situ clumps, formed by VDI, we identify ex situ mergers 
as part of the off centre clump population. RP greatly reduces the relative number 
of ex situ clumps, likely by reducing their central density prior to merging with 
the main galaxy and making them more susceptible to tidal disruption and stripping. 
Ex situ clumps tend to be more massive than in situ clumps. Among off-centre clumps 
in the RP simulations with $\Mc>10^7\msun$, ex situ clumps are $\sim 5\%$ in number, 
$\sim 37\%$ in mass and $\sim 29\%$ in SFR.

\smallskip
\item The distribution of clump masses and SFRs normalized to their host disc is 
very similar at all redshifts. They exhibit a power-law at the bright end, with 
a slope slightly shallower than $-2$ and a cutoff at high masses and SFRs. The 
cutoff is at a higher mass at higher redshifts, so individual clumps become less 
dominant dynamically as time goes on. The shape of these distributions, including 
their slopes and the trend of cutoff value with redshift, are similar to observed 
luminosity functions of high-$z$ clumps \citep{Guo15}. 

\smallskip
\item The fraction of clumpy galaxies is $\sim 40-60\%$ over a broad range of redhshift 
and stellar mass, with a tendency to decrease towards lower redshift and for low mass 
galaxies. The general trends and percentages are similar to observations, within the 
uncertainties. Among clumpy galaxies, clumps contribute $\sim 3-30\%$ to the total SFR, 
and $\sim 0.1-3\%$ to the total stellar mass, though the contribution for massive galaxies 
can be higher. There is only a very weak dependence of these fractions with redshift, where 
clumps contribute a larger fraction of the stellar mass and a smaller fraction of the SFR 
at higher redshift.

\smallskip
\item Long-lived-clumps tend to migrate towards the disc centre over a migration 
time of $\sim 10-30$ disc dynamical times, and can be found at distances as small 
as $\sim 0.1\Rd$. SLCs are mostly located near the disc edge, and this is predicted 
to be the case in post-compaction discs. The two populations 
are well separated in the plane of age-sSFR, where LLCs are older and with lower 
sSFR values. LLCs exhibit radial gradients in age, mass, sSFR, gas fraction and 
metallicity, becoming older, more massive and metal-enriched but with lower gas 
fraction and sSFR closer to the disc centre. The SLCs show flatter gradients in 
age and gas fraction, steeper gradients in metallicity, and no significant gradient 
in mass or sSFR. Ex situ clumps tend to have older stellar ages and higher masses 
than the in situ clumps, especially near the disc edge. 
However, their gas fraction, metallicity and sSFR values are similar to the LLCs. 
These differences are in principle observable, and can be used to distinguish 
long-lived migrating clumps from short-lived, rapidly disrupting clumps and ex 
situ mergers, and can place strong constraints on the origin and fate of observed 
giant clumps at high-$z$. 
\end{enumerate}

%%%%%%%%%%%%%%%%%%%%%%%%%%%%%%%%% 
\section*{Acknowledgments} 
We thank the anonymous referee for 
comments that have greatly improved this 
manuscript. The simulations were performed 
at the National Energy Research Scientific 
Computing centre (NERSC), Lawrence Berkeley 
National Laboratory, and at NASA Advanced 
Supercomputing (NAS) at NASA Ames Reserach 
centre. The analysis was performed on the 
Astric cluster at HU. This work was supported 
by ISF grant 24/12, by BSF grant 2014-273, 
by GIF grant G-1052-104.7/2009, by the I-CORE 
Program of the PBC, ISF grant 1829/12, by 
CANDELS grant HST-GO-12060.12-A, and by NSF
grants AST-1010033 and AST-1405962. DC 
has been partly funded by the ERC Advanced Grant, 
STARLIGHT: Formation of the First Stars (project 
number 339177).

%%%%%%%%%%%%%%%%%%%%%%%%%%%%%%%%%%%% 
\bibliographystyle{mn2e} 
%\bibliography{biblio}

\begin{thebibliography}{120}
\expandafter\ifx\csname natexlab\endcsname\relax\def\natexlab#1{#1}\fi

\bibitem[{{Agertz}, {Teyssier} \& {Moore}(2009){Agertz}, {Teyssier}, \&
  {Moore}}]{Agertz09}
{Agertz} O., {Teyssier} R., {Moore} B., 2009, \mnras, 397, L64

\bibitem[{{Aubert}, {Pichon} \& {Colombi}(2004){Aubert}, {Pichon}, \&
  {Colombi}}]{Aubert04}
{Aubert} D., {Pichon} C., {Colombi} S., 2004, \mnras, 352, 376

\bibitem[{{Behrendt}, {Burkert} \& {Schartmann}(2015){Behrendt}, {Burkert}, \&
  {Schartmann}}]{Behrendt15}
{Behrendt} M., {Burkert} A., {Schartmann} M., 2015, \mnras, 448, 1007

\bibitem[{{Behroozi}, {Conroy} \& {Wechsler}(2010){Behroozi}, {Conroy}, \&
  {Wechsler}}]{Behroozi10}
{Behroozi} P.~S., {Conroy} C., {Wechsler} R.~H., 2010, \apj, 717, 379

\bibitem[{{Behroozi}, {Wechsler} \& {Conroy}(2013){Behroozi}, {Wechsler}, \&
  {Conroy}}]{Behroozi13}
{Behroozi} P.~S., {Wechsler} R.~H., {Conroy} C., 2013, \apj, 770, 57

\bibitem[{{Birnboim} \& {Dekel}(2003)}]{bd03}
{Birnboim} Y., {Dekel} A., 2003, \mnras, 345, 349

\bibitem[{{Bond} {et~al}\mbox{.}(1991){Bond}, {Cole}, {Efstathiou}, \&
  {Kaiser}}]{Bond91}
{Bond} J.~R., {Cole} S., {Efstathiou} G., {Kaiser} N., 1991, \apj, 379, 440

\bibitem[{{Bournaud} {et~al}\mbox{.}(2011){Bournaud}, {Dekel}, {Teyssier},
  {Cacciato}, {Daddi}, {Juneau}, \& {Shankar}}]{Bournaud11}
{Bournaud} F., {Dekel} A., {Teyssier} R., {Cacciato} M., {Daddi} E., {Juneau}
  S., {Shankar} F., 2011, \apjl, 741, L33

\bibitem[{{Bournaud} \& {Elmegreen}(2009)}]{Bournaud09}
{Bournaud} F., {Elmegreen} B.~G., 2009, \apjl, 694, L158

\bibitem[{{Bournaud}, {Elmegreen} \& {Elmegreen}(2007){Bournaud}, {Elmegreen},
  \& {Elmegreen}}]{Bournaud07}
{Bournaud} F., {Elmegreen} B.~G., {Elmegreen} D.~M., 2007, \apj, 670, 237

\bibitem[{{Bournaud} {et~al}\mbox{.}(2012){Bournaud}, {Juneau}, {Le Floc'h},
  {Mullaney}, {Daddi}, {Dekel}, {Duc}, {Elbaz}, {Salmi}, \&
  {Dickinson}}]{Bournaud12}
{Bournaud} F. {et~al.}, 2012, \apj, 757, 81

\bibitem[{{Bournaud} {et~al}\mbox{.}(2014){Bournaud}, {Perret}, {Renaud},
  {Dekel}, {Elmegreen}, {Elmegreen}, {Teyssier}, {Amram}, {Daddi}, {Duc},
  {Elbaz}, {Epinat}, {Gabor}, {Juneau}, {Kraljic}, \& {Le Floch'}}]{Bournaud14}
{Bournaud} F. {et~al.}, 2014, \apj, 780, 57

\bibitem[{{Bryan} \& {Norman}(1998)}]{Bryan98}
{Bryan} G.~L., {Norman} M.~L., 1998, \apj, 495, 80

\bibitem[{{Burkert} {et~al}\mbox{.}(2015){Burkert}, {F{\"o}rster Schreiber},
  {Genzel}, {Lang}, {Tacconi}, {Wisnioski}, {Wuyts}, {Bandara}, {Beifiori},
  {Bender}, {Brammer}, {Chan}, {Davies}, {Dekel}, {Fabricius}, {Fossati},
  {Kulkarni}, {Lutz}, {Mendel}, {Momcheva}, {Nelson}, {Naab}, {Renzini},
  {Saglia}, {Sharples}, {Sternberg}, {Wilman}, \& {Wuyts}}]{Burkert15}
{Burkert} A. {et~al.}, 2015, ArXiv e-prints

\bibitem[{{Cacciato}, {Dekel} \& {Genel}(2012){Cacciato}, {Dekel}, \&
  {Genel}}]{Cacciato12}
{Cacciato} M., {Dekel} A., {Genel} S., 2012, \mnras, 421, 818

\bibitem[{{Ceverino} {et~al}\mbox{.}(2016{\natexlab{a}}){Ceverino}, {Arribas},
  {Colina}, {Rodr{\'{\i}}guez Del Pino}, {Dekel}, \& {Primack}}]{Ceverino16a}
{Ceverino} D., {Arribas} S., {Colina} L., {Rodr{\'{\i}}guez Del Pino} B.,
  {Dekel} A., {Primack} J., 2016{\natexlab{a}}, \mnras, 460, 2731

\bibitem[{{Ceverino}, {Dekel} \& {Bournaud}(2010){Ceverino}, {Dekel}, \&
  {Bournaud}}]{CDB}
{Ceverino} D., {Dekel} A., {Bournaud} F., 2010, \mnras, 404, 2151

\bibitem[{{Ceverino} {et~al}\mbox{.}(2012){Ceverino}, {Dekel}, {Mandelker},
  {Bournaud}, {Burkert}, {Genzel}, \& {Primack}}]{Ceverino12}
{Ceverino} D., {Dekel} A., {Mandelker} N., {Bournaud} F., {Burkert} A.,
  {Genzel} R., {Primack} J., 2012, \mnras, 420, 3490

\bibitem[{{Ceverino} {et~al}\mbox{.}(2015){Ceverino}, {Dekel}, {Tweed}, \&
  {Primack}}]{Ceverino15a}
{Ceverino} D., {Dekel} A., {Tweed} D., {Primack} J., 2015, \mnras, 447, 3291

\bibitem[{{Ceverino} \& {Klypin}(2009)}]{ceverino09}
{Ceverino} D., {Klypin} A., 2009, \apj, 695, 292

\bibitem[{{Ceverino} {et~al}\mbox{.}(2014){Ceverino}, {Klypin}, {Klimek},
  {Trujillo-Gomez}, {Churchill}, {Primack}, \& {Dekel}}]{Ceverino14}
{Ceverino} D., {Klypin} A., {Klimek} E.~S., {Trujillo-Gomez} S., {Churchill}
  C.~W., {Primack} J., {Dekel} A., 2014, \mnras, 442, 1545

\bibitem[{{Ceverino}, {Primack} \& {Dekel}(2015){Ceverino}, {Primack}, \&
  {Dekel}}]{Ceverino15b}
{Ceverino} D., {Primack} J., {Dekel} A., 2015, \mnras, 453, 408

\bibitem[{{Ceverino} {et~al}\mbox{.}(2016{\natexlab{b}}){Ceverino},
  {S{\'a}nchez Almeida}, {Mu{\~n}oz Tu{\~n}{\'o}n}, {Dekel}, {Elmegreen},
  {Elmegreen}, \& {Primack}}]{Ceverino16b}
{Ceverino} D., {S{\'a}nchez Almeida} J., {Mu{\~n}oz Tu{\~n}{\'o}n} C., {Dekel}
  A., {Elmegreen} B.~G., {Elmegreen} D.~M., {Primack} J., 2016{\natexlab{b}},
  \mnras, 457, 2605

\bibitem[{{Chabrier}(2003)}]{Chabrier03}
{Chabrier} G., 2003, \pasp, 115, 763

\bibitem[{{Conroy} \& {Wechsler}(2009)}]{Conroy09}
{Conroy} C., {Wechsler} R.~H., 2009, \apj, 696, 620

\bibitem[{{Cresci} {et~al}\mbox{.}(2009){Cresci}, {Hicks}, {Genzel},
  {Schreiber}, {Davies}, {Bouch{\'e}}, {Buschkamp}, \& {et al.,}}]{Cresci09}
{Cresci} G., {Hicks} E.~K.~S., {Genzel} R., {Schreiber} N.~M.~F., {Davies} R.,
  {Bouch{\'e}} N., {Buschkamp} P., {et al.,}, 2009, \apj, 697, 115

\bibitem[{{Daddi} {et~al}\mbox{.}(2010){Daddi}, {Bournaud}, {Walter},
  {Dannerbauer}, {Carilli}, {Dickinson}, {Elbaz}, {Morrison}, {Riechers},
  {Onodera}, {Salmi}, {Krips}, \& {Stern}}]{Daddi10}
{Daddi} E. {et~al.}, 2010, \apj, 713, 686

\bibitem[{{Davis} {et~al}\mbox{.}(2014){Davis}, {Jiang}, {Stone}, \&
  {Murray}}]{Davis14}
{Davis} S.~W., {Jiang} Y.-F., {Stone} J.~M., {Murray} N., 2014, \apj, 796, 107

\bibitem[{{Dekel} \& {Birnboim}(2006)}]{db06}
{Dekel} A., {Birnboim} Y., 2006, \mnras, 368, 2

\bibitem[{{Dekel} {et~al}\mbox{.}(2009){Dekel}, {Birnboim}, {Engel},
  {Freundlich}, {Goerdt}, {Mumcuoglu}, {Neistein}, {Pichon}, {Teyssier}, \&
  {Zinger}}]{Dekel09}
{Dekel} A. {et~al.}, 2009, \nat, 457, 451

\bibitem[{{Dekel} \& {Burkert}(2014)}]{DekelBurkert14}
{Dekel} A., {Burkert} A., 2014, \mnras, 438, 1870

\bibitem[{{Dekel} \& {Krumholz}(2013)}]{DK13}
{Dekel} A., {Krumholz} M.~R., 2013, \mnras, 432, 455

\bibitem[{{Dekel}, {Sari} \& {Ceverino}(2009){Dekel}, {Sari}, \&
  {Ceverino}}]{DSC}
{Dekel} A., {Sari} R., {Ceverino} D., 2009, \apj, 703, 785

\bibitem[{{Dekel} \& {Silk}(1986)}]{Dekel86}
{Dekel} A., {Silk} J., 1986, \apj, 303, 39

\bibitem[{{Elmegreen}(2011)}]{Elmegreen11}
{Elmegreen} B.~G., 2011, \apj, 731, 61

\bibitem[{{Elmegreen}, {Bournaud} \& {Elmegreen}(2008){Elmegreen}, {Bournaud},
  \& {Elmegreen}}]{Elmegreen08}
{Elmegreen} B.~G., {Bournaud} F., {Elmegreen} D.~M., 2008, \apj, 688, 67

\bibitem[{{Elmegreen} \& {Elmegreen}(2005)}]{Elmegreen05b}
{Elmegreen} B.~G., {Elmegreen} D.~M., 2005, \apj, 627, 632

\bibitem[{{Elmegreen} {et~al}\mbox{.}(2009){Elmegreen}, {Elmegreen},
  {Fernandez}, \& {Lemonias}}]{Elmegreen09}
{Elmegreen} B.~G., {Elmegreen} D.~M., {Fernandez} M.~X., {Lemonias} J.~J.,
  2009, \apj, 692, 12

\bibitem[{{Elmegreen} {et~al}\mbox{.}(2007){Elmegreen}, {Elmegreen},
  {Ravindranath}, \& {Coe}}]{Elmegreen07}
{Elmegreen} D.~M., {Elmegreen} B.~G., {Ravindranath} S., {Coe} D.~A., 2007,
  \apj, 658, 763

\bibitem[{{Federrath}, {Klessen} \& {Schmidt}(2008){Federrath}, {Klessen}, \&
  {Schmidt}}]{Federrath08}
{Federrath} C., {Klessen} R.~S., {Schmidt} W., 2008, \apjl, 688, L79

\bibitem[{{Ferland} {et~al}\mbox{.}(1998){Ferland}, {Korista}, {Verner},
  {Ferguson}, {Kingdon}, \& {Verner}}]{Ferland98}
{Ferland} G.~J., {Korista} K.~T., {Verner} D.~A., {Ferguson} J.~W., {Kingdon}
  J.~B., {Verner} E.~M., 1998, \pasp, 110, 761

\bibitem[{{Forbes}, {Krumholz} \& {Burkert}(2012){Forbes}, {Krumholz}, \&
  {Burkert}}]{Forbes12}
{Forbes} J., {Krumholz} M., {Burkert} A., 2012, \apj, 754, 48

\bibitem[{{Forbes} {et~al}\mbox{.}(2014){Forbes}, {Krumholz}, {Burkert}, \&
  {Dekel}}]{Forbes13}
{Forbes} J.~C., {Krumholz} M.~R., {Burkert} A., {Dekel} A., 2014, \mnras, 438,
  1552

\bibitem[{{F{\"o}rster Schreiber} {et~al}\mbox{.}(2009){F{\"o}rster Schreiber},
  {Genzel}, {Bouch{\'e}}, {Cresci}, {Davies}, {Buschkamp}, {Shapiro}, \& {et
  al.,}}]{Forster09}
{F{\"o}rster Schreiber} N.~M., {Genzel} R., {Bouch{\'e}} N., {Cresci} G.,
  {Davies} R., {Buschkamp} P., {Shapiro} K., {et al.,}, 2009, \apj, 706, 1364

\bibitem[{{F{\"o}rster Schreiber} {et~al}\mbox{.}(2006){F{\"o}rster Schreiber},
  {Genzel}, {Lehnert}, {Bouch{\'e}}, {Verma}, {Erb}, {Shapley}, \& {et
  al.,}}]{Forster06}
{F{\"o}rster Schreiber} N.~M., {Genzel} R., {Lehnert} M.~D., {Bouch{\'e}} N.,
  {Verma} A., {Erb} D.~K., {Shapley} A.~E., {et al.,}, 2006, \apj, 645, 1062

\bibitem[{{F{\"o}rster Schreiber} {et~al}\mbox{.}(2011){F{\"o}rster Schreiber},
  {Shapley}, {Genzel}, {Bouch{\'e}}, {Cresci}, {Davies}, {Erb}, {Genel},
  {Lutz}, {Newman}, {Shapiro}, {Steidel}, {Sternberg}, \&
  {Tacconi}}]{Forster11b}
{F{\"o}rster Schreiber} N.~M. {et~al.}, 2011, \apj, 739, 45

\bibitem[{{Freundlich} {et~al}\mbox{.}(2013){Freundlich}, {Combes}, {Tacconi},
  {Cooper}, {Genzel}, {Neri}, {Bolatto}, {Bournaud}, {Burkert}, {Cox}, {Davis},
  {F{\"o}rster Schreiber}, {Garcia-Burillo}, {Gracia-Carpio}, {Lutz}, {Naab},
  {Newman}, {Sternberg}, \& {Weiner}}]{Freundlich13}
{Freundlich} J. {et~al.}, 2013, \aap, 553, A130

\bibitem[{{Gammie}(2001)}]{Gammie01}
{Gammie} C.~F., 2001, \apj, 553, 174

\bibitem[{{Genel} {et~al}\mbox{.}(2012){Genel}, {Naab}, {Genzel}, {F{\"o}rster
  Schreiber}, {Sternberg}, {Oser}, {Johansson}, {Dav{\'e}}, {Oppenheimer}, \&
  {Burkert}}]{Genel12a}
{Genel} S. {et~al.}, 2012, \apj, 745, 11

\bibitem[{{Genzel} {et~al}\mbox{.}(2008){Genzel}, {Burkert}, {Bouch{\'e}},
  {Cresci}, {F{\"o}rster Schreiber}, {Shapley}, {Shapiro}, \& {et
  al.,}}]{Genzel08}
{Genzel} R., {Burkert} A., {Bouch{\'e}} N., {Cresci} G., {F{\"o}rster
  Schreiber} N.~M., {Shapley} A., {Shapiro} K., {et al.,}, 2008, \apj, 687, 59

\bibitem[{{Genzel} {et~al}\mbox{.}(2014){Genzel}, {F{\"o}rster Schreiber},
  {Lang}, {Tacchella}, {Tacconi}, {Wuyts}, {Bandara}, {Burkert}, {Buschkamp},
  {Carollo}, {Cresci}, {Davies}, {Eisenhauer}, {Hicks}, {Kurk}, {Lilly},
  {Lutz}, {Mancini}, {Naab}, {Newman}, {Peng}, {Renzini}, {Shapiro Griffin},
  {Sternberg}, {Vergani}, {Wisnioski}, {Wuyts}, \& {Zamorani}}]{Genzel14}
{Genzel} R. {et~al.}, 2014, \apj, 785, 75

\bibitem[{{Genzel} {et~al}\mbox{.}(2006){Genzel}, {Tacconi}, {Eisenhauer},
  {F{\"o}rster Schreiber}, {Cimatti}, {Daddi}, {Bouch{\'e}}, \& {et
  al.,}}]{Genzel06}
{Genzel} R., {Tacconi} L.~J., {Eisenhauer} F., {F{\"o}rster Schreiber} N.~M.,
  {Cimatti} A., {Daddi} E., {Bouch{\'e}} N., {et al.,}, 2006, \nat, 442, 786

\bibitem[{{Genzel} {et~al}\mbox{.}(2015){Genzel}, {Tacconi}, {Lutz},
  {Saintonge}, {Berta}, {Magnelli}, {Combes}, {Garc{\'{\i}}a-Burillo}, {Neri},
  {Bolatto}, {Contini}, {Lilly}, {Boissier}, {Boone}, {Bouch{\'e}}, {Bournaud},
  {Burkert}, {Carollo}, {Colina}, {Cooper}, {Cox}, {Feruglio}, {F{\"o}rster
  Schreiber}, {Freundlich}, {Gracia-Carpio}, {Juneau}, {Kovac}, {Lippa},
  {Naab}, {Salome}, {Renzini}, {Sternberg}, {Walter}, {Weiner}, {Weiss}, \&
  {Wuyts}}]{Genzel15}
{Genzel} R. {et~al.}, 2015, \apj, 800, 20

\bibitem[{{Genzel} {et~al}\mbox{.}(2011){Genzel} {et~al.}}]{Genzel11}
{Genzel} R., {et~al.}, 2011, \apj, 733, 101

\bibitem[{{Green} {et~al}\mbox{.}(2014){Green}, {Glazebrook}, {McGregor},
  {Damjanov}, {Wisnioski}, {Abraham}, {Colless}, {Sharp}, {Crain}, {Poole}, \&
  {McCarthy}}]{Green14}
{Green} A.~W. {et~al.}, 2014, \mnras, 437, 1070

\bibitem[{{Grogin} {et~al}\mbox{.}(2011){Grogin}, {Kocevski}, {Faber},
  {Ferguson}, {Koekemoer}, {Riess}, {Acquaviva}, {Alexander}, {Almaini},
  {Ashby}, {Barden}, {Bell}, {Bournaud}, {Brown}, {Caputi}, {Casertano},
  {Cassata}, {Castellano}, {Challis}, {Chary}, {Cheung}, {Cirasuolo},
  {Conselice}, {Roshan Cooray}, {Croton}, {Daddi}, {Dahlen}, {Dav{\'e}}, {de
  Mello}, {Dekel}, {Dickinson}, {Dolch}, {Donley}, {Dunlop}, {Dutton}, {Elbaz},
  {Fazio}, {Filippenko}, {Finkelstein}, {Fontana}, {Gardner}, {Garnavich},
  {Gawiser}, {Giavalisco}, {Grazian}, {Guo}, {Hathi}, {H{\"a}ussler},
  {Hopkins}, {Huang}, {Huang}, {Jha}, {Kartaltepe}, {Kirshner}, {Koo}, {Lai},
  {Lee}, {Li}, {Lotz}, {Lucas}, {Madau}, {McCarthy}, {McGrath}, {McIntosh},
  {McLure}, {Mobasher}, {Moustakas}, {Mozena}, {Nandra}, {Newman}, {Niemi},
  {Noeske}, {Papovich}, {Pentericci}, {Pope}, {Primack}, {Rajan},
  {Ravindranath}, {Reddy}, {Renzini}, {Rix}, {Robaina}, {Rodney}, {Rosario},
  {Rosati}, {Salimbeni}, {Scarlata}, {Siana}, {Simard}, {Smidt}, {Somerville},
  {Spinrad}, {Straughn}, {Strolger}, {Telford}, {Teplitz}, {Trump}, {van der
  Wel}, {Villforth}, {Wechsler}, {Weiner}, {Wiklind}, {Wild}, {Wilson},
  {Wuyts}, {Yan}, \& {Yun}}]{Grogin11}
{Grogin} N.~A. {et~al.}, 2011, \apjs, 197, 35

\bibitem[{{Guo} {et~al}\mbox{.}(2015){Guo}, {Ferguson}, {Bell}, {Koo},
  {Conselice}, {Giavalisco}, {Kassin}, {Lu}, {Lucas}, {Mandelker}, {McIntosh},
  {Primack}, {Ravindranath}, {Barro}, {Ceverino}, {Dekel}, {Faber}, {Fang},
  {Koekemoer}, {Noeske}, {Rafelski}, \& {Straughn}}]{Guo15}
{Guo} Y. {et~al.}, 2015, \apj, 800, 39

\bibitem[{{Guo} {et~al}\mbox{.}(2012){Guo}, {Giavalisco}, {Ferguson},
  {Cassata}, \& {Koekemoer}}]{Guo12}
{Guo} Y., {Giavalisco} M., {Ferguson} H.~C., {Cassata} P., {Koekemoer} A.~M.,
  2012, \apj, 757, 120

\bibitem[{{Haardt} \& {Madau}(1996)}]{HaardtMadau96}
{Haardt} F., {Madau} P., 1996, \apj, 461, 20

\bibitem[{{Hopkins} \& {Beacom}(2006)}]{hopkins06}
{Hopkins} A.~M., {Beacom} J.~F., 2006, \apj, 651, 142

\bibitem[{{Hopkins}(2013)}]{Hopkins13}
{Hopkins} P.~F., 2013, \mnras, 430, 1653

\bibitem[{{Hopkins} {et~al}\mbox{.}(2012){Hopkins}, {Kere{\v s}}, {Murray},
  {Quataert}, \& {Hernquist}}]{Hopkins12a}
{Hopkins} P.~F., {Kere{\v s}} D., {Murray} N., {Quataert} E., {Hernquist} L.,
  2012, \mnras, 427, 968

\bibitem[{{Hopkins}, {Quataert} \& {Murray}(2011){Hopkins}, {Quataert}, \&
  {Murray}}]{Hopkins11}
{Hopkins} P.~F., {Quataert} E., {Murray} N., 2011, \mnras, 417, 950

\bibitem[{{Hopkins}, {Quataert} \& {Murray}(2012){Hopkins}, {Quataert}, \&
  {Murray}}]{Hopkins12b}
{Hopkins} P.~F., {Quataert} E., {Murray} N., 2012, \mnras, 421, 3488

\bibitem[{{Immeli} {et~al}\mbox{.}(2004{\natexlab{a}}){Immeli}, {Samland},
  {Gerhard}, \& {Westera}}]{Immeli04a}
{Immeli} A., {Samland} M., {Gerhard} O., {Westera} P., 2004{\natexlab{a}},
  \aap, 413, 547

\bibitem[{{Immeli} {et~al}\mbox{.}(2004{\natexlab{b}}){Immeli}, {Samland},
  {Westera}, \& {Gerhard}}]{Immeli04b}
{Immeli} A., {Samland} M., {Westera} P., {Gerhard} O., 2004{\natexlab{b}},
  \apj, 611, 20

\bibitem[{{Inoue} {et~al}\mbox{.}(2016){Inoue}, {Dekel}, {Mandelker},
  {Ceverino}, {Bournaud}, \& {Primack}}]{Inoue16}
{Inoue} S., {Dekel} A., {Mandelker} N., {Ceverino} D., {Bournaud} F., {Primack}
  J., 2016, \mnras, 456, 2052

\bibitem[{{Jonsson}(2006)}]{Jonsson06}
{Jonsson} P., 2006, \mnras, 372, 2

\bibitem[{{Kennicutt}(1998)}]{Kennicutt98}
{Kennicutt}, Jr. R.~C., 1998, \apj, 498, 541

\bibitem[{{Kere{\v s}} {et~al}\mbox{.}(2005){Kere{\v s}}, {Katz}, {Weinberg},
  \& {Dav{\'e}}}]{Keres05}
{Kere{\v s}} D., {Katz} N., {Weinberg} D.~H., {Dav{\'e}} R., 2005, \mnras, 363,
  2

\bibitem[{{Koekemoer} {et~al}\mbox{.}(2011){Koekemoer}, {Faber}, {Ferguson},
  {Grogin}, {Kocevski}, {Koo}, {Lai}, {Lotz}, {Lucas}, {McGrath}, {Ogaz},
  {Rajan}, {Riess}, {Rodney}, {Strolger}, {Casertano}, {Castellano}, {Dahlen},
  {Dickinson}, {Dolch}, {Fontana}, {Giavalisco}, {Grazian}, {Guo}, {Hathi},
  {Huang}, {van der Wel}, {Yan}, {Acquaviva}, {Alexander}, {Almaini}, {Ashby},
  {Barden}, {Bell}, {Bournaud}, {Brown}, {Caputi}, {Cassata}, {Challis},
  {Chary}, {Cheung}, {Cirasuolo}, {Conselice}, {Roshan Cooray}, {Croton},
  {Daddi}, {Dav{\'e}}, {de Mello}, {de Ravel}, {Dekel}, {Donley}, {Dunlop},
  {Dutton}, {Elbaz}, {Fazio}, {Filippenko}, {Finkelstein}, {Frazer}, {Gardner},
  {Garnavich}, {Gawiser}, {Gruetzbauch}, {Hartley}, {H{\"a}ussler},
  {Herrington}, {Hopkins}, {Huang}, {Jha}, {Johnson}, {Kartaltepe},
  {Khostovan}, {Kirshner}, {Lani}, {Lee}, {Li}, {Madau}, {McCarthy},
  {McIntosh}, {McLure}, {McPartland}, {Mobasher}, {Moreira}, {Mortlock},
  {Moustakas}, {Mozena}, {Nandra}, {Newman}, {Nielsen}, {Niemi}, {Noeske},
  {Papovich}, {Pentericci}, {Pope}, {Primack}, {Ravindranath}, {Reddy},
  {Renzini}, {Rix}, {Robaina}, {Rosario}, {Rosati}, {Salimbeni}, {Scarlata},
  {Siana}, {Simard}, {Smidt}, {Snyder}, {Somerville}, {Spinrad}, {Straughn},
  {Telford}, {Teplitz}, {Trump}, {Vargas}, {Villforth}, {Wagner}, {Wandro},
  {Wechsler}, {Weiner}, {Wiklind}, {Wild}, {Wilson}, {Wuyts}, \&
  {Yun}}]{Koekemoer11}
{Koekemoer} A.~M. {et~al.}, 2011, \apjs, 197, 36

\bibitem[{{Komatsu} {et~al}\mbox{.}(2009){Komatsu}, {Dunkley}, {Nolta},
  {Bennett}, {Gold}, {Hinshaw}, {Jarosik}, {Larson}, {Limon}, {Page},
  {Spergel}, {Halpern}, {Hill}, {Kogut}, {Meyer}, {Tucker}, {Weiland},
  {Wollack}, \& {Wright}}]{WMAP5}
{Komatsu} E. {et~al.}, 2009, \apjs, 180, 330

\bibitem[{{Kravtsov}(2003)}]{kravtsov03}
{Kravtsov} A.~V., 2003, \apjl, 590, L1

\bibitem[{{Kravtsov}, {Klypin} \& {Khokhlov}(1997){Kravtsov}, {Klypin}, \&
  {Khokhlov}}]{kravtsov97}
{Kravtsov} A.~V., {Klypin} A.~A., {Khokhlov} A.~M., 1997, \apjs, 111, 73

\bibitem[{{Krumholz} \& {Burkert}(2010)}]{krumholz_burkert10}
{Krumholz} M., {Burkert} A., 2010, \apj, 724, 895

\bibitem[{{Krumholz} \& {Dekel}(2010)}]{KrumholzDekel}
{Krumholz} M.~R., {Dekel} A., 2010, \mnras, 406, 112

\bibitem[{{Krumholz} \& {Dekel}(2012)}]{KD12}
{Krumholz} M.~R., {Dekel} A., 2012, \apj, 753, 16

\bibitem[{{Krumholz} \& {Thompson}(2012)}]{KT12}
{Krumholz} M.~R., {Thompson} T.~A., 2012, \apj, 760, 155

\bibitem[{{Krumholz} \& {Thompson}(2013)}]{KT13}
{Krumholz} M.~R., {Thompson} T.~A., 2013, \mnras, 434, 2329

\bibitem[{{Leitherer} {et~al}\mbox{.}(1999){Leitherer}, {Schaerer}, {Goldader},
  {Delgado}, {Robert}, {Kune}, {de Mello}, {Devost}, \&
  {Heckman}}]{Starburst99}
{Leitherer} C. {et~al.}, 1999, \apjs, 123, 3

\bibitem[{{Livermore} {et~al}\mbox{.}(2015){Livermore}, {Jones}, {Richard},
  {Bower}, {Swinbank}, {Yuan}, {Edge}, {Ellis}, {Kewley}, {Smail}, {Coppin}, \&
  {Ebeling}}]{Livermore15}
{Livermore} R.~C. {et~al.}, 2015, \mnras, 450, 1812

\bibitem[{{Madau} \& {Dickinson}(2014)}]{Madau14}
{Madau} P., {Dickinson} M., 2014, \araa, 52, 415

\bibitem[{{Madau}, {Pozzetti} \& {Dickinson}(1998){Madau}, {Pozzetti}, \&
  {Dickinson}}]{Madau98}
{Madau} P., {Pozzetti} L., {Dickinson} M., 1998, \apj, 498, 106

\bibitem[{{Mandelker} {et~al}\mbox{.}(2014){Mandelker}, {Dekel}, {Ceverino},
  {Tweed}, {Moody}, \& {Primack}}]{M14}
{Mandelker} N., {Dekel} A., {Ceverino} D., {Tweed} D., {Moody} C.~E., {Primack}
  J., 2014, \mnras, 443, 3675

\bibitem[{{Moody} {et~al}\mbox{.}(2014){Moody}, {Guo}, {Mandelker}, {Ceverino},
  {Mozena}, {Koo}, {Dekel}, \& {Primack}}]{Moody14}
{Moody} C.~E., {Guo} Y., {Mandelker} N., {Ceverino} D., {Mozena} M., {Koo}
  D.~C., {Dekel} A., {Primack} J., 2014, \mnras, 444, 1389

\bibitem[{{Moster}, {Naab} \& {White}(2013){Moster}, {Naab}, \&
  {White}}]{Moster13}
{Moster} B.~P., {Naab} T., {White} S.~D.~M., 2013, \mnras, 428, 3121

\bibitem[{{Moster} {et~al}\mbox{.}(2010){Moster}, {Somerville}, {Maulbetsch},
  {van den Bosch}, {Macci{\`o}}, {Naab}, \& {Oser}}]{Moster10}
{Moster} B.~P., {Somerville} R.~S., {Maulbetsch} C., {van den Bosch} F.~C.,
  {Macci{\`o}} A.~V., {Naab} T., {Oser} L., 2010, \apj, 710, 903

\bibitem[{{Murray}, {Quataert} \& {Thompson}(2010){Murray}, {Quataert}, \&
  {Thompson}}]{murray10}
{Murray} N., {Quataert} E., {Thompson} T.~A., 2010, \apj, 709, 191

\bibitem[{{Newman} {et~al}\mbox{.}(2012){Newman}, {Shapiro Griffin}, {Genzel},
  {Davies}, {F{\"o}rster-Schreiber}, {Tacconi}, {Kurk}, {Wuyts}, {Genel},
  {Lilly}, {Renzini}, {Bouch{\'e}}, {Burkert}, {Cresci}, {Buschkamp},
  {Carollo}, {Eisenhauer}, {Hicks}, {Lutz}, {Mancini}, {Naab}, {Peng}, \&
  {Vergani}}]{Newman12}
{Newman} S.~F. {et~al.}, 2012, \apj, 752, 111

\bibitem[{{Noguchi}(1999)}]{Noguchi99}
{Noguchi} M., 1999, \apj, 514, 77

\bibitem[{{Ocvirk}, {Pichon} \& {Teyssier}(2008){Ocvirk}, {Pichon}, \&
  {Teyssier}}]{Ocvirk08}
{Ocvirk} P., {Pichon} C., {Teyssier} R., 2008, \mnras, 390, 1326

\bibitem[{{Oklopcic} {et~al}\mbox{.}(2016){Oklopcic}, {Hopkins}, {Feldmann},
  {Keres}, {Faucher-Giguere}, \& {Murray}}]{Oklopcic16}
{Oklopcic} A., {Hopkins} P.~F., {Feldmann} R., {Keres} D., {Faucher-Giguere}
  C.-A., {Murray} N., 2016, ArXiv e-prints

\bibitem[{{Padoan}, {Nordlund} \& {Jones}(1997){Padoan}, {Nordlund}, \&
  {Jones}}]{Padoan97}
{Padoan} P., {Nordlund} A., {Jones} B.~J.~T., 1997, \mnras, 288, 145

\bibitem[{{Press} \& {Schechter}(1974)}]{Press74}
{Press} W.~H., {Schechter} P., 1974, \apj, 187, 425

\bibitem[{{Price}, {Federrath} \& {Brunt}(2011){Price}, {Federrath}, \&
  {Brunt}}]{Price11}
{Price} D.~J., {Federrath} C., {Brunt} C.~M., 2011, \apjl, 727, L21

\bibitem[{{Rosdahl} {et~al}\mbox{.}(2015){Rosdahl}, {Schaye}, {Teyssier}, \&
  {Agertz}}]{Roshdal15}
{Rosdahl} J., {Schaye} J., {Teyssier} R., {Agertz} O., 2015, \mnras, 451, 34

\bibitem[{{Rosolowsky}(2005)}]{Rosolowski05}
{Rosolowsky} E., 2005, \pasp, 117, 1403

\bibitem[{{Saintonge} {et~al}\mbox{.}(2011){Saintonge}, {Kauffmann}, {Kramer},
  {Tacconi}, {Buchbender}, {Catinella}, {Fabello}, {Graci{\'a}-Carpio}, {Wang},
  {Cortese}, {Fu}, {Genzel}, {Giovanelli}, {Guo}, {Haynes}, {Heckman},
  {Krumholz}, {Lemonias}, {Li}, {Moran}, {Rodriguez-Fernandez}, {Schiminovich},
  {Schuster}, \& {Sievers}}]{Saintonge11}
{Saintonge} A. {et~al.}, 2011, \mnras, 415, 32

\bibitem[{{Scalo} {et~al}\mbox{.}(1998){Scalo}, {V{\'a}zquez-Semadeni},
  {Chappell}, \& {Passot}}]{Scalo98}
{Scalo} J., {V{\'a}zquez-Semadeni} E., {Chappell} D., {Passot} T., 1998, \apj,
  504, 835

\bibitem[{{Shapiro}, {Genzel} \& {F{\"o}rster Schreiber}(2010){Shapiro},
  {Genzel}, \& {F{\"o}rster Schreiber}}]{Shapiro10}
{Shapiro} K.~L., {Genzel} R., {F{\"o}rster Schreiber} N.~M., 2010, \mnras, 403,
  L36

\bibitem[{{Shapiro} {et~al}\mbox{.}(2008){Shapiro}, {Genzel}, {F{\"o}rster
  Schreiber}, {Tacconi}, {Bouch{\'e}}, {Cresci}, {Davies}, {Eisenhauer},
  {Johansson}, {Krajnovi{\'c}}, {Lutz}, {Naab}, {Arimoto}, {Arribas}, \& {et
  al.,}}]{Shapiro08}
{Shapiro} K.~L. {et~al.}, 2008, \apj, 682, 231

\bibitem[{{Shibuya} {et~al}\mbox{.}(2016){Shibuya}, {Ouchi}, {Kubo}, \&
  {Harikane}}]{Shibuya16}
{Shibuya} T., {Ouchi} M., {Kubo} M., {Harikane} Y., 2016, \apj, 821, 72

\bibitem[{{Snyder} {et~al}\mbox{.}(2015){Snyder}, {Lotz}, {Moody}, {Peth},
  {Freeman}, {Ceverino}, {Primack}, \& {Dekel}}]{Snyder15}
{Snyder} G.~F., {Lotz} J., {Moody} C., {Peth} M., {Freeman} P., {Ceverino} D.,
  {Primack} J., {Dekel} A., 2015, \mnras, 451, 4290

\bibitem[{{Stark} {et~al}\mbox{.}(2008){Stark}, {Swinbank}, {Ellis}, {Dye},
  {Smail}, \& {Richard}}]{Stark08}
{Stark} D.~P., {Swinbank} A.~M., {Ellis} R.~S., {Dye} S., {Smail} I.~R.,
  {Richard} J., 2008, \nat, 455, 775

\bibitem[{{Tacchella} {et~al}\mbox{.}(2016{\natexlab{a}}){Tacchella}, {Dekel},
  {Carollo}, {Ceverino}, {DeGraf}, {Lapiner}, {Mandelker}, \&
  {Primack}}]{Tacchella16a}
{Tacchella} S., {Dekel} A., {Carollo} C.~M., {Ceverino} D., {DeGraf} C.,
  {Lapiner} S., {Mandelker} N., {Primack} J.~R., 2016{\natexlab{a}}, \mnras,
  458, 242

\bibitem[{{Tacchella} {et~al}\mbox{.}(2016{\natexlab{b}}){Tacchella}, {Dekel},
  {Carollo}, {Ceverino}, {DeGraf}, {Lapiner}, {Mandelker}, \& {Primack
  Joel}}]{Tacchella16b}
{Tacchella} S., {Dekel} A., {Carollo} C.~M., {Ceverino} D., {DeGraf} C.,
  {Lapiner} S., {Mandelker} N., {Primack Joel} R., 2016{\natexlab{b}}, \mnras,
  457, 2790

\bibitem[{{Tacconi} {et~al}\mbox{.}(2010){Tacconi}, {Genzel}, {Neri}, {Cox},
  {Cooper}, {Shapiro}, {Bolatto}, \& {et al.,}}]{Tacconi10}
{Tacconi} L.~J., {Genzel} R., {Neri} R., {Cox} P., {Cooper} M.~C., {Shapiro}
  K., {Bolatto} A., {et al.,}, 2010, \nat, 463, 781

\bibitem[{{Tacconi} {et~al}\mbox{.}(2013){Tacconi}, {Neri}, {Genzel}, {Combes},
  {Bolatto}, {Cooper}, {Wuyts}, {Bournaud}, {Burkert}, {Comerford}, {Cox},
  {Davis}, {F{\"o}rster Schreiber}, {Garc{\'{\i}}a-Burillo}, {Gracia-Carpio},
  {Lutz}, {Naab}, {Newman}, {Omont}, {Saintonge}, {Shapiro Griffin}, {Shapley},
  {Sternberg}, \& {Weiner}}]{Tacconi13}
{Tacconi} L.~J. {et~al.}, 2013, \apj, 768, 74

\bibitem[{{Tamburello} {et~al}\mbox{.}(2015){Tamburello}, {Mayer}, {Shen}, \&
  {Wadsley}}]{Tamburello15}
{Tamburello} V., {Mayer} L., {Shen} S., {Wadsley} J., 2015, \mnras, 453, 2490

\bibitem[{{Tomassetti} {et~al}\mbox{.}(2016){Tomassetti}, {Dekel}, {Mandelker},
  {Ceverino}, {Lapiner}, {Faber}, {Kneller}, {Primack}, \&
  {Sai}}]{Tomassetti16}
{Tomassetti} M. {et~al.}, 2016, \mnras, 458, 4477

\bibitem[{{Toomre}(1964)}]{toomre64}
{Toomre} A., 1964, \apj, 139, 1217

\bibitem[{{Tweed} {et~al}\mbox{.}(2009){Tweed}, {Devriendt}, {Blaizot},
  {Colombi}, \& {Slyz}}]{Tweed09}
{Tweed} D., {Devriendt} J., {Blaizot} J., {Colombi} S., {Slyz} A., 2009, \aap,
  506, 647

\bibitem[{{Vazquez-Semadeni}(1994)}]{VS94}
{Vazquez-Semadeni} E., 1994, \apj, 423, 681

\bibitem[{{V{\'a}zquez-Semadeni} {et~al}\mbox{.}(2008){V{\'a}zquez-Semadeni},
  {Gonz{\'a}lez}, {Ballesteros-Paredes}, {Gazol}, \& {Kim}}]{VS08}
{V{\'a}zquez-Semadeni} E., {Gonz{\'a}lez} R.~F., {Ballesteros-Paredes} J.,
  {Gazol} A., {Kim} J., 2008, \mnras, 390, 769

\bibitem[{{Williams} \& {McKee}(1997)}]{Williams97}
{Williams} J.~P., {McKee} C.~F., 1997, \apj, 476, 166

\bibitem[{{Wisnioski} {et~al}\mbox{.}(2015){Wisnioski}, {F{\"o}rster
  Schreiber}, {Wuyts}, {Wuyts}, {Bandara}, {Wilman}, {Genzel}, {Bender},
  {Davies}, {Fossati}, {Lang}, {Mendel}, {Beifiori}, {Brammer}, {Chan},
  {Fabricius}, {Fudamoto}, {Kulkarni}, {Kurk}, {Lutz}, {Nelson}, {Momcheva},
  {Rosario}, {Saglia}, {Seitz}, {Tacconi}, \& {van Dokkum}}]{wisnioski15}
{Wisnioski} E. {et~al.}, 2015, \apj, 799, 209

\bibitem[{{Wuyts} {et~al}\mbox{.}(2012){Wuyts}, {F{\"o}rster Schreiber},
  {Genzel}, {Guo}, {Barro}, {Bell}, {Dekel}, {Faber}, {Ferguson}, {Giavalisco},
  {Grogin}, {Hathi}, {Huang}, {Kocevski}, {Koekemoer}, {Koo}, \& {et
  al.,}}]{Wuyts12}
{Wuyts} S. {et~al.}, 2012, \apj, 753, 114

\bibitem[{{Wuyts} {et~al}\mbox{.}(2013){Wuyts}, {F{\"o}rster Schreiber},
  {Nelson}, {van Dokkum}, {Brammer}, {Chang}, {Faber}, {Ferguson}, {Franx},
  {Fumagalli}, {Genzel}, {Grogin}, {Kocevski}, {Koekemoer}, {Lundgren}, {Lutz},
  {McGrath}, {Momcheva}, {Rosario}, {Skelton}, {Tacconi}, {van der Wel}, \&
  {Whitaker}}]{Wuyts13}
{Wuyts} S. {et~al.}, 2013, \apj, 779, 135

\bibitem[{{Zanella} {et~al}\mbox{.}(2015){Zanella}, {Daddi}, {Le Floc'h},
  {Bournaud}, {Gobat}, {Valentino}, {Strazzullo}, {Cibinel}, {Onodera},
  {Perret}, {Renaud}, \& {Vignali}}]{Zanella15}
{Zanella} A. {et~al.}, 2015, \nat, 521, 54

\bibitem[{{Zolotov} {et~al}\mbox{.}(2015){Zolotov}, {Dekel}, {Mandelker},
  {Tweed}, {Inoue}, {DeGraf}, {Ceverino}, {Primack}, {Barro}, \&
  {Faber}}]{Zolotov15}
{Zolotov} A. {et~al.}, 2015, \mnras, 450, 2327

\end{thebibliography}

\appendix 
 %%%%%%%%%%%%%%%%%%%%%%%%%%%%%%%%%%%%%%%%%%%%%%%%  
\section{Parameter Dependence}
\label{sec:params} 
% Dependence on parameters
\smallskip
The three main parameters in the clump finding algorithm are 
the resolution of the uniform grid, $\Delta=70\pc$, the width 
of the Gaussian used to smooth the density field, 
$\Fw={\rm min}(2.5\kpc, 0.5\Rd)$, and the residual threshold 
to assign a cell to a clump $\delmin=10$. In \citet{M14} we 
used a similar algorithm with $\Fw=2\kpc$, $\delmin=15$ and 
$\Delta=70\pc$, together with an additional narrow Gaussian 
having a FWHM of $\Fn=140\pc$, the residual defined as 
$\delrho = (\rho_{\rm N}-\rho_{\rm W})/\rho_{\rm W}$. 
We find both the narrow smoothing and $\Delta$ to have no 
noticeable effect on any but the smallest clumps.

\smallskip
We examine the sensitivity of our clump finder to variations 
in $\Fw$ and $\delmin$ using one of our clumpiest RP simulations, 
V13, shown face on at redshift $z=1.5$ in \fig{clump_finder}. 
We experimented with values of $\Fw=2.0, 2.5\kpc$ and $\delmin=5,10,15$. 
Since using $\delmin=5$ results in many small, low-mass, low-density 
clumps, and in any case we are only concerned with the effect on 
larger clumps, we only considered for the purposes of this test 
clumps with at least 27 uni-grid cells, rather than the fiducial 
value of 8 used throughout the paper. We examined all in situ clumps 
found in this way in the redshift range $1.3\lsim z\lsim 5$.

\begin{figure*}
\centering
\subfloat{\includegraphics[width =0.93 \textwidth]{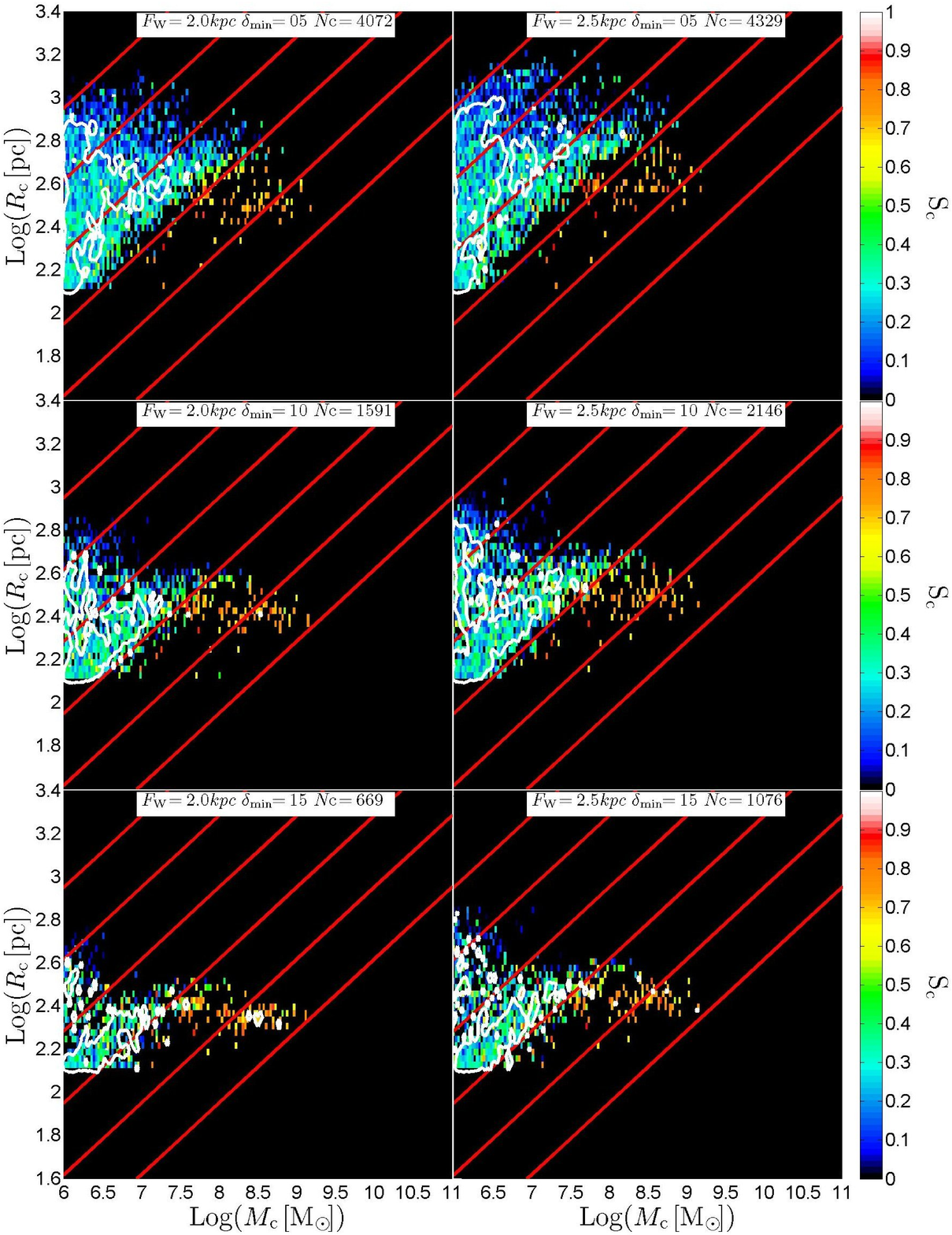}}
\vspace{-3mm}
\caption{Dependence of clump masses, sizes and shapes, on the parameters 
of the clump finder. We show the distribution of in situ clumps identified 
in the RP version of V13 at $1.3\lsim z\lsim 5$, in the mass-radius plane. 
The color is the median clump shape, $\Sc$, per pixel. The white contours 
enclose the densest pixels in terms of number of clumps per pixel, that 
contain $50\%$ of the clumps. The red lines are lines of constant density, 
${\rm log}(\nc \cdot \cmc)=-2,-1,0,1,2,3$ from top to bottom. The different 
panels are for different combinations of the parameters $\Fw$ and $\delmin$, 
as written at the top of each panel. The left (right) column refers to 
$\Fw=2.0\kpc\:(2.5\kpc)$ while from top to bottom we have $\delmin=5,10,15$. 
Each panel also lists the total number of identified clumps. Lowering $\delmin$ 
or raising $\Fw$ makes clumps larger and less dense by adding additional background 
material to the clumps, and it increases the number of low-mass, low-density clumps. 
As long as $\delmin \gsim 10$, the population of massive, dense, bound clumps is 
not terribly sensitive to changes in the parameters. However, $\delmin=5$ embeds 
these clumps in large background features, making them appear much larger and less 
separated from the bulk of the population. 
}
\label{fig:M_R_params} 
\end{figure*}

\smallskip
\Fig{M_R_params} shows the distribution of clumps in the mass-size 
plane, for each of the six combinations of $\Fw$ and $\delmin$ 
examined. The color represents the median value within each pixel 
of the clump shape, $\Sc$, the red lines represent constant baryonic 
density, ranging from $\nc=0.01-1000\cmc$ in factors of 10, and the 
contour in each panel represents the number density of clumps per 
pixel, enclosing 50\% of the total clump population. In each panel 
we list the parameters used and the total number of in situ clumps 
(including ZLCs). 

\smallskip
Focusing on the bottom two rows, we see that increasing $\Fw$ 
from $2.0$ to $2.5\kpc$, or decreasing $\delmin$ from $15$ to 
$10$, makes clumps larger and less dense as more background 
material is added to the clump. It also adds many low-mass, 
low-density clumps to the sample. However, in all cases, we 
learn from the contour that the bulk of the clump population 
follows a sequence of roughly constant density, $\nc \sim 1-10 \cmc$, 
and has prolate shapes with $\Sc \lsim 0.35$. In addition, we 
see a well defined population of dense, $\nc \sim 10-1000\cmc$, 
and oblate, $\Sc\gsim 0.6$ clumps, with radii $\Rc\sim 300-400\pc$ 
and masses $\Mc\gsim 10^8\msun$. These are the long-lived clumps, 
that have undergone gravitational collapse and remain bound. While 
raising $\Fw$ or lowering $\delmin$ makes these clumps more extended, 
they remain well separated from the bulk of the population and their 
mass does not vary by much.

\begin{figure*}
\centering
\subfloat{\includegraphics[width =0.45 \textwidth]{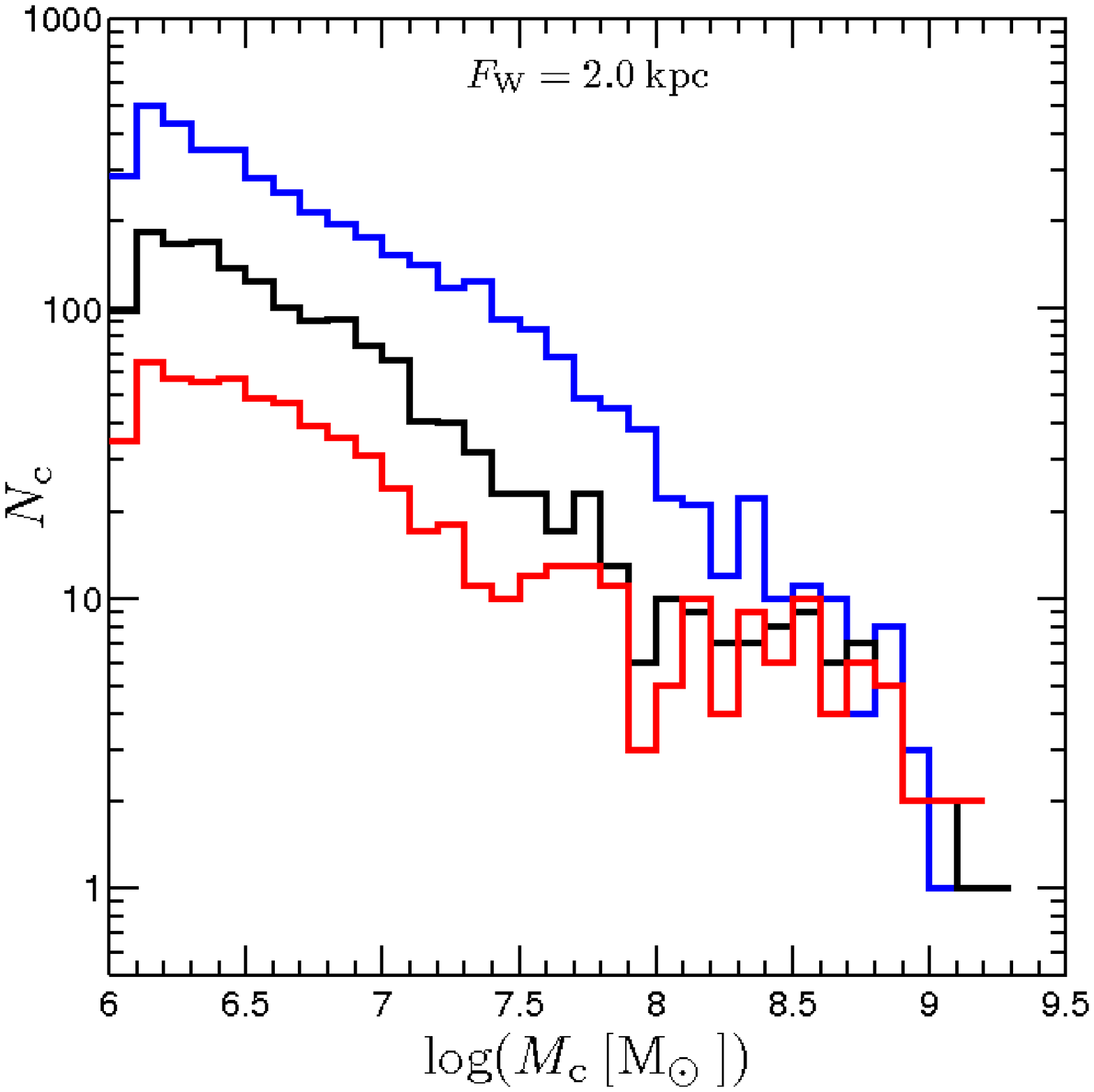}}
\subfloat{\includegraphics[trim={1.65cm 0 0 0}, clip, width =0.406 \textwidth]{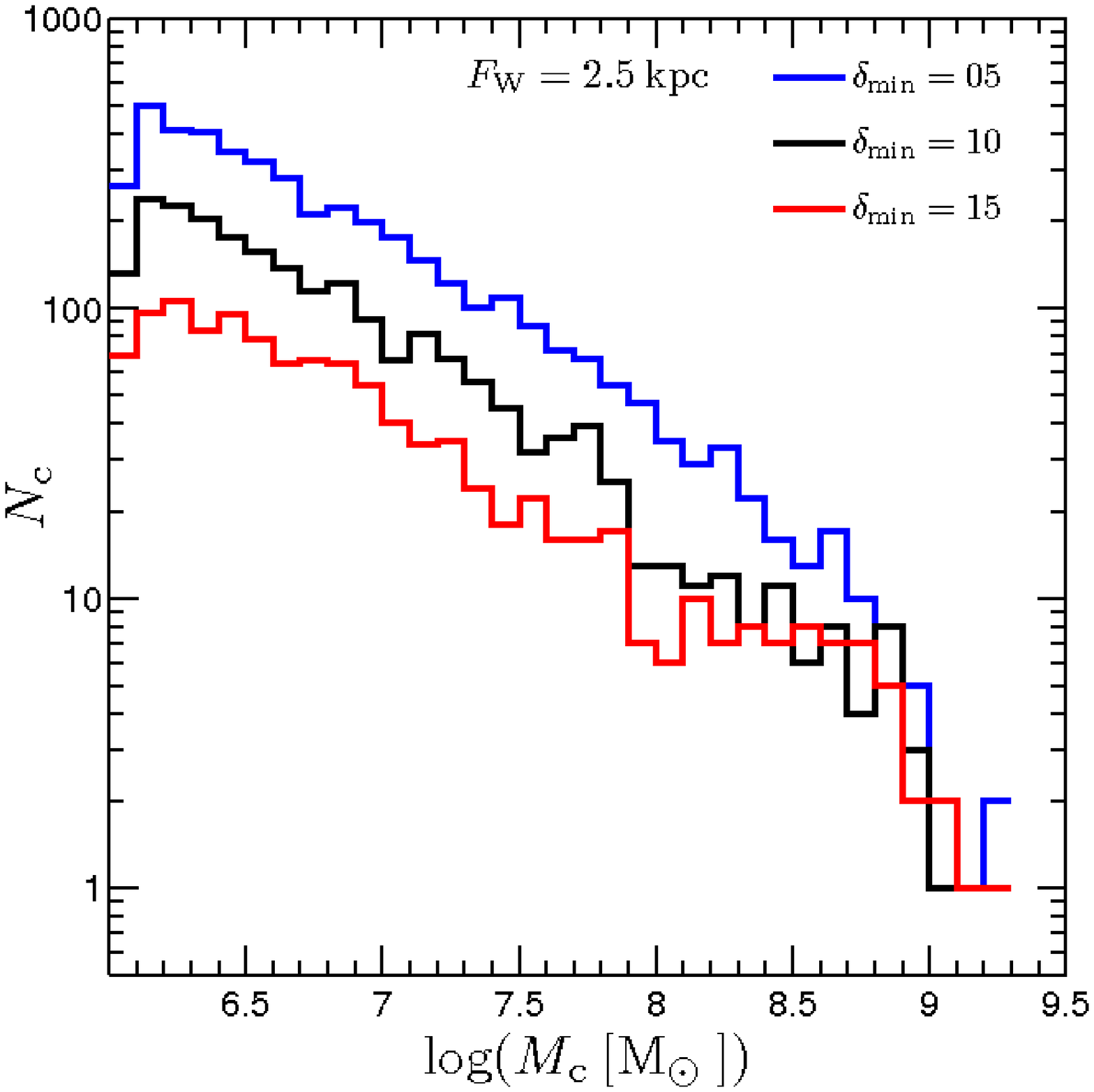}}
\caption{Mass function of in situ clumps in V13 for different combinations of 
parameters. The $X$ axis shows the clump baryonic mass and the $Y$ axis shows 
the total number of clumps per logartihmic mass bin. The left (right) panel 
represents $\Fw=2.0\:(2.5)\kpc$, and the blue, black and red lines represent 
$\delmin=5,10$, and $15$ respectively. The high-end of the mass function is 
insensitive to changes in the rang $\Fw=2-2.5\kpc$, $\delmin=10-15$, but a 
value of $\delmin=5$ changes both the slope and the normalization. At the 
low-mass end, decreasing $\delmin$ or increasing $\Fw$ results in more low-mass, 
low-density clumps, but does not change the slope of the mass function.
}
\label{fig:mass_function_params} 
\end{figure*} 

\smallskip
Using $\delmin=5$ reduces the median clump density to below $1\cmc$ 
and makes even the densest clumps much larger, significantly lowering 
their density. With such a low value for the residual threshold, most 
of these clumps become embedded in extended features of the ISM which 
are not bound. This blurs the distinction between the populations, and 
appears to be too low a value. This is to be expected, since most of the 
ISM volume is at densities $n <1 \cmc$, but gas only becomes self 
gravitating at densities $n>10\cmc$ (\fig{dens_PDF}).

\smallskip
In \fig{mass_function_params} we examine the mass functions for these 6 
parameter combinations. The massive end, $\Mc\gsim 10^8 \msun$, is virtually 
unaffected by increasing $\Fw$ from $2.0$ to $2.5\kpc$, and the effect on 
the low-mass end is less than a factor of $\sim 1.5$ in normalization, while 
the slope remains the same. Massive, oblate, bound clumps are unaffected by 
changing $\delmin$ in the rangs $10-15$. However, for $\delmin=5$ massive, 
compact clumps become embedded in extended, low density features that are 
not bound to the clump. This results in many more high mass, low-density 
clumps, which increase both the slope and the normalization of the mass 
function.

%\section{Mass Functions for Ex Situ and Bulge Clumps}
%\label{sec:Es_mass} 

\section{Selection Effects}
\label{sec:bias} 

\smallskip
Many observational studies of clumps \citep[e.g.][]{Guo12,Guo15,Wuyts12,Wuyts13,Shibuya16} 
target UV bright clumps. \citet{Guo15} is perhaps the most directly relatable study 
to our own. They observed a mass complete sample of several thousand SFGs in the 
CANDELS/GOODS-S and UDS fields, and used an automated clump finder very similar in 
spirit to the one used in this work, which detects local overdensities in the galaxy 
UV luminosity. They defined clumps as objects detedted by their algorithm (``blobs") 
containing at least $8\%$ of the total galaxy UV luminosity. In order to compare with 
their results, we must modify our clump selection to mimic the observational bias of 
UV bright clumps.

\smallskip
Naively, a clump containing $8\%$ of the galaxy UV light should contain $\sim 8\%$ 
of its SFR. However, clumps in their sample typically had lower dust extinction than 
the underlying discs, as was also pointed out by \citet{Wuyts13}. Moreover, comparison 
of the clumps found in this work to clumps found in mock-CANDELS images of our simulations 
reveals that for bright clumps, with $\gsim 5\%$ of the galaxy SFR, the contamination in 
the observed SFR from nearby sources is $\sim 30\%$, increasing a little towards higher 
redshift and brighter clumps (Guo et al, in preparation). This motivates selecting 
clumps with a fractional contribution to the galaxy SFR less than $8\%$. Comparing 
our clump SFR function (\fig{functions}) to the UV luminosity functions of ``blobs" 
for galaxies in a similar mass and redshift range (figures 7 and 8 from \citet{Guo15}), 
we estimate that a threshold of $s={\rm log}(SFR_{\rm c}/SFR_{\rm d})\simeq -1.5$ 
is a good approximation to the observational selection. In any case, our results are 
insensitive to changes of $\lsim 0.2~{\rm dex}$ in this threshold. This is much higher 
than the threshold of $s=-3$ that was used in our analysis in \se{grad} and \se{gal_clump} 
and it is worth checking how our results might change as a result of this selection. 

\smallskip
It turns out that most of our results remain unchanged. The only major change is in 
the mass gradient. Selecting UV bright clumps introduces a mass gradient for SLCs 
as well as for LLCs. The clumps with the highest SFR are typically the most massive 
clumps, so this result suggests that the maximal fragmentation mass increases closer  
to the disc centre, presumably because of higher velocity dispersions can lead to higher 
Jeans masses. While a mass complete sample of clumps, as was studied above, avoids 
this issue, a UV flux (or SFR) limited sample will suffer from it, making the mass 
gradient a poor indicator of the longevity of clumps. However, all our other results 
are qualitatively unchanged, and LLCs can still be distinguished from SLCs by their 
gas fractions, sSFRs and stellar ages.

\begin{figure}
\subfloat{\includegraphics[width =0.45 \textwidth]{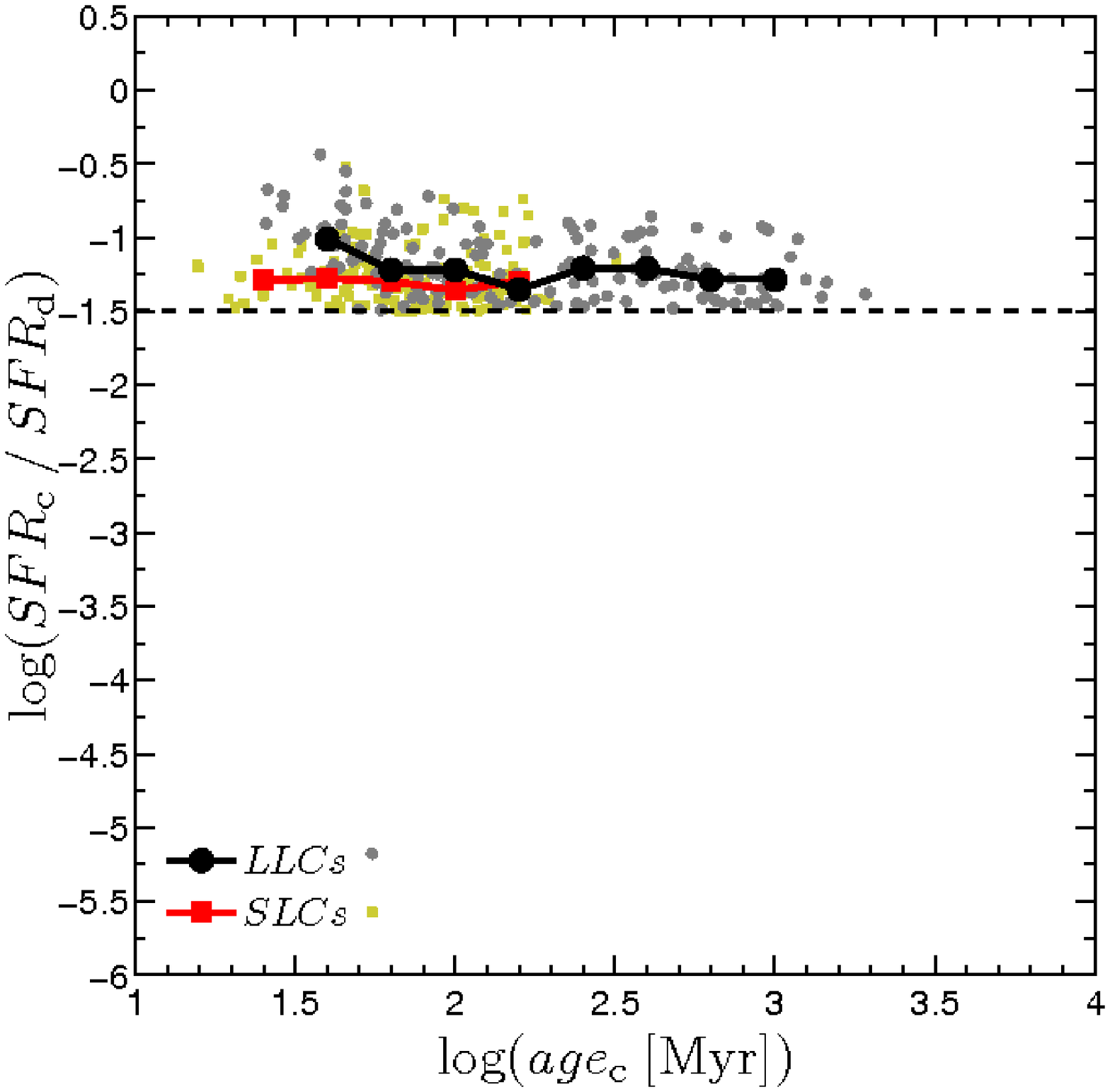}}
\caption{Same as \fig{age_sfr}, but using the ``observational", rather than the ``clean", 
sample of clumps.
}
\label{fig:age_sfr2}
\end{figure}

\begin{figure} 
\subfloat{\includegraphics[width =0.45 \textwidth]{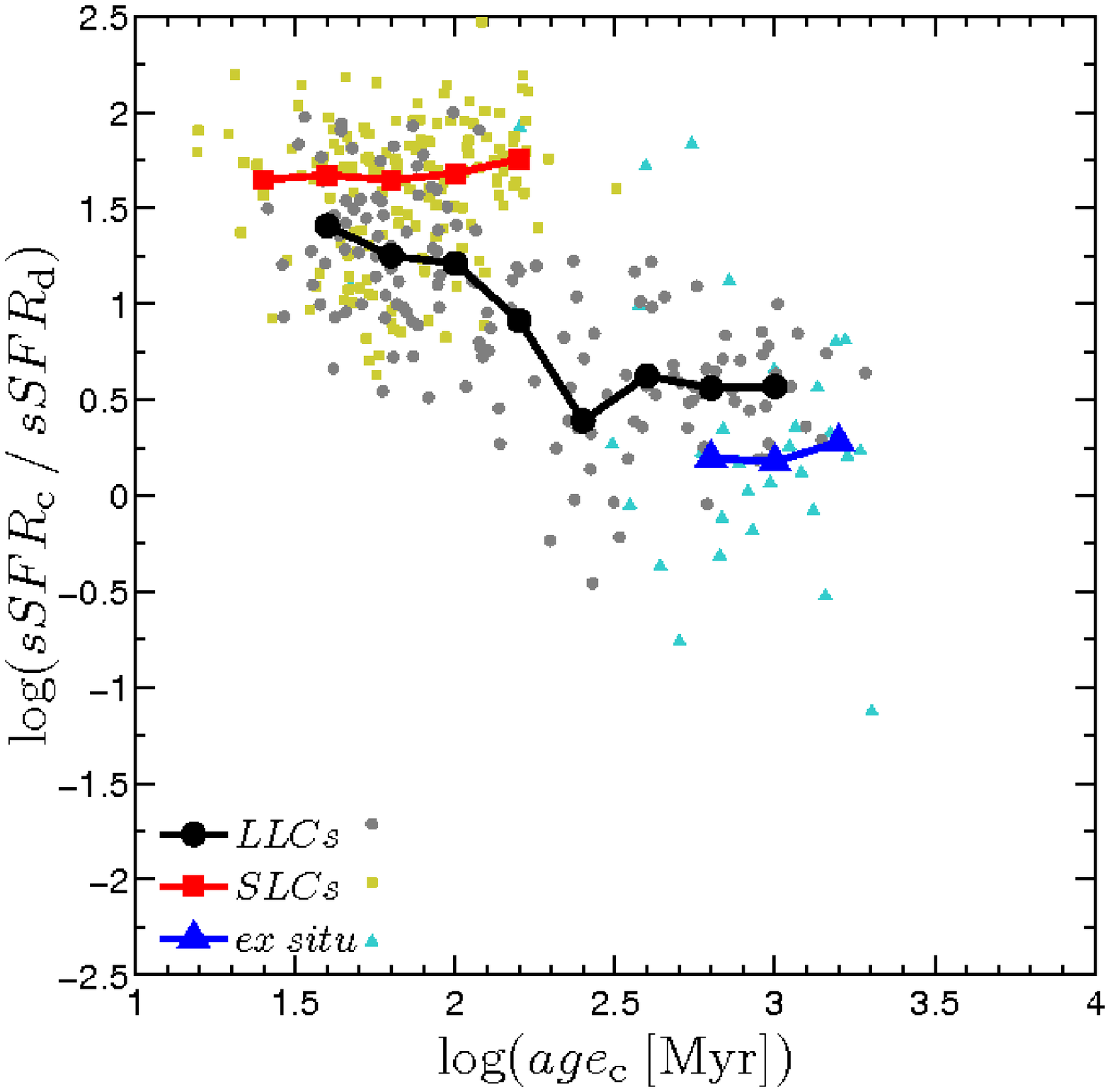}}
\caption{Same as \fig{age_ssfr}, but using the ``observational", rather than the ``clean", 
sample of clumps.
}
\label{fig:age_ssfr2} 
\end{figure}

\begin{figure} 
\subfloat{\includegraphics[width =0.45 \textwidth]{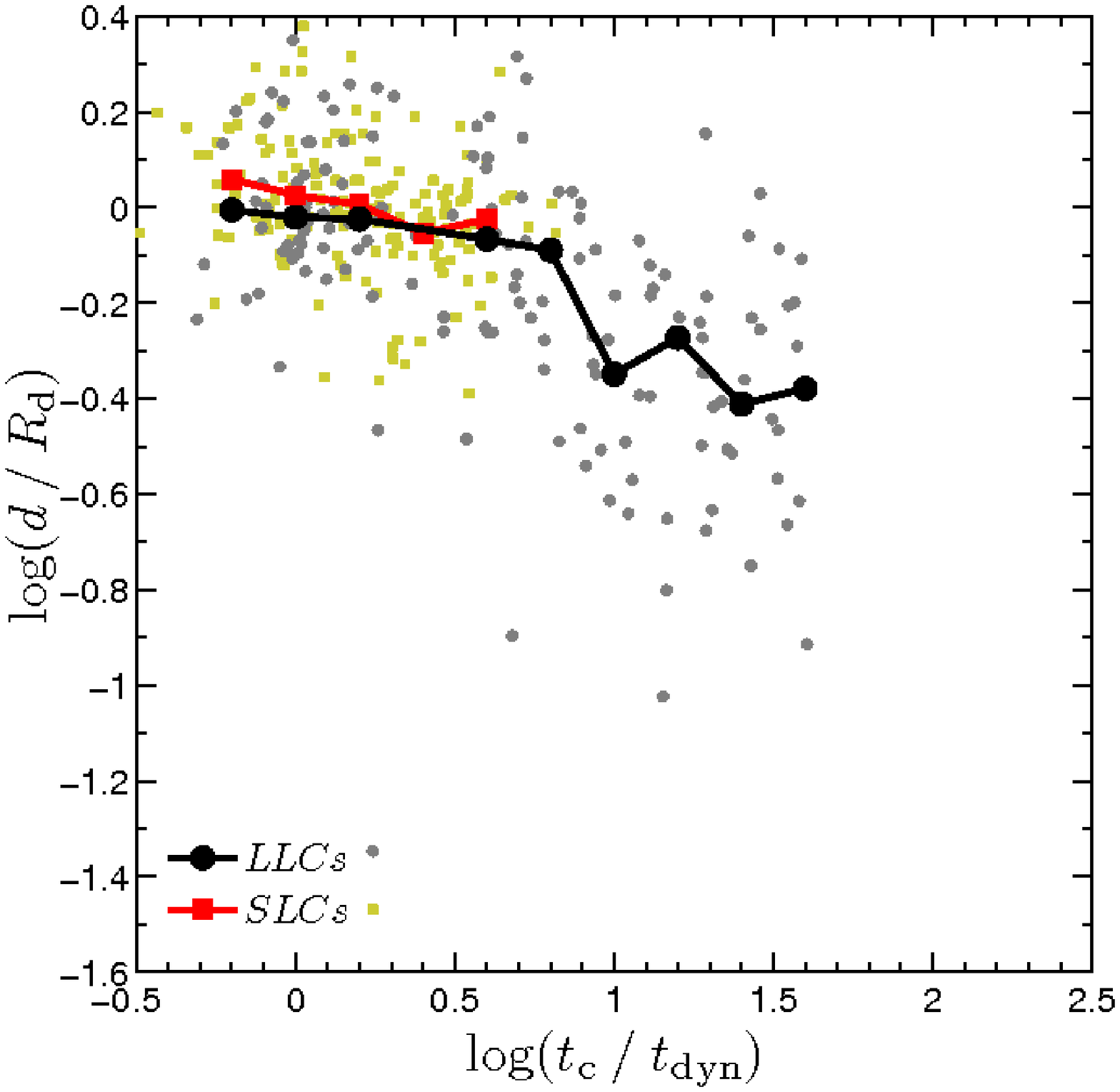}}
\caption{Same as \fig{age_time}, but using the ``observational", rather than the ``clean", 
sample of clumps.
}
\label{fig:age_time2} 
\end{figure}

\begin{figure*}
\centering
\subfloat{\includegraphics[trim={0 1.75cm 0 0}, clip, width =0.3 \textwidth]{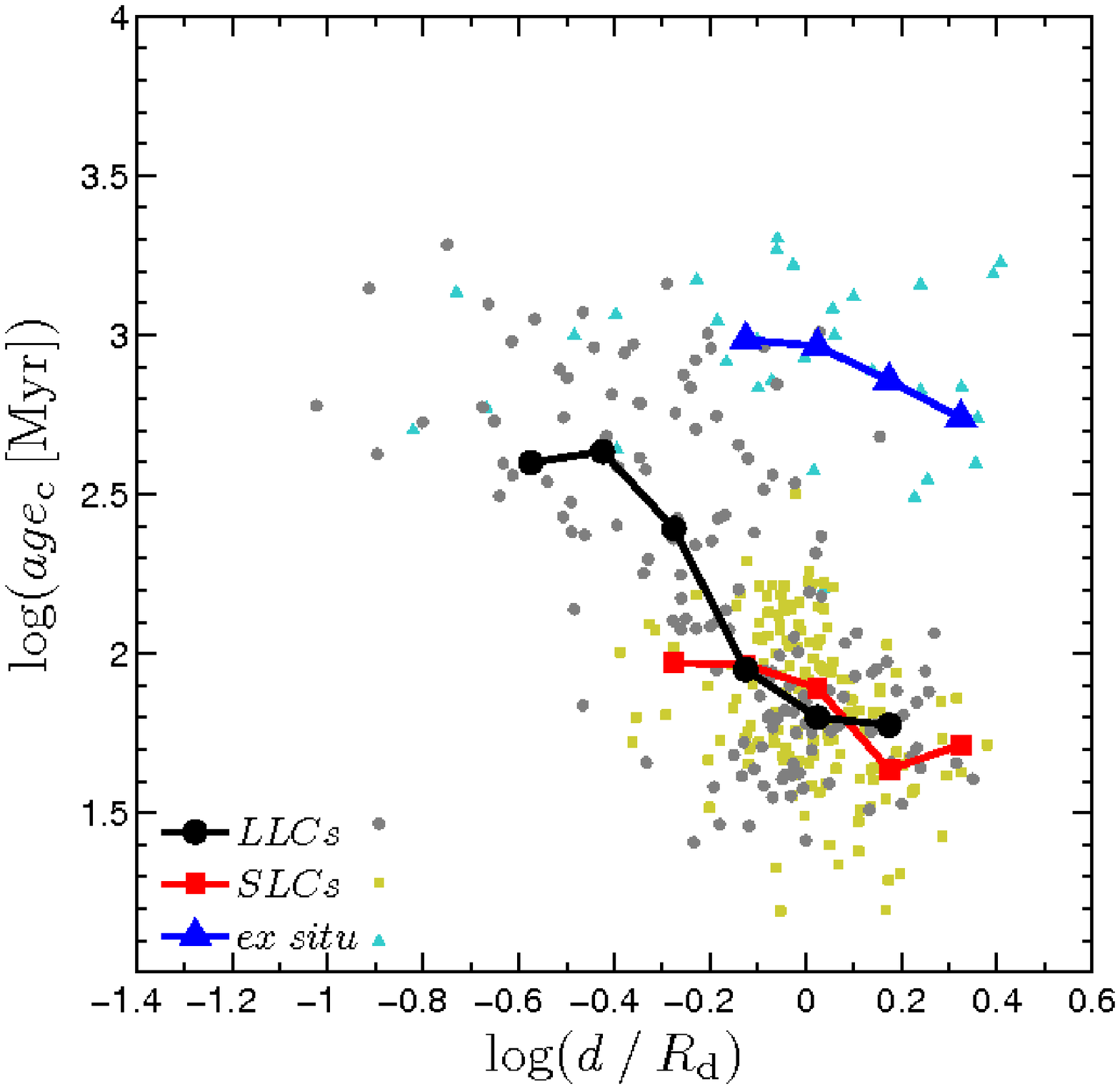}}
\subfloat{\includegraphics[trim={0 1.75cm 0 0}, clip, width =0.3 \textwidth]{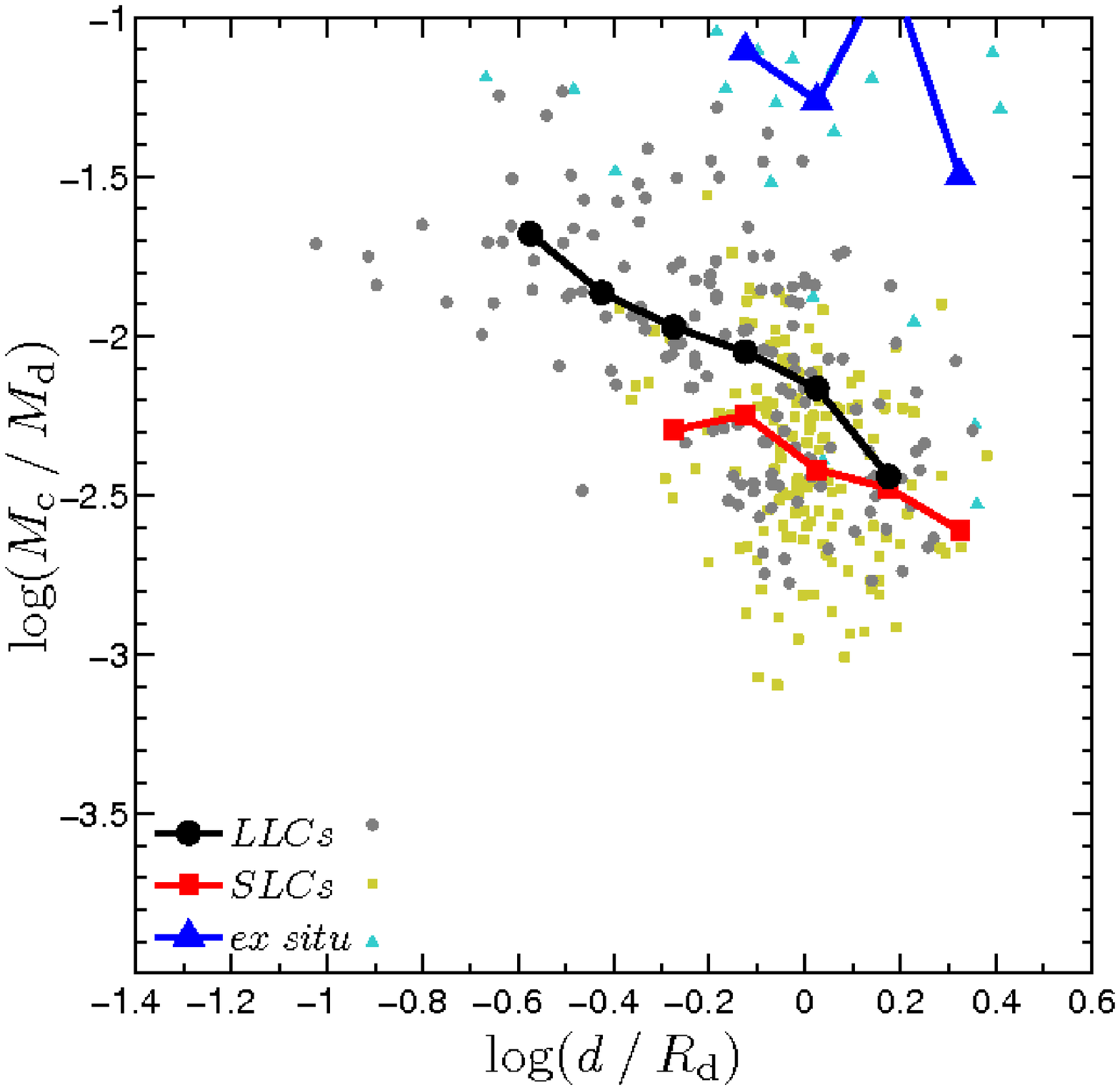}}
\subfloat{\includegraphics[trim={0 1.75cm 0 0}, clip, width =0.3 \textwidth]{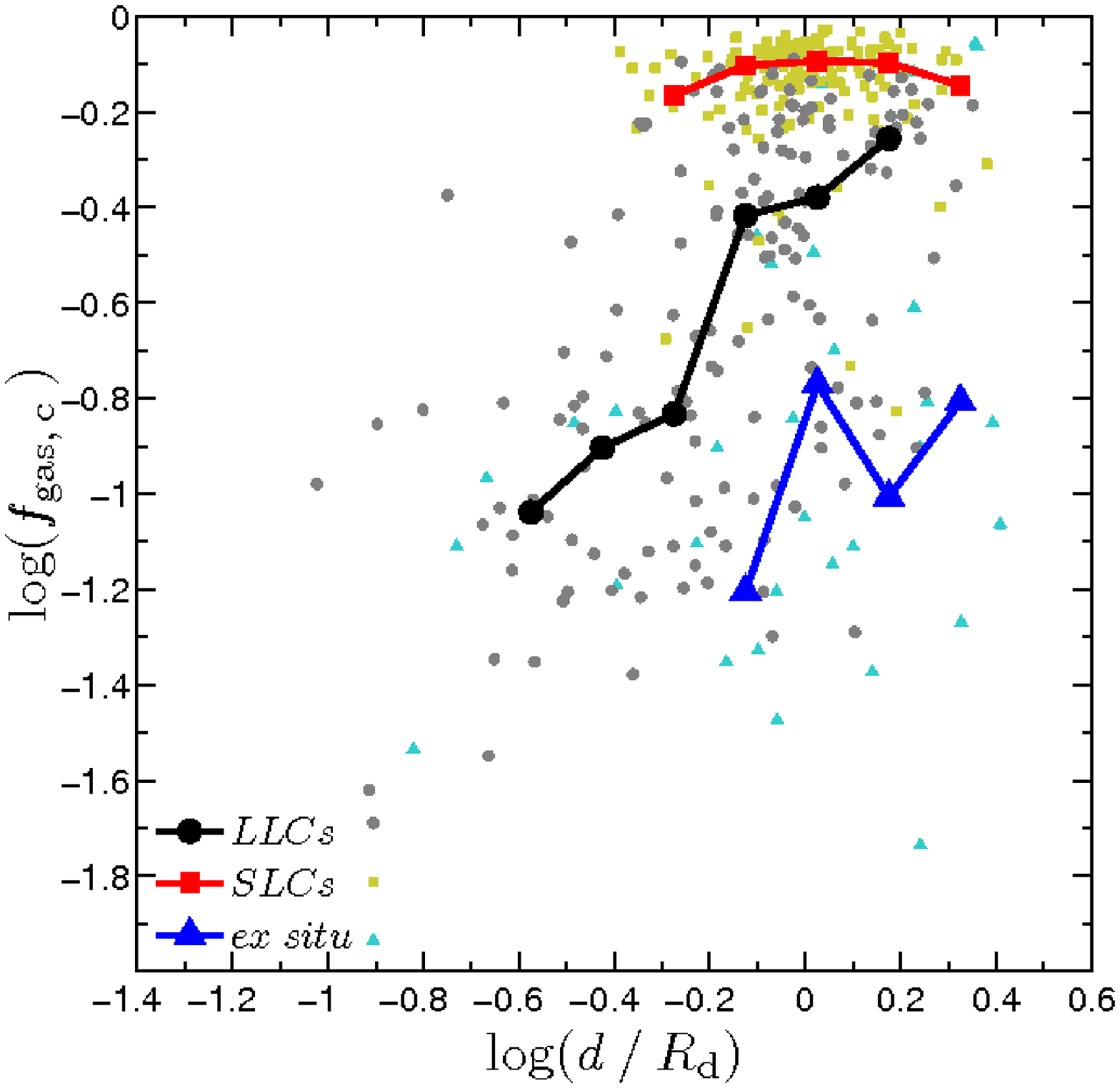}}\\
\vspace{-4.0mm}
\subfloat{\includegraphics[width =0.3 \textwidth]{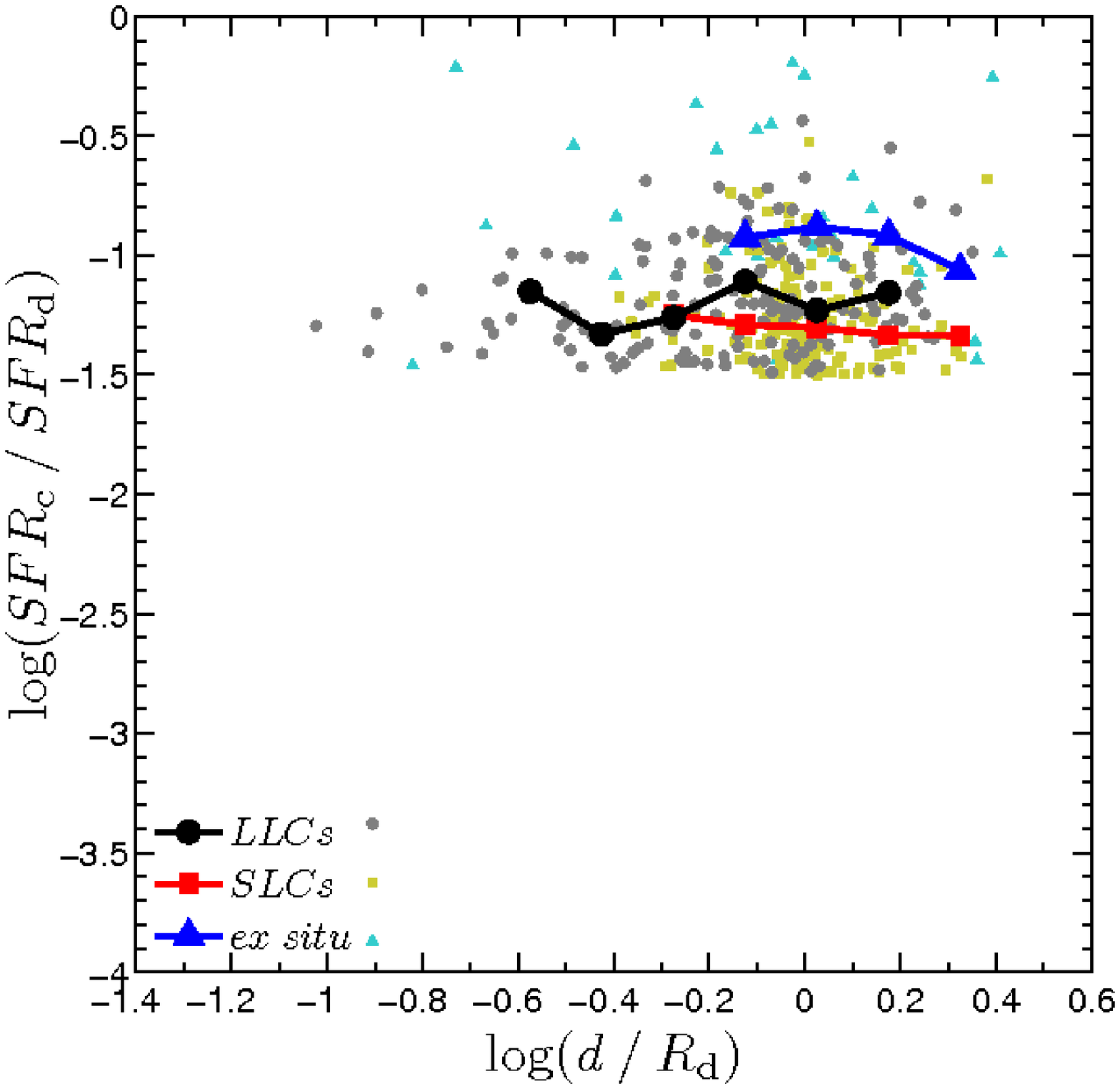}}
\subfloat{\includegraphics[width =0.3 \textwidth]{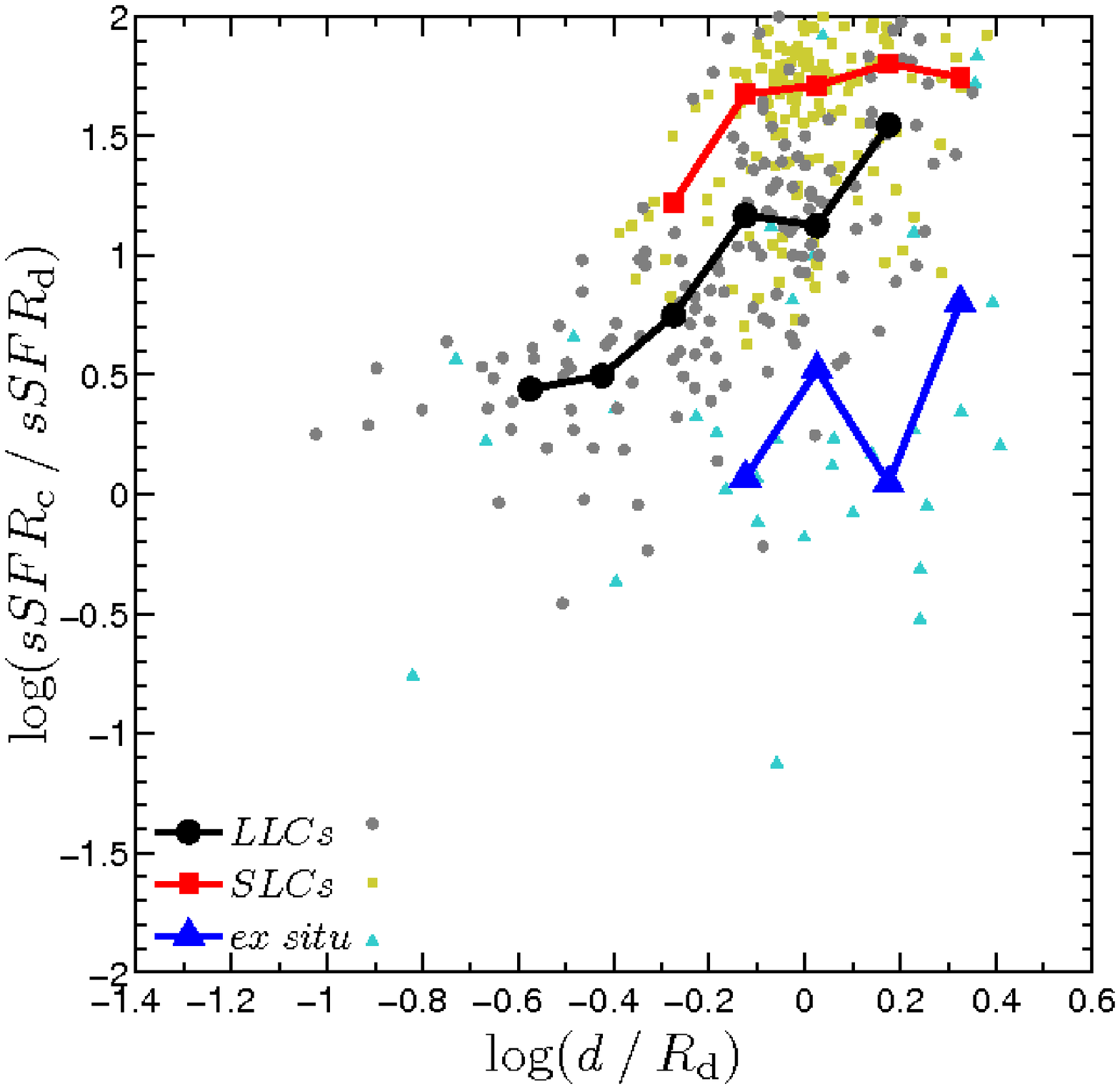}}
\subfloat{\includegraphics[width =0.3 \textwidth]{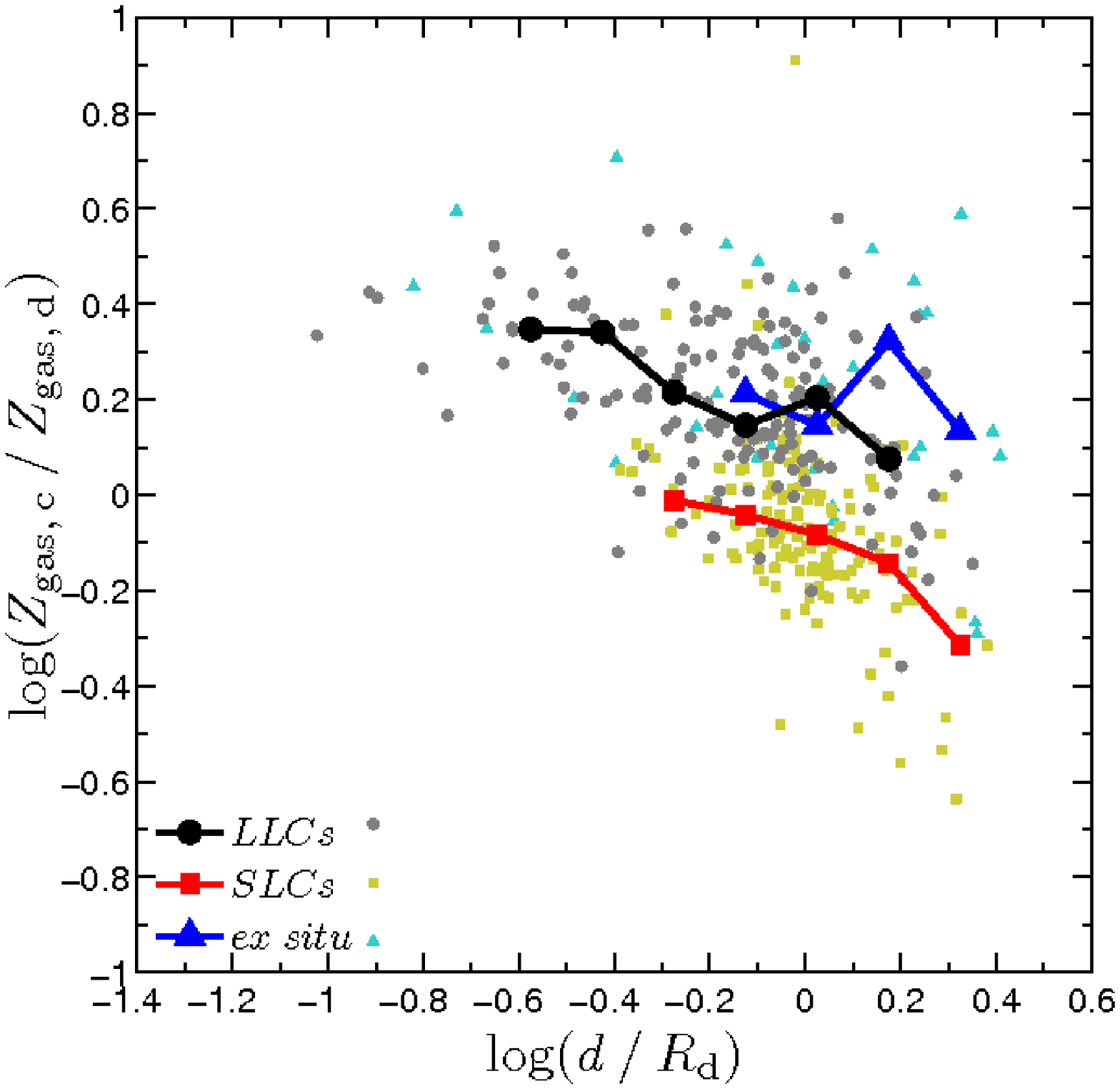}}\\
\caption{Same as \fig{gradients}, but using the ``observational", rather than the ``clean", 
sample of clumps.
}
\label{fig:gradients2} 
\end{figure*}

\begin{figure*}
\centering
\hspace{-1.4mm}
\subfloat{\includegraphics[trim={0 -0.45cm 0 0}, clip, width =0.3070 \textwidth]{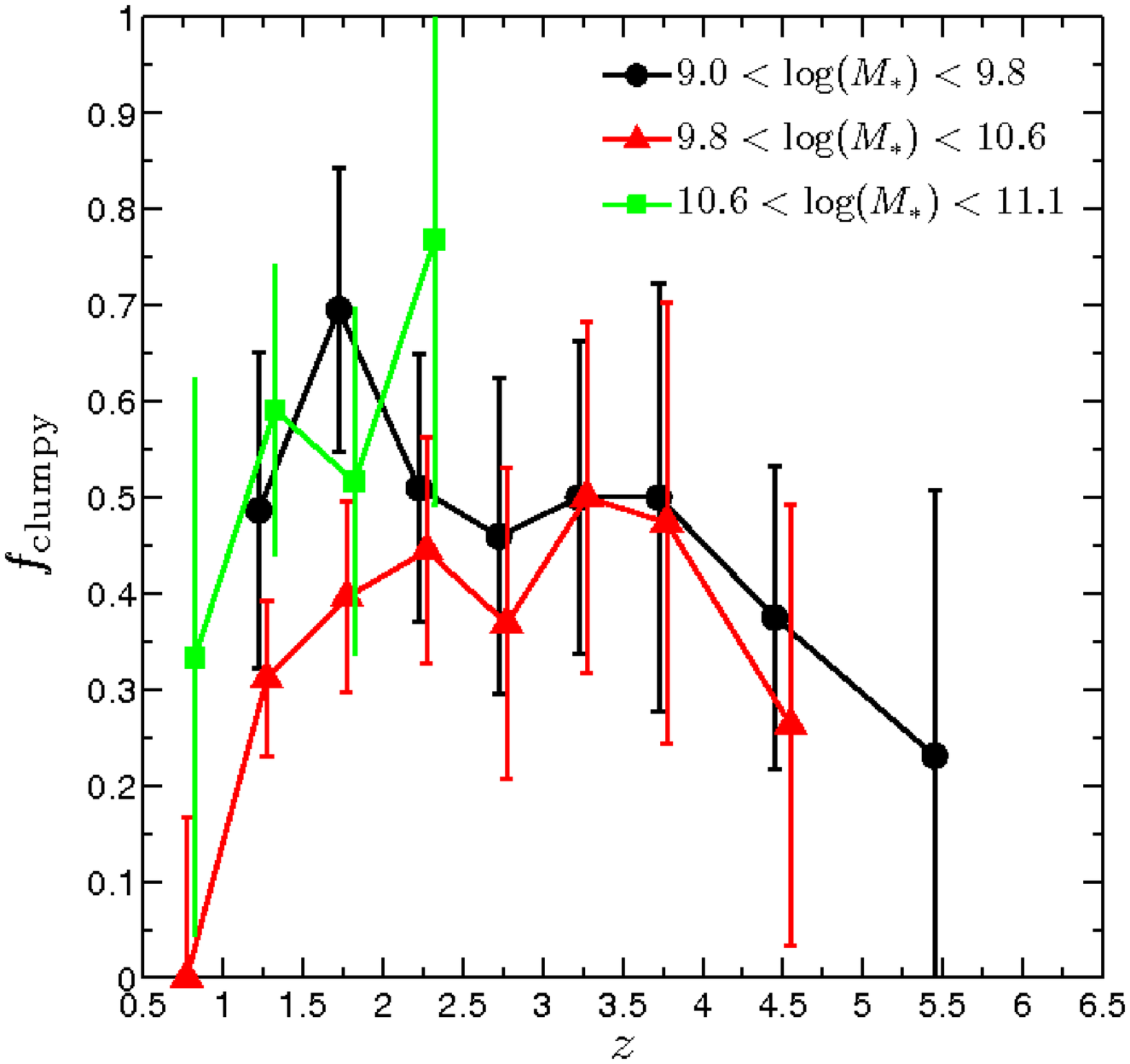}}
\subfloat{\includegraphics[width =0.3 \textwidth]{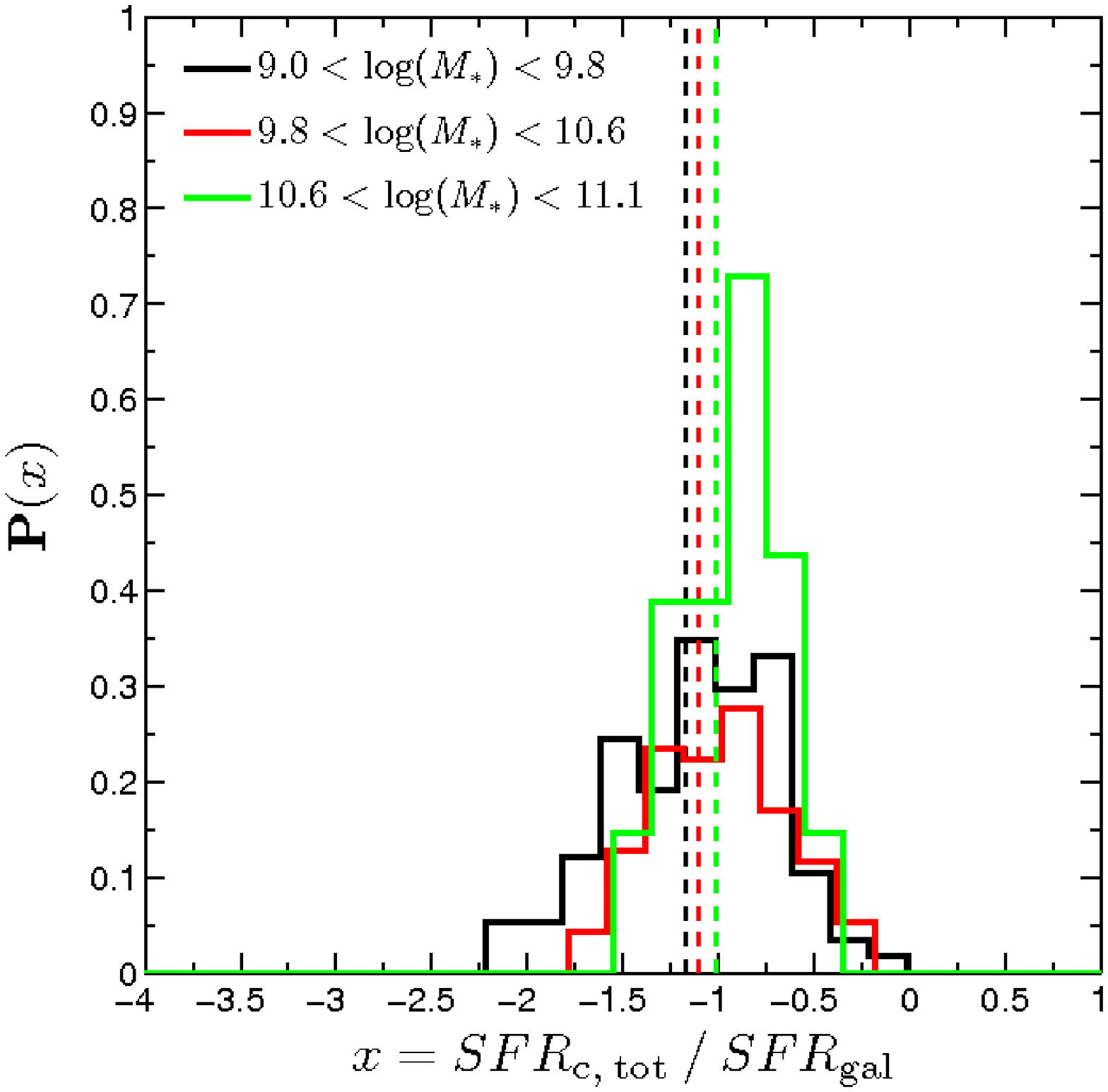}}
\subfloat{\includegraphics[width =0.3 \textwidth]{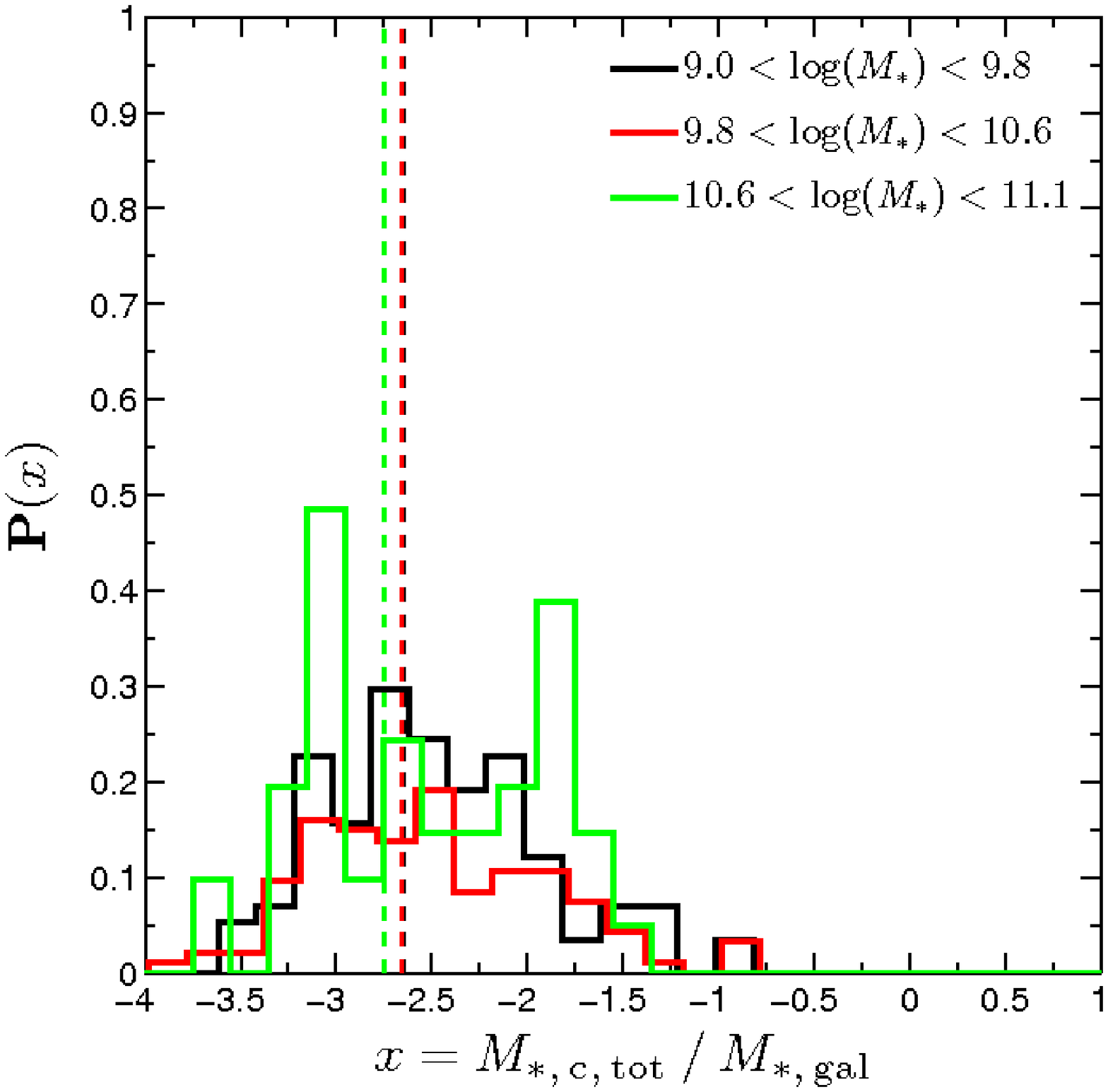}}\\
\vspace{-4.5mm}
\subfloat{\includegraphics[trim={0 -0cm 0 0}, clip, width =0.3070 \textwidth]{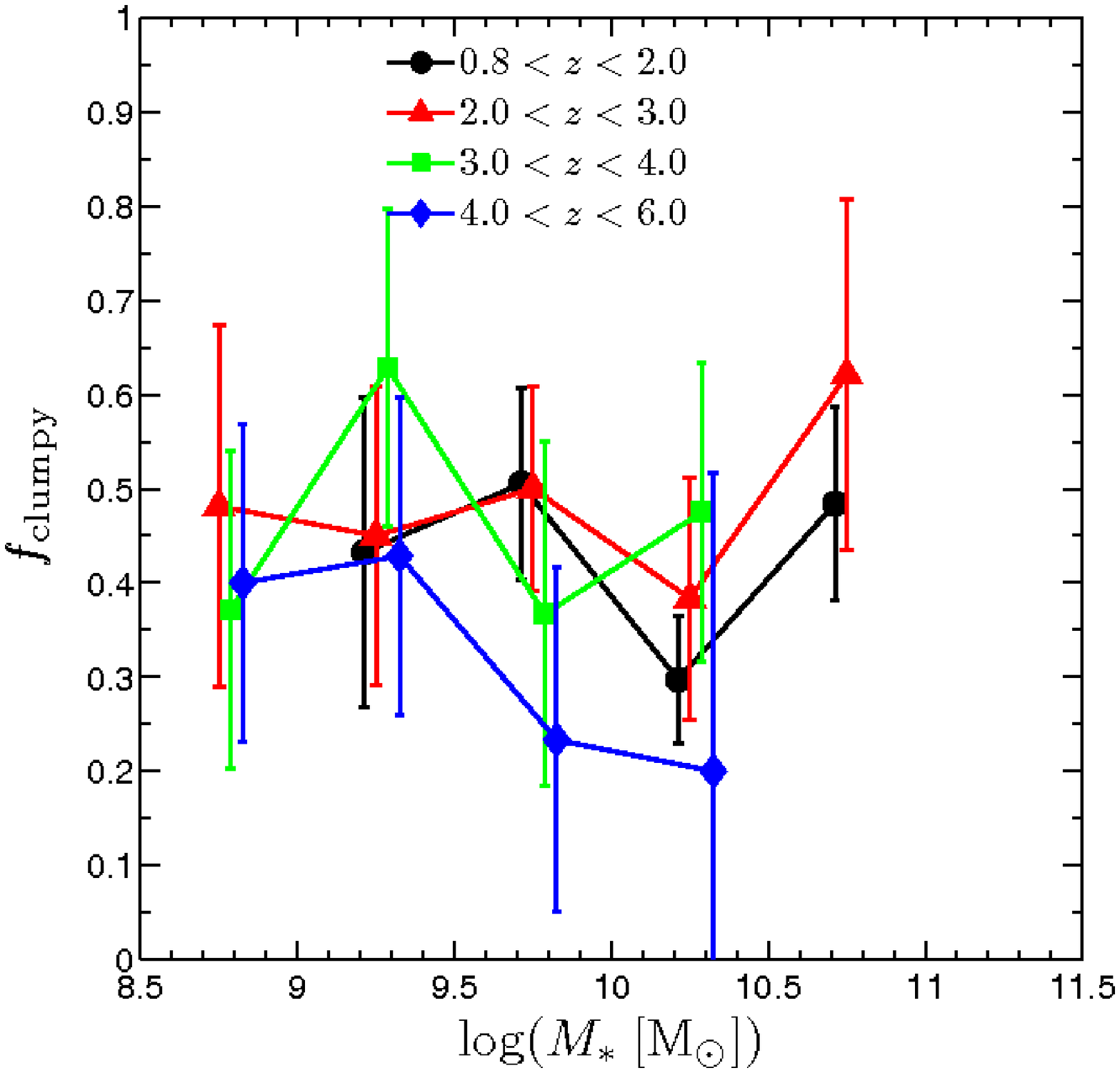}}
\subfloat{\includegraphics[width =0.3 \textwidth]{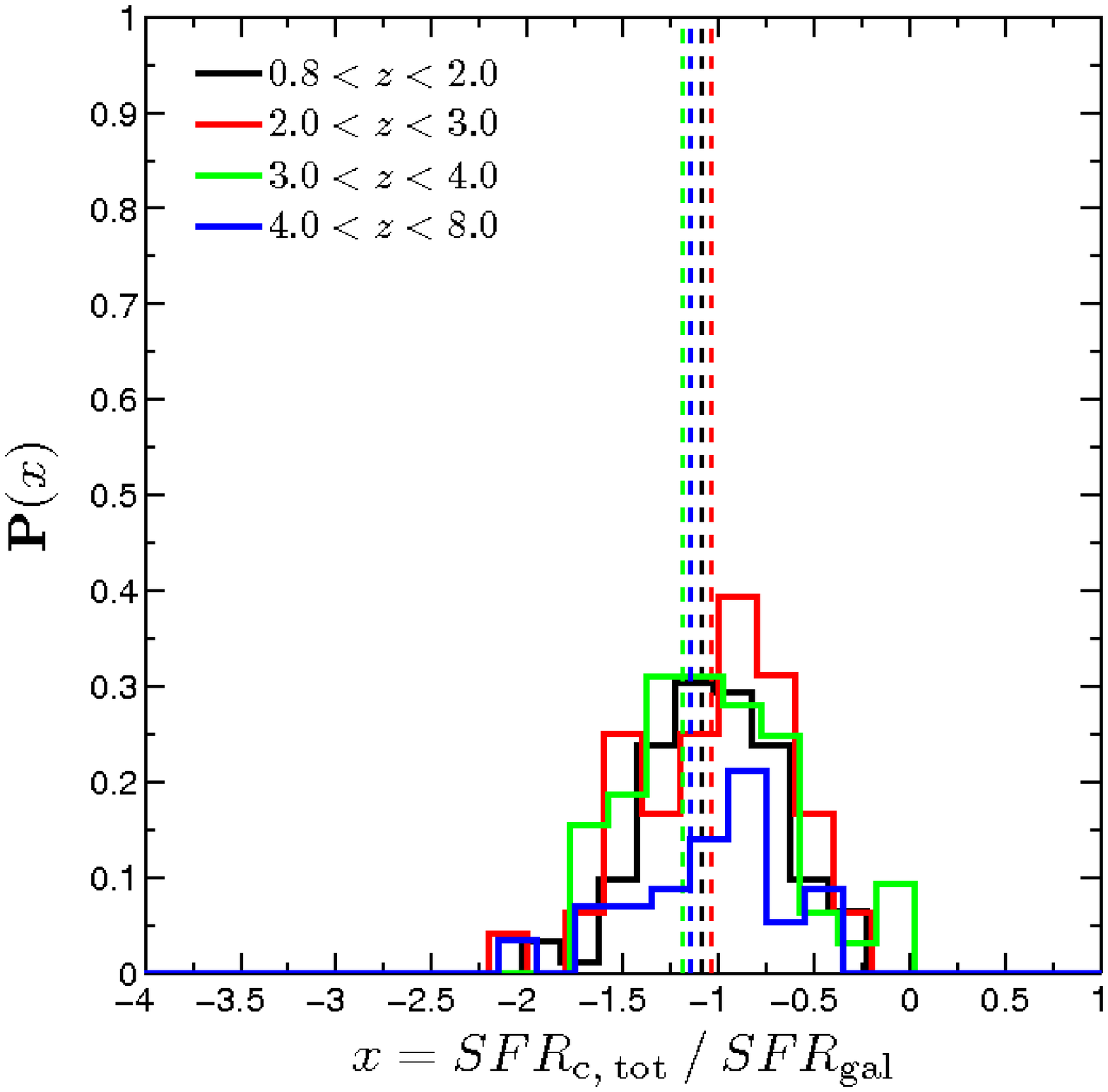}}
\subfloat{\includegraphics[width =0.3 \textwidth]{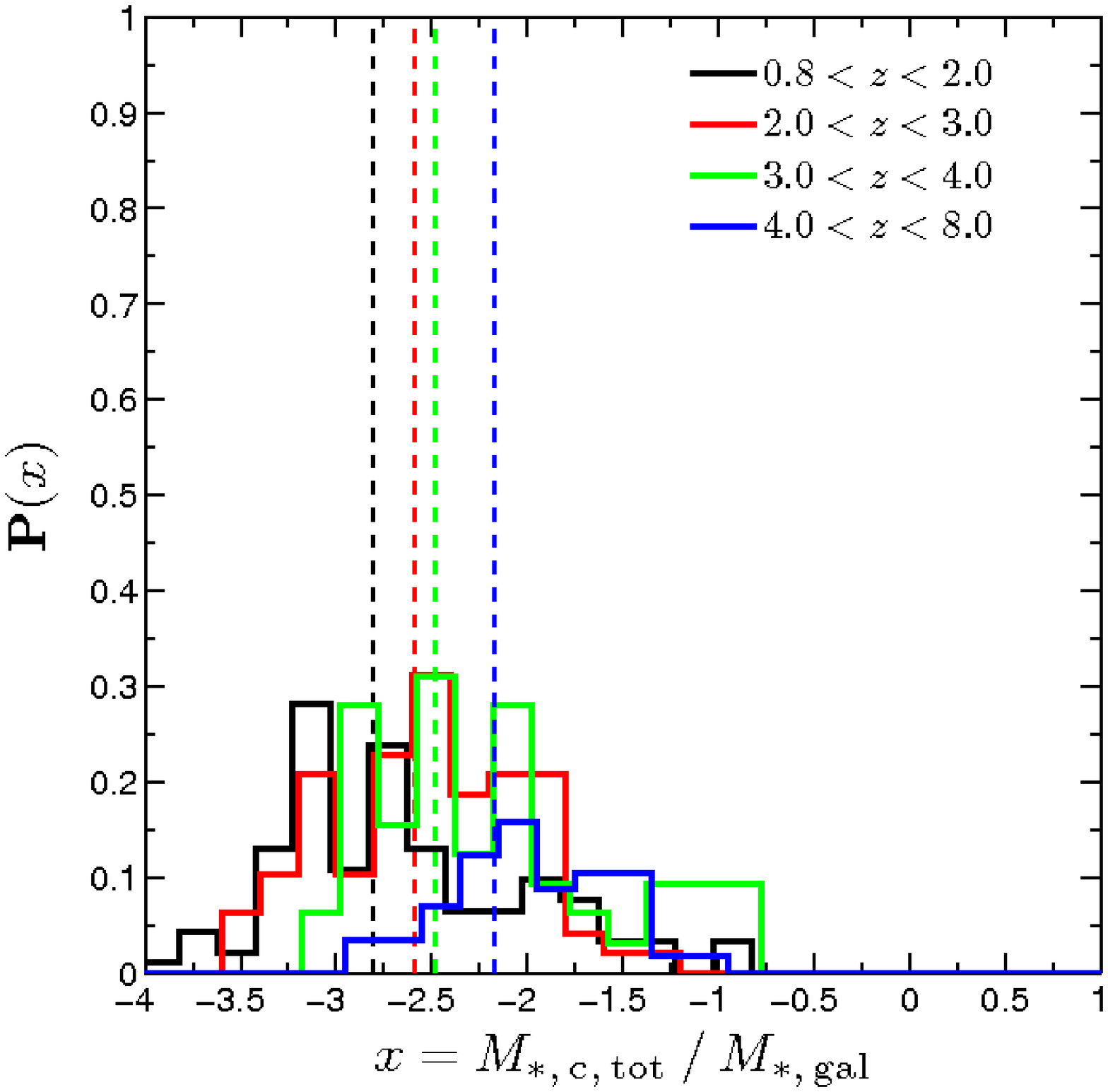}}\\
\caption{Same as \fig{clumpiness}, but using the ``observational", rather than the ``clean", 
sample of clumps.
}
\label{fig:clumpiness2} 
\end{figure*}

\smallskip
\fig{age_sfr2}-\fig{clumpiness2} are the same as \fig{age_sfr} - \fig{clumpiness}, 
but only using clumps with $s>-1.5$ rather than $s>-3$. The same mass threshold 
of $\Mc>10^7 \msun$ was used, as this is both our completeness limit for clumps, 
and roughly matches the lowest observable clump mass. We refer to this as our 
``observational" sample of clumps.

\newpage

\label{lastpage} 
 
\end{document}